\documentstyle[12pt]{article}
\newcommand{\be}[1]{\begin{equation}\label{#1}}
\newcommand{\ba}[1]{\begin{eqnarray}\label{#1}}
\newcommand{\ee}{\end{equation}}
\newcommand{\ea}{\end{eqnarray}}
\newcommand{\non}{\nonumber\\\rule{0pt}{30pt}}
\newcommand{\nona}[1]{\nonumber\\\rule{0pt}{#1pt}}
\newcommand{\num}{\\\rule{0pt}{20pt}}
\newcommand{\numa}[1]{\\\rule{0pt}{#1pt}}
\newcommand{\dis}{\displaystyle}
\newcommand{\eq}[1]{(\ref{#1})}
\newcommand{\tV}{\widetilde V}
\newcommand{\Ne}{{\cal N}_\epsilon}
\newcommand{\hG}{\hat G}
\newcommand{\hD}{\hat D}
\newcommand{\hB}{\hat B}
\newcommand{\hP}{\hat\Phi}
\newcommand{\hI}{\hat I}
\newcommand{\hPP}{\hat\Pi}
\newcommand{\hPo}{\hat\Phi^{(0)}}

\newcommand{\h}{\hat}
\newcommand{\Rho}{\hat{\cal R}}

\newcommand{\hAo}[1]{\hat{\cal A}_0(#1)}
\newcommand{\hAop}[1]{\hat{\cal A}_0^{\|}(#1)}
\newcommand{\hAoo}[1]{\hat{\cal A}_0^{\bot}(#1)}
\newcommand{\hW}{\hat{\cal W}}
\newcommand{\hi}{\hat\imath}
\newcommand{\hx}{\hat\chi}
\newcommand{\hb}{\hat b}
\newcommand{\odr}{|1\rangle}
\newcommand{\dvr}{|2\rangle}
\newcommand{\odl}{\langle1|}
\newcommand{\dvl}{\langle2|}

\renewcommand{\Im}{\mathop{\rm Im}}
\newcommand{\sign}{\mathop{\rm sign}}
\newcommand{\Tr}{\mathop{\rm Tr}}
\newcommand{\tr}{\mathop{\rm tr}}
\newcommand{\stint}{\int_{-\infty}^\infty}
\newcommand{\brad}{(0|}
\newcommand{\ketd}{|0)}
\newcommand{\freop}{\Bigl(\tilde I+\widetilde V\Bigr)}

\newcommand{\OD}{\hspace{4pt}\mbox{\raisebox{10pt}
            {$\scriptscriptstyle\circ$}}
            \hspace{-7.5pt}|1\rangle\hspace{4pt}}
\newcommand{\DV}{\hspace{4pt}\mbox{\raisebox{10pt}
            {$\scriptscriptstyle\circ$}}
            \hspace{-7.5pt}|2\rangle\hspace{4pt}}
\newcommand{\DOD}{\hspace{6pt}\mbox{\raisebox{10pt}
            {$\scriptscriptstyle\circ$}}
            \hspace{-8.5pt}\langle1|\hspace{2pt}}
\newcommand{\DDV}{\hspace{6pt}\mbox{\raisebox{10pt}
            {$\scriptscriptstyle\circ$}}
            \hspace{-8.5pt}\langle2|\hspace{2pt}}
\newcommand{\sss}{\hspace{-6pt}}

\newcommand{\ODo}{\hspace{4pt}\mbox{\raisebox{10pt}
             {$\scriptscriptstyle 1$}}
            \hspace{-7.5pt}|1\rangle\hspace{2pt}}

\newcommand{\DDVo}{\hspace{6pt}\mbox{\raisebox{10pt}
            {$\scriptscriptstyle 1$}}
            \hspace{-8.5pt}\langle2|\hspace{2pt}}

\newcommand{\DVd}{\hspace{4pt}\mbox{\raisebox{10pt}
            {$\scriptscriptstyle 2$}}
            \hspace{-7.5pt}|2\rangle\hspace{2pt}}
\newcommand{\DODd}{\hspace{6pt}\mbox{\raisebox{10pt}
            {$\scriptscriptstyle 2$}}
            \hspace{-8.5pt}\langle1|\hspace{2pt}}

\newcommand{\ODi}{\hspace{4pt}\mbox{\raisebox{10pt}
            {$\scriptscriptstyle i$}}
            \hspace{-7.5pt}|1\rangle\hspace{2pt}}
\newcommand{\DVi}{\hspace{4pt}\mbox{\raisebox{10pt}
            {$\scriptscriptstyle i$}}
            \hspace{-7.5pt}|2\rangle\hspace{2pt}}
\newcommand{\DODi}{\hspace{6pt}\mbox{\raisebox{10pt}
             {$\scriptscriptstyle i$}}
            \hspace{-8.5pt}\langle1|\hspace{2pt}}
\newcommand{\DDVi}{\hspace{6pt}\mbox{\raisebox{10pt}
            {$\scriptscriptstyle i$}}
            \hspace{-8.5pt}\langle2|\hspace{2pt}}
\newcommand{\rc}{\h\varrho_0^{(c)}}
\newcommand{\rct}{\h\varrho_0^{(c)T}}
\newcommand{\rca}[1]{\h\varrho_{0{#1}}^{(c)}}
\newcommand{\rcta}[1]{\h\varrho_{0{#1}}^{(c)T}}
\newcommand{\pa}{{\scriptscriptstyle\|}}
\newlength{\HFPP}       \HFPP5.4mm
\makeatletter
\@addtoreset{equation}{section}
\makeatother

\begin{document}

\begin{center}
{\large\bf On the Riemann-Hilbert approach to the asymptotic 
analysis of the correlation functions of the Quantum
Nonlinear Schr\"odinger equation. Non-free fermionic case.}

\vspace{2cm}

{\normalsize 
A.~R.~Its\raisebox{2mm}{{\scriptsize 1}} 
and 
N.~A.~Slavnov\raisebox{2mm}{{\scriptsize 2}}}\\
\vskip .5em
\vspace{1cm}
~\raisebox{2mm}{{\scriptsize 1}}
{\it Department of Mathematical Sciences IUPUI\\
402 N. Blackford st., Indianapolis IN 46202-3216}\\
itsa@math.iupui.edu
\vskip1.5em
~\raisebox{2mm}{{\scriptsize 2}}
{\it Steklov Mathematical Institute,\\
Gubkina 8, 117966,  Moscow, Russia\\}
nslavnov@mi.ras.ru
\end{center}

\vskip4pt

\begin{abstract}
\noindent
We consider the local field dynamical temperature correlation function of the Quantum
Nonlinear Schr\"odinger equation with the finite coupling constant. This correlation function admits a Fredholm determinant representation.
The related operator-valued Riemann--Hilbert
problem is used for analysing the leading term of the large time and long distance asymptotics of the correlation function.
\end{abstract}

\section{Introduction\label{I}}

This work continues the study of the 
 correlation functions of the 
Quantum Nonlinear Schr\"odinger equation (QNLS) out off free 
fermionic point which was originated in the papers \cite{KKS1}--\cite{KS2}. 

We consider the temperature correlation function of the local fields 
of the QNLS equation,
\be{Itempcorrel}
\langle\Psi(0,0)\Psi^\dagger(x,t)\rangle_T=
\frac{\tr\left( e^{-\frac HT}\Psi(0,0)\Psi^\dagger(x,t)\right)
}
{\tr e^{-\frac HT}}.
\ee
Here $\Psi(x,t),~\Psi^{\dagger}(x,t),~(x,t \in {\bf R})$
are the canonical Bose fields obeying the standard
equal-time commutation relations
\be{Icom}
[\Psi(x,t), \Psi^{\dagger}(y,t)]=\delta(x-y),
\ee
and acting in the Fock space as
\be{Ivac}
\Psi(x,t)|0\rangle =0,\qquad\qquad
\langle0|\Psi^\dagger(x,t) =0.
\ee
The evolution of the field $\Psi$ with respect to time $t$
is usual
\be{Ievol}
\Psi(x,t)= e^{iHt}\Psi(x,0)e^{-iHt}.
\ee
The Hamiltonian of the model  $H$  in \eq{Itempcorrel} and \eq{Ievol}
is equal to
\be{IHamilton}
{H}={\dis\int \,dx
\left({\partial_x}\Psi^{\dagger}(x)
{\partial_x} \Psi(x)+
c\Psi^{\dagger}(x)\Psi^{\dagger}(x)\Psi(x)\Psi(x)
-h \Psi^{\dagger}(x)\Psi(x)\right).}
\ee
Here $0<c<\infty$ is the coupling constant and $h$
is the chemical potential. The parameter $T$ in \eq{Itempcorrel} is a
temperature. 

The quantum Nonlinear Schr\"odinger equation describes 
one-di\-men\-si\-o\-nal Bose gas with delta-function interactions. 
The basic thermodynamic  equation of the model is the Yang--Yang 
equation \cite{YY} for the energy of an one-particle excitation 
$\varepsilon(\lambda)$ in thermal equilibrium

\be{IYY}
\varepsilon(\lambda)=\lambda^2-h-\frac{T}{2\pi}
\int_{-\infty}^{\infty}
\frac{2c}{c^2+(\lambda-\mu)^2}\ln
\left(1+e^{-\frac{\varepsilon(\mu)}{T}}
\right)d\mu.
\ee
The function $\varepsilon(\lambda)$ behaves as $\lambda^2-h$ at
$\lambda\to\infty$. It is positive for $\lambda\in R$, if
$h<0$, and it has two real roots $\varepsilon(\pm q)=0$, if
$h>0$.

It is worth mentioning also the integral equation for the total
spectral density of vacancies in the gas $\rho_t(\lambda)$:
\be{Irhot}
2\pi\rho_t(\lambda)=1+\int_{-\infty}^{\infty}
\frac{2c}{c^2+(\lambda-\mu)^2}\vartheta(\mu)
\rho_t(\mu)\,d\mu,
\ee
where
\be{IFermi}
\vartheta(\lambda)=\left(1+\exp\left[
\frac{\varepsilon(\lambda)}{T}\right]\right)^{-1}
\ee
is the Fermi weight. The value $\vartheta(\lambda)\rho_t(\lambda)$
defines the spectral density of particles in the gas. Due to the
properties of $\varepsilon(\lambda)$ the Fermi weight $\vartheta
(\lambda)$ decays as $\exp\{-\lambda^2/T\}$ at $\lambda\to\infty$.

\vskip .2in 

In this work we analyse a large time and long distance behavior of
the correlation function \eq{Itempcorrel}.

The asymptotic evaluation of the correlation functions is one
of the most challenging analytic problems in the theory of
exactly solvable quantum field  models.
For the case of zero temperature the leading asymptotic term can be 
found via conformal field theory \cite{BPZ}. The small temperature 
limit also can be considered in the framework of this approach, 
although, strictly speaking, increasing temperature destroys 
conformal properties of the model. In the present paper we develop 
the method which is based on the Fredholm determinant representations of the correlation functions  and which allows
to remove the small temperature restriction. 

A systematic exposition of the determinant representation method is 
given in \cite{KBI}. For the reader's convinience, we shall outline
the principal features of the scheme together with a brief
historical review concerning the asymptotic
analysis of the correlation functions.

\vskip .2in 

The determinant representation method is based on the remarkable fact that the
correlation functions of the $1+1$ exactly solvable quantum models 
can be represented as Fredholm
determinants of the integral operators $V$ acting in $L^{2}(\Sigma, 
d\lambda)$ and whose kernels, $V(\lambda_{1}, \lambda_{2})$, have the
following special form :
\begin{equation}\label{iio}
V(\lambda_{1}, \lambda_{2}) =
\frac{\sum_{j=1}^{N}e_{j}(\lambda_{1})f_{j}(\lambda_{2})}
{\lambda_{1}-\lambda_{2}},\qquad
\sum_{j} e_{j}(\lambda)f_{j}(\lambda)=0, 
\end{equation}
where functions $e_{j}(\lambda), f_{j}(\lambda)$, and, in fact, the
oriented contour of integration $\Sigma$, depend on the model under consideration. 
The first representation of this type was
obtained in \cite{L} for equal-time correlation functions in one-dimensional
impenetrable ($c=\infty $) bosons.  Later on, determinant formulae 
were derived for
a majority of  exactly solvable statistical mechanics and quantum field models (for the principal references we refer the reader
to the book \cite{KBI}). In particular, the Fredholm determinant 
for the time-dependent correlations in one-dimensional
impenetrable bosons was constructed 
in \cite{KS3}. A generalization of the results of \cite{L}  for non-free fermionic
case was obtained in \cite{KS4}. The determinant representation for the 
most general case of  non - free fermionic time-dependent temperature 
correlation function \eq{Itempcorrel} was found in \cite{KKS1}.

The determinant formulae can be used to obtain nonlinear differential
equations for  quantum correlation functions.  These nonlinear equations turn out to be {\it
classical} integrable systems.  Mor exactly, zero-temperature and equal-time two-point correlators are described by integrable ODEs of the Painlev\'e type (see \cite{BMW} -- \cite{MPS1}), while time-dependent and 
temperature or/and multi -
point correlation functions appear to be the $\tau$-functions of 
integrable PDEs 
(see \cite{JMMS}, \cite{MPS2} -- \cite{IIKS} ). 
It is interesting to notice (see e.g. \cite{IIKS1}, \cite{LLSS},
and section \ref{DE} of this paper) that if the quantum 
system is a result of quantization of a classical integrable system, as  
is the case for the QNLS model, then the integrable PDEs
describing the correlation functions belong to the underlying classical
integrable hierarchy.

The determinant representations for correlation functions can also be  
used to study their asymptotics.
In the zero-temperature case, a  comprehensive asymptotic analysis of the correlation functions
related to $XXO$ and impenetrable Bose gas models  has been
carried out in \cite{VT1}, \cite{VT2}, \cite{MPS1}, \cite{MPS2}, \cite{MT}.  
For the two-dimensional Ising
 model the analogous results were obtained earlier in \cite{MTW}
(see also \cite{BT} and \cite{T}).

A further development of the  determinant representation method was
achieved in the series of papers \cite{IIK1}, \cite{IIKS0} 
(see also \cite{KBI}).
The approach of \cite{IIK1}, \cite{IIKS0} is based on the use
of the Riemann-Hilbert method of the theory
of {\it classical} integrable systems for the asymptotic
evaluation of the Fredholm determinants describing the
correlation functions of the {\it quantum} integrable systems.
The Riemann-Hilbert method allowed to extend the mentioned
above zero-temperature results for the $XXO$ and impenetrable Bose
gas models to the general finite-temperature case (see \cite{IIK1},
\cite{IIKV}, \cite{IIKS}, and \cite{DZ3}).

The Riemann-Hilbert asymptotic scheme is the principal analytic tool
which is used in the present paper. 
In what follows, we describe its basic ideas
in some details. 

\vskip .2in
 
Riemann-Hilbert problem (RHP) appears in the theory of correlation functions
due to a simple yet
important fact  
that the resolvent kernel  corresponding to kernel (\ref{iio})
 can be explicitly evaluated in terms
of the solution of the matrix Riemann-Hilbert problem with the jump
matrix $G(\lambda)$ given by the equation 
(cf. \cite{IIK1}, \cite{IIKS0}, \cite{IIKS1}): 
$$
G_{jk}(\lambda)=\delta_{jk} + 2\pi i e_{j}(\lambda)f_{k}(\lambda).
$$
More exactly, let $\chi(\lambda)$ be a $N\times N$ matrix function
which solves the following {\it Riemann-Hilbert} problem:
\ba{defnormcond0}
&&{\dis
1.~\chi(\lambda)\to I,
\quad \lambda\to\infty,
\quad\mbox{(normalization condition)}}\num
&&{\dis
\label{defanalytprop0}
2.~\chi(\lambda)\quad\mbox{is analytic function of $\lambda$ 
if}~ \lambda\notin \Sigma,}\num
&&{\dis
\label{defjumpcond0}
3.~\chi_-(\lambda)=\chi_+(\lambda) G(\lambda),\qquad
\lambda\in \Sigma, \qquad\mbox{(jump condition)},}
\ea
where $\chi_{\pm}(\lambda)$ denote the $(\pm)$ - boundary values of the
function $\chi(\lambda)$ on $\Sigma$, i.e.
$$
\chi_{\pm}(\lambda) = \lim_{\lambda' \rightarrow \lambda}\chi(\lambda'),
\quad \lambda' \in (\pm) - \quad \mbox{side of}\quad \Sigma.
$$
Then, the resolvent kernel $R(\lambda_1, \lambda_2)$ corresponding to the
kernel (\ref{iio}) ($1 - R = (1+V)^{-1}$) is given by the following 
explicit formulae, 
\begin{equation}\label{riio} 
R(\lambda_{1}, \lambda_{2}) = 
\frac{\sum_{j=1}^{N}E_{j}(\lambda_{1})F_{j}(\lambda_{2})}
{\lambda_{1}-\lambda_{2}},
\end{equation}
$$
E_{j}(\lambda) = 
\sum_{k=1}^{N}(\chi_{+}(\lambda))_{jk}e_{k}(\lambda),\quad
F_{j}(\lambda) = \sum_{k=1}^{N}f_{k}(\lambda)
(\chi^{-1}_{+}(\lambda))_{kj}.
$$
 
The dynamical parameters, i.e. distance $x$ and time $t$, 
enter the jump matrix $G(\lambda)$ through the transformation,
\begin{equation}\label{gt}
G(\lambda) \rightarrow e^{D(\lambda; x,t)}G(\lambda)e^{-D(\lambda; x,t)},
\end{equation}
where the rational in $\lambda$ and linear in $x, t$ diagonal 
matrix function $D(\lambda; x,t)$ represents the dispersion law of 
the underlying classical model (e.g. $D(\lambda; x,t) = 
\mathop{\rm diag}(it\lambda^{2}- ix\lambda,\quad  
-it\lambda^{2}+ix\lambda )$ for the NLS equation).  It can be shown 
(see e.g. \cite{KBI}; see also \cite{TW}) that the logariphmic 
derivatives of $\det (1+V)$ with respect to $x$ and $t$ (and, in 
 fact, with respect to any other physical parameter) are explicitly 
expressible in terms of the resolvent $R$. Hence the asymptotic 
analysis of the original Fredholm determinat is reduced by 
(\ref{riio}) to the asymptotic analysis of the {\it oscillatory} 
Riemann-Hilbert problem (\ref{defnormcond0}-\ref{defjumpcond0}, 
\ref{gt}).

In the theory of classical integrable systems, the Riemann-Hilbert problems
 of the type (\ref{defnormcond0}-\ref{defjumpcond0}, \ref{gt}) represent
 solutions of the  Cauchy problems for integrable
PDEs. In this context, the development of the relevant apparatus for
 the asymptotic
analysis of the oscillatory matrix RHPs  was originated in  
1973-1977 in the  works \cite{Sh} -- \cite{AS1}. 
It was essentially completed
 (for a detailed historical
review see \cite{DIZ1}) in 1993 in the
 paper \cite{DZ} where a nonlinear analog
of the classical steepest descent method for oscillatory Riemann-Hilbert
 problems was suggested.

{\it The Deift-Zhou nonlinear steepest descent method} consists 
of three basic steps (see \cite{DZ}; see also \cite{DZ2} and \cite{DIZ1}): 
\begin{enumerate}
\item A deformation of the original jump contour $\Sigma$
 to the steepest descent (with respect to the dispersion 
exponent $D(\lambda; x,t)$) contours and asymptotic evaluation
of the solution away from the corresponding saddle points.
\item The use of the relevant Lax pair and  certain model Riemann-Hilbert problems to construct
a parametrix for the solution 
near the saddle points.
\item Assembling the above pieces into a {\it uniform} asymptotic
solution which makes it possible to justify the 
whole construction by standard estimates \cite{GK} of the
 theory of  singular integral operators on
the complicated contours.
\end{enumerate}
Each of these steps has its natural analog in the classical steepest
descent method for oscillatory contour integrals yet exploits much
more sophisticated technics and  analysis. In contrast with the 
classical steepest descent method, 
which is used
for asymptotic evaluation of oscillatory  {\it integral representations},
the nonlinear steepest descent method deals with a special type of oscillatory 
(singular) {\it integral equations}. In fact, the RHP 
(\ref{defnormcond0}-\ref{defjumpcond0}) is equivalent to the following
singular integral equation:
\begin{equation}\label{sie}
\chi_{+}(\lambda) = I +
\frac{1}{2\pi i}\int_{\Sigma}\chi_{+}(\mu)(I-G(\mu))\frac{d\mu}{\mu-
\lambda_{+}}\, .
\end{equation}
The nonlinear steepest descent method provides a regular way of
finding the proper transformation (which is highly 
nontrivial and virtually impossible to be seen
directly!) of the original
singular equation (\ref{sie}) to an equivalent one with uniformly
small kernel.

Some of the principal ideas  involved in the steps 1, 2 of the
nonlinear steepest descent method
(e.g. explicit solutions for the
model Riemann-Hilbert problems) go back to
the earlier works \cite{Man} and \cite{I1}. 
These earlier versions of the
Riemann-Hilbert approach
had already been successfully exploited in the asymptotic
analysis of the various temperature correlation functions (see
\cite{IIK1}, \cite{IIKV}, \cite{IIKS}). The
use of the nonlinear steepest descent method increases considerably
the power of the original scheme of \cite{IIK1}, \cite{IIKS0}.
In partricular, in \cite{DZ3} the use of the nonlinear steepest descent method
allowed  to compute the long-time asymptotics of the autocorrelation
function of the transverse Ising chain at the critical magnetic field
for the first time at finite temperature. The method has also produced
solutions for some long-standing problems in the
theory of random matrices and orthogonal
polynomials  (see e.g. \cite{DIZ2}, \cite{BI},
and \cite{DKMVZ}).

\vskip .2in
The $XXO$ magnet and impenetrable Bose gas are free fermionic models.
The non-free fermionic case is much more complicated because of 
several reasons. First, the Riemann-Hilbert problems, which
 describe the corresponding Fredholm
determinants, become operator-valued (cf. \cite{KBI}, \cite{KS1}): the
matrix elements $G_{j,k}(\lambda)$ turn into integral operators
$\hat G_{j,k}(\lambda)$ acting in an auxilary $L_{2}$  space
($G_{j,k}(\lambda) \rightarrow G_{j,k}(\lambda|u,v)$).
In other words, an  infinite-dimensional environment,
absent in the free fermionic problems, arises. This transforms
 the associated Lax pairs and classical nonlinear
PDEs into their non-Abelian analogies and obviously provides new significant difficulties for 
the asymptotic analysis. In fact, up to now the only attempt to solve an 
operator-valued RHP, which is related to correlation function of $XXZ$ magnet,
had been done in \cite{FIK}.

The second difficulty is more subtle than the first one, and it is
related to the presence of another 
infinite-dimensional context in the non-free fermionic  problems;
 this time -- of the quantum field
nature.  The fact of the 
matter is that out off free fermionic point the correlation functions
can not be presented directly in the determinant form. Due to 
existence of non-trivial $S$-matrix, one needs to introduce \cite{Ko} auxiliary 
quantum operators --- Korepin's dual fields --- in order to find such a 
representation. As a result, the determinants obtained depend on 
these operators, and the correlation functions are equal to the vacuum 
expectation values of the Fredholm determinants in an
 auxiliary Fock space. In particular, for the correlation function
\eq{Itempcorrel} the following equation  takes place (cf. \cite{KKS1}),
\be{main1}
\langle\Psi(0,0)\Psi^\dagger(x,t)\rangle_T= const \cdot e^{-iht}
\brad{\cal B}\Bigl([\psi,\phi_A,\phi_D],x,t\Bigr) \ketd\, ,
\ee
where the factor {\it const} only
depends on the temperature, coupling constant and
chemical potential, and ${\cal B}$ is an operator acting in the auxilary 
Fock space with the vacuum vector $\ketd$. It involves the Fredholm determinant 
which functionally
depends on the three basic quantum operators (dual fields) $\psi(\lambda),
\phi_A(\lambda)$, and $\phi_D(\lambda)$. 
The exact definitions of the quantum operator ${\cal B}$ is given in
appendix A. The dual
fields  $\psi(\lambda),\phi_A(\lambda),\phi_D(\lambda)$ are defined
in section \ref{B} by equations (\ref{Bdualfields}) - (\ref{Bcommutators}).
  
The operator-valued RHPs allow to evaluate the Fredholm 
determinants, but not their vacuum mean values. The two
infinite-dimensional contexts, i.e.
the operator nature of the RHP and the presence of the
dual quantum fields, are completely unrelated. 
 Therefore, in order to 
find the large time and long distance asymptotics of a non-free
fermionic correlation 
function one needs to be sure that the asymptotics of the mean value 
is equal to the mean value of the asymptotics.

It had been shown in \cite{S1} that a naive asymptotic analysis of the 
Fredholm determinants containing dual fields does not provide a
satisfactory result, i.e. the asymptotics is not uniform with
respest to the avaraging over the dual fields. Thus, the problem
arises --- to obtain the asymptotic description of the 
determinant related to \eq{Itempcorrel} which would be stable with respect to 
the procedure of 
averaging over the dual fields. This is the problem which we deal with 
in the present paper.
\vskip.2in

Let us describe now the content of the paper. 

In section 
\ref{B}, following \cite{KKS1} - \cite{KS1}, we give
the basic formul\ae~and definitions 
concerning the non-free fermionic version of the
 determinant representation -- Riemann-Hilbert approach to
the asymptotic analysis of the correlation function \eq{Itempcorrel}.

 In section \ref{F}, following \cite{KS1}, we formulate the
central object of the analysis, i.e. the operator-valued RHP 
 associated with the correlation function \eq{Itempcorrel}
(see (\ref{Fovrhp}) and (\ref{Fjumpmatreg}) below). The jump contour
$\Sigma$ of the RHP coincides with the real line, and the corresponding
jump operator $\h G(\lambda)$ is realized as a $2\times 2$ matrix whose
entries $\h G_{jk}(\lambda)$ are the integral operators 
in an auxilary $L_{2}(-\infty, \infty)$
space. A principal feature of this RHP is that the operators
$\h G_{jk}(\lambda)$ have the following special structure:

\begin{equation}\label{jumpop}
\h G_{jk}(\lambda) = \hi \, \delta_{jk} +
{|j\rangle}(G_{jk}-\delta_{jk}){\langle k|}, \quad j,k =1,2,
\end{equation}
\vskip .1in
\noindent
where $\hi$ is the identical operator in $L_{2}(-\infty, \infty)$,
i.e. its kernel is a delta - function: $\hi (u,v) = \delta(u-v)$. 
The symbols $|j\rangle \equiv |j, u\rangle$
and $\langle k| \equiv \langle k, v| $ denote certain elements of $L_{2}(-\infty, \infty)$
(see definition (\ref{Fdefvectors}) below) satisfying the normalization condition,
\be{norm0}
\odl1\rangle\equiv\stint\langle1,u|1,u\rangle\,du=\dvl2\rangle
\equiv\stint\langle2,u|2,u\rangle\,du=1.
\ee
The numerical matrix $G(\lambda)$ in (\ref{jumpop}) is closely related to the jump matrix
of the free fermionic impenetrable Bose gas model (cf. \cite{IIKV}). 

Equations similar to (\ref{jumpop}) take place for other
 non-free fermionic  correlation functions as well, and they have
a very important algebraic meaning. In fact, as it can be easiely
verified, formulae (\ref{jumpop}) and (\ref{norm0}) define a representation,
\begin{equation}\label{repr}
G \rightarrow \hG \equiv \hat{\cal A}(G),
\end{equation}
of the group ${\bf GL}(2,\rm C)$ in the group of Fredholm invertible 
operators in $L_{2}((-\infty, \infty), {\rm C}^{2})$. It also can be shown
(see \cite{IKW}) that representation  (\ref{repr}) preserves 
determinants, i.e.
\begin{equation}\label{det}
\det G = \det \hG .
\end{equation}
The link (\ref{repr}) between the non-free fermionic jump operators 
and  their
free fermionic counterparts   
was first noticed by Korepin, and it has already proved very usefull
in the asymptotic analysis of the non-free fermionic correlators
(see \cite{IKW} and \cite{FIK}). The mapping (\ref{repr})
plays a crucial role
in the following 
sections where we develop the operator-valued version of the 
Deift-Zhou nonlinear steepest descent method.

The main obstacle  in taking the full advantage of the
group-representation nature of equation (\ref{jumpop}) is that the vectors
$|j\rangle $ and $\langle k|$ {\it depend on the parameter $\lambda$}, i.e.
$|j\rangle \equiv |j, u, \lambda \rangle$
and $\langle k| \equiv \langle k, v, \lambda| $ 
(see  (\ref{Fdefvectors}) below); moreover,
they become singular for complex $\lambda$. Therefore,  
one can not solve a non-free fermionic (operator ) Riemann-Hilbert problem by just applying
the representation (\ref{repr}) to the solution of the corresponding
free fermionic (matrix) Riemann Hilbert problem. Nevertheless, this
representation helps to perform the first step of the nonlinear
steepest descent method (the contour deformation) 
and evaluate the leading term of
the asymptotic solution of the RHP. This is done in sections \ref{NCP}, \ref{PCP}
where we closely follow the methodology of \cite{FIK}. The leading
asymptotic term we found coincide with the one obtained
earlier in \cite{S1} via the direct asymptotic analysis of the
Fredholm determinant. 

The objective of sections \ref{L}--\ref{R}
is an order $t^{-1/2}$ correction. This is the second step 
of the nonlinear steepest descent method. 
In the free-fermionic case (see e.g. \cite{DZ2} and \cite{DIZ1})
this step constitutes the reduction of the deformed RHP to
a model problem associated with the saddle point. The model
problem is then solved explicitly in terms of the
parabolic cylinder functions (see again  \cite{DZ2}
 and \cite{DIZ1}; see also \cite{I1}).
In our case, the model problem is an operator-valued version
of the classical model problem. In fact, the jump operator
of the model problem is the image of a certain jump matrix, 
closely related to the corresponding free fermionic
model jump matrix, under the {\it modified} mapping $\hat {\cal A}_{0}$.
The latter is the representation $\hat {\cal A}$ with the vectors
$|j\rangle $ and $\langle k|$ replaced by their values evaluated
at the saddle point $\lambda_{0}$,

\begin{equation}\label{vecmod1}
{|\stackrel{\scriptscriptstyle\circ}{j}, u\rangle} = |j, u, \lambda_{0}\rangle,
\qquad
{\langle\stackrel{\scriptscriptstyle\circ}{k}, v|} = \langle k, v, \lambda_{0}|.
\end{equation}
The representation $\hat {\cal A}_{0}$ does not depend 
on  $\lambda$ and hence can be used to transform a free
fermionic matrix model solution into the operator model
solution. The transfortmation though is not quite straightforward,
but it allows eventually (see sections \ref{A}, \ref{R})
to solve the model RHP explicitly 
(more exactly, up to the inversion of an integral operator
which is independent on $\lambda$, $x$, and $t$). 
In achieving this result, the following `operator - indexed' 
generalization $D_{\hat \nu}(\xi)$
of the classical parabolic cylinder functions $D_{\nu}(\xi)$,
 
\begin{equation}
D_{\hat \nu}(\xi) = D_{0}(\xi) \left( \hi -
|\stackrel{\scriptscriptstyle\circ}{j}\rangle 
\langle\stackrel{\scriptscriptstyle\circ}{j}|\right)
+ |\stackrel{\scriptscriptstyle\circ}{j}\rangle 
\langle\stackrel{\scriptscriptstyle\circ}{j}|\, 
D_{\nu}(\xi),
\qquad \hat \nu =
\nu |\stackrel{\scriptscriptstyle\circ}{j}\rangle 
\langle\stackrel{\scriptscriptstyle\circ}{j}|\, ,
\end{equation}
plays an important role.

The order $t^{-1/2}$ correction is a main threshold
in the asymptotic analysis of the temperature correlation
functions (cf. e.g. \cite{IIKV}). After it is passed,
one can, in principle, obtain a total asymptotic expansion
using the elementary algebra only, just substituting
a proper asymptotic series into the nonlinear PDEs
associated with the correlation function under consideration.
In our case, the relevant nonlinear PDE  is the mentioned 
above non-Abelian version of the classical NLS equations 
(see (\ref{Bopeq}) below).
In section \ref{DE} we use this non-Abelian NLS for exact
evaluation of a few terms next to the order $t^{-1/2}$
term and for evalution of the pre-exponential factor
in the asymptotics of the Fredholm determinant. The
non-Abelian NLS also allow us to extract a valuable information about
the full asymptotic expansion. We analyse this information in section
\ref{ID} and conclude that to make the asymptotic expansion
obtained stable with respect to the dual field avaraging a certain 
modification of our construction is needed. The modification
concerns the equation determing the saddle point $\lambda_{0}$,
and it is made in the last section \ref{MA}.

\vskip .2in 

As the {\bf main result} of the paper, we suggest the
following asymptotic formulae
for the quantum operator ${\cal B}$ in the r.h.s. of equation (\ref{main1}),

\ba{main2}
&&\hspace{-11mm}
{\cal B}={\dis
C_{-}(\phi_{D},\phi_{A}|\lambda_0,
T, h, c)(2t)^{-\frac{(\nu(\Lambda)+1)^2}{2}}
e^{\psi(\Lambda)+it{\Lambda}^{2}-ix\Lambda}}\non
&&\hspace{-6mm}{\dis\times\exp\left\{
\frac{1}{2\pi}\stint\biggl(x-2\lambda
t+i\psi'(\lambda)\biggr)\sign(\Lambda-\lambda)\right.}\non
&&\hspace{-6mm}{\dis\left.\vphantom{\stint}\times
\ln\Bigl\{1-\vartheta(\lambda)
\biggl(1+e^{\phi(\lambda)\sign(\lambda-\Lambda)}\biggr)\Bigr\}
\,d\lambda\right\}\left(1+{\cal O}
\left(\frac{\ln^{2}(t)}{t}\right) \right),}
\ea
$$
t\rightarrow \infty, \qquad \frac{x}{2t} \equiv \lambda_{0} = O(1),
$$
for the case of  negative chemical potential $h$, and

\ba{main3}
&&\hspace{-11mm}
{\cal B}={\dis
C_{+}(\phi_{D},\phi_{A}|\lambda_0,
T, h, c)(2t)^{-\frac{\nu^{2}(\Lambda)}{2}}
e^{\psi(\Lambda_{1})+it{\Lambda_{1}}^{2}-ix\Lambda_{1}}}\non
&&\hspace{-6mm}{\dis\times\exp\left\{
\frac{1}{2\pi}\int_{\Gamma}\biggl(x-2\lambda
t+i\psi'(\lambda)\biggr)\sign(\Lambda-\Re \lambda)\right.}\non
&&\hspace{-6mm}{\dis\left.\vphantom{\stint}\times
\ln\Bigl\{1-\vartheta(\lambda)
\biggl(1+e^{\phi(\lambda)\sign(\Re \lambda-\Lambda)}\biggr)\Bigr\}
\,d\lambda\right\}\left(1+{\cal O}
\left(\frac{1}{\sqrt t}\right) \right),}
\ea
$$
t\rightarrow \infty, \qquad \frac{x}{2t} \equiv \lambda_{0} = O(1),
$$
for the case of positive chemical potential $h$.

A few remarks explaining the notations which are used in 
equations (\ref{main2}), (\ref{main3}) and the meaning of
the equations themselves are needed:

\begin{enumerate} 
\item The function $\vartheta(\lambda)$ is the Fermi
weight (\ref{IFermi}).
\item The dual field $\phi(\lambda)$ equals the difference
$\phi_{A}(\lambda) - \phi_{D}(\lambda)$. 
\item The exponent $\nu(\Lambda)$ is defined
by the formula,
$$
\nu(\Lambda) = -\frac{1}{2\pi i}\ln \left[ \left( 1 - \vartheta(\Lambda)
(1+e^{-\phi(\Lambda)})\right)\left( 1 - \vartheta(\Lambda)
(1+e^{\phi(\Lambda)})\right)\right], 
$$ 
\item $\Lambda_{1}$ and $\Lambda$ are the roots of the equations
\begin{equation}\label{Lambda1}
1 - \vartheta(\Lambda_1)
(1+e^{-\phi(\Lambda_1)}) = 0,\quad h>0,
\end{equation}
and
\begin{equation}\label{Lambda}
\Lambda = \lambda_{0} + \frac{i}{2t}\psi'(\Lambda),
\end{equation}
respectively. Since $\Lambda$ is not necessary real one should
understand the integrals \eq{main2}, \eq{main3} as
\be{Esign}
\int_{-\infty}^\infty
F\Bigl(\sign(\Lambda-\lambda),\lambda\Bigr)\,d\lambda=
\int_{-\infty}^\Lambda
F(1,\lambda)\,d\lambda+\int_\Lambda^{\infty}
F(-1,\lambda)\,d\lambda.
\ee
\item The undetermined factors $C_{\pm}(\phi_A,\phi_D | \lambda_{0},
T, h, c)$ are 
functionals of the dual fields $\phi_A$, $\phi_D$,
and they are functions of 
the temperature $T$, the chemical potential $h$, and
the coupling constant $c$. They do not depend on the
dual field  $\psi$, and they only depend on distance $x$
and time $t$ through the ratio $\lambda_{0} = x/2t$.
\item All the commutation relations 
involving the dual fields $\phi_{A}(\lambda)$, $\phi_{D}(\lambda)$,
and $\psi(\lambda)$ are trivial (see (\ref{BAbel})).
Therefore, when  deriving  (\ref{main2}) and (\ref{main3})
we treat the dual fields as the complex-valued functions
analytic  in the strip $|\Im \lambda |
< c/2$; the functions $\exp{\phi_{A,D}}$ and  $\exp{\psi}$
are supposed to be bounded in the strip. 
It is also assumed
that equation (\ref{Lambda1}) has no real solutions if $h<0$,
and it has exactly two real roots, i.e. $\Lambda_1$, $\Lambda_2$,
if $h>0$. Moreover, it is supposed that 
\begin{equation}\label{condLambda}
\Lambda_{1} < \Lambda_{2} < \Lambda,
\end{equation}
and that the same inequalities, in the case $h>0$,
 take place for the roots
of the function $1 - \vartheta(\lambda)
(1+e^{\phi(\lambda)})$.
These assumptions are justified in sections \ref{B}, \ref{F} and \ref{PCP}.
Accordingly, the contour of integration $\Gamma$ in (\ref{main3})
is choosen as it is indicated in Figure 2 below, so that the
integral in (\ref{main3}) makes sense.
\item The point $\Lambda$, which plays the role of the `shifted' saddle
point, is defined by equation (\ref{Lambda})
as its root which approches $\lambda_{0}$ as $t\rightarrow
\infty$. In other words, $\Lambda - \lambda_{0} = {\cal O}
(t^{-1})$. Nevertheless, we do not replace $\Lambda$ by $\lambda_{0}$
in the asymptotic formulae (\ref{main2} - \ref{main3}). As it
is explained in sections \ref{ID} and \ref{MA}, to make the
asymptotics of $\cal B$ compatible with the avaraging over
the dual fields we need to trace the presence
of the field $\psi$ in all the corrections to the leading
terms indicated in (\ref{main2} - \ref{main3}). In fact, we
need all the corrections to be of the order ${\cal O}(\psi^{m}/
t^{n})$ with $0<m<n$. The replacement $\Lambda \mapsto \lambda_{0}$,
if made, would introduce  the corrections of the order ${\cal O}(\psi^{m}/
t^{n})$ with $m\geq n$ which produce the positive
powers of $t$ after the avaraging in
(\ref{main1}) (see sections \ref{ID} and \ref{MA} for
more details). In other words, when the operator nature
of equation (\ref{Lambda}) is taking into account one has
to keep in mind that, although the factor $i/2t$ is small,
the quantity $\psi'(\lambda)$ is an unbounded operator of
the derivative type.
\item Because of the quantum operator nature of the
dual fields $\phi_{A}(\lambda)$, $\phi_{D}(\lambda)$,
and $\psi(\lambda)$, the r.h.s. of  equations (\ref{main2})
and (\ref{main3}) have, strictly speaking, a symbolic meaning.
One needs  the method of avaraging of the
functionals of the dual fields  developed in \cite{KS2}
in order to evaluate the mean values of
the objects indicated in (\ref{main2} - \ref{main3}).
The results of these calculations, which would finalize
the asymptotic analysis of the leading terms of
the correlation function \eq{Itempcorrel}, will be
given in the forthcoming paper.
\end{enumerate}

\vskip .2in

We conclude the introduction by emphasizing that in
this paper we do not address the questions related
to the third step
of the nonlinear steepest descend method,
i.e. the questions related to the rigorous
justification of the asymptotic
equations (\ref{main2} - \ref{main3}). Some of these
questions, including the rigorous setting 
and the general theory of the operator-valued Riemann-
Hilbert problems arising in the non-free fermionic
models, are still under the investigation.
The purpose of this paper is to  show that, using
the operator-valued Riemann-Hilbert
problems, it is posible to carry out an effective 
asymptotic analysis of the general
time-dependent temperature non-free
fermionic correlation functions based on their determinant
representation.

\section{The basic formul\ae~and definitions\label{B}}

This sections contains the basic definitions and notations, which
are used in the paper. The reader may find the details in the papers
\cite{KKS1}, \cite{KS1}. Here we restrict our selves with only
necessary formul\ae.

Consider the operators of the following form
\be{hatoperators}
\hat A\equiv \hat A(u,v) = \left(
\begin{array}{cc}
{\dis \hat A_{11}(u,v)} &{\dis  \hat A_{12}(u,v)}\num
{\dis  \hat A_{21}(u,v)} &{\dis   \hat A_{22}(u,v)}
\end{array}\right)
\ee
The entries of this $2\times 2$ matrix  are integral kernels
$A_{jk}(u,v)$ acting on functions from the $L_2(-\infty,
\infty)$ space. We shall mark these operators (as well as their
entries) with the symbol `hat'. Also, we will 
frequently use the equation,
$$
\hat M \equiv \hat M(u,v),
$$
to indicate that $\hat M(u,v)$ is the kernel
of an operator $\hat M$.

 The product of two operators of this
type is defined as
\be{BProd}
(\hat A\hat B)_{jk}(u,v)=\sum_{l=1}^{2}\stint
\hat A_{jl}(u,w)\hat B_{lk}(w,v)\,dw.
\ee
We shall also consider the products of separate matrix elements:
\be{Bprod}
(\hat A_{jk}\hat B_{lm})(u,v)=\stint
\hat A_{jk}(u,w)\hat B_{lm}(w,v)\,dw.
\ee
The definition of trace of `hat'-operators is standard
\be{Trace}
\Tr\hat A=\tr\hat A_{11}+\tr\hat A_{22}=
\stint\left(\hat A_{11}(u,u)+\hat A_{22}(u,u)\right)\,du,
\ee
where
\be{trace}
\tr\hat A_{jk}=\stint\hat A_{jk}(u,u)\,du.
\ee

The main object of study in this paper is an operator-valued
Riemann--Hilbert problem. Here we give the general formulation
of this problem. 

The RHP consists of the following: find an operator $\hx(\lambda)$
depending on complex
parameter $\lambda$ and satisfying the following conditions
(cf. (\ref{defnormcond0}) -- (\ref{defjumpcond0})):
\ba{defnormcond}
&&{\dis
1^{(0)}.~\hx(\lambda)\to\h I,
\quad \lambda\to\infty,
\quad\mbox{(normalization condition)}}\num
&&{\dis
\label{defanalytprop}
2^{(0)}.~\hx(\lambda)\quad\mbox{is analytical function of $\lambda$ 
if}~ \lambda\notin R,}\num
&&{\dis
\label{defjumpcond}
3^{(0)}.~\hx_-(\lambda)=\hx_+(\lambda)\h G(\lambda),\qquad
\lambda\in R, \qquad\mbox{(jump condition)}.}
\ea
Here $\h I$ is the identity operator
\be{hatident}
\h I=\left(
\begin{array}{cc}
1&0\\0&1
\end{array}\right)\delta(u-v).
\ee
Operators $\hx_\pm(\lambda)$ are boundary values of the function
$\hx(\lambda)$  from the left (from up half-plane) and
from the right (from low half plane) of the real axis respectively
(compare again with (\ref{defnormcond0}) -- (\ref{defjumpcond0})).
$\h G(\lambda)$ is a given operator. A particular
choice of this operator which corresponds to
the correlation function \eq{Itempcorrel} will be
described in detail in the next section.

 Following the tradition we call
the operator $\h G(\lambda)$ `jump matrix', although this matrix evidently
is infinite dimensional. We hope that this terminology will not cause a
confusion.

Although the RHP $(0)$ looks like $2\times2$ matrix RHP, here
we deal with infinite-dimensional integral operators. In more
detail, the jump condition $3^{(0)}$ can be written as
\be{jumpconddet}
\hx_{jk-}(\lambda|u,v)=
\sum_{l=1}^{2}\stint\hx_{jl+}(\lambda|u,w)\h G_{lk}(\lambda|w,v)\,dw,
\qquad
\lambda\in R.
\ee
The analyticity of the operator $\hx (\lambda)$ is understood as
the $\lambda$-analyticity of its kernel $\hx (\lambda|u,v)$. 
 All the limits and expansions involving $\hx (\lambda)
\equiv \hx (\lambda|u,v)$
are supposed to be uniform with respect to $u$ and $v$. We
will not go further in the specification of the setting
of the RHP (\ref{defnormcond}) -- (\ref{defjumpcond}).  As it has
already been indicated in the introduction, in this paper we
do not discuss the general theory of the
operator-valued RHPs. In what follows, we assume
that the unique solutions of the operator-valued RHPs we are
dealing with always exist and satisfy all the natural
properties. For instance, we shall assume that
if $\hat G(\lambda)$ is a Fredholm operator, i.e. the determinant
$\det \hat G(\lambda)$ exists, then $\hx (\lambda)$ is a
Fredholm operator, and its determinant $\det \hx(\lambda)$
is a unique solution of the following {\it scalar} RHP,
\ba{detnormcond}
&&{\dis
1.~\det \hx(\lambda)\to 1,
\quad \lambda\to\infty,}\num
&&{\dis
\label{detanalytprop}
2.~\det \hx(\lambda)\quad\mbox{is analytical function of $\lambda$ 
if}~ \lambda\notin R,}\num
&&{\dis
\label{detjumpcond}
3.~\det \hx_-(\lambda)=\det \hx_+(\lambda)\det \h G(\lambda),\quad
\lambda\in R .}
\ea
In particular, this means that the function $\det \hx(\lambda)$
is given by the explicit formula,
\begin{equation}\label{detcauchy}
\det \hx(\lambda) =
\exp \left\{-\frac{1}{2\pi i}\int_{-\infty}^{\infty}
\frac{d\mu}{\mu - \lambda}\ln \det \h G(\mu)\right\}\, ,
\end{equation}
subject the zero-index solvability condition,
\be{scalsolv}
\arg \det \hat G(\lambda) \to 0, \quad \lambda \to \pm \infty\,.
\ee
\vskip .2in

The asymptotic expansion of the solution of the RHP  at 
$\lambda\to\infty$ will play the central role 
\be{Basyexpan} 
\hx(\lambda)=\h I+\frac{\hb}{\lambda}+\frac{\h c}{\lambda^2} 
+\dots~.  
\ee 
The first two coefficients of this expansion $\hb$ and $\h c$ allow 
one to reconstruct the Fredholm determinant, i.e the
$\tau$-function corresponding to the solution $\hx(\lambda)$,
which in turn directly 
related to the correlation function of the local fields
\eq{Itempcorrel}. This correlation function can be presented as
\be{Bdetrepprel}
\langle\Psi(0,0)\Psi^\dagger(x,t)\rangle_T= const\cdot e^{-iht}
\brad{\cal B}\Bigl([\psi,\phi_A,\phi_D],x,t\Bigr) \ketd.
\ee
We have extracted in this formula the  factor $\exp\{-iht\}$, which
trivially depends on $t$, and a constant, which does not depend on
$x$ and $t$. Strictly speaking this `constant' depends on
the temperature, coupling constant and chemical potential. 
The most important object in \eq{Bdetrepprel}             
is ${\cal B}\Bigl([\psi,\phi_A,\phi_D],x,t\Bigr)$, which, in 
particular, includes a Fredholm determinant, and which is 
an operator in auxiliary Fock space.  This operator is a function of 
$x$ and $t$ and it functionally depends on three quantum operators 
(dual fields) $\psi(\lambda)$, $\phi_A(\lambda)$ and 
$\phi_D(\lambda)$.  The last ones act in the auxiliary Fock space 
having the vacuum vector $\ketd$ and dual vector $\brad$ (one should 
not confuse the auxiliary vacuum vector $\ketd$ with the vector 
$|0\rangle$ in \eq{Ivac}). Thus, the correlation function is 
proportional to the vacuum expectation value of the operator $\cal 
B$. The detailed definition and the properties of the dual fields will 
be given later in this section.  First we want to focus our attention on the 
relationship between the operator $\cal B$ and the operator-valued 
RHP.

The following representation for the operator $\cal B$ had been found
in \cite{KKS1}, \cite{KKS2}
\be{Bdetrep}
{\cal B}\Bigl([\psi,\phi_A,\phi_D],x,t\Bigr)=
\det\left(\tilde{I}+\widetilde{V}\right)
\cdot \stint\hb_{12}(u,v)\,du dv. 
\ee
We call \eq{Bdetrep} `determinant representation', since the
r.h.s. of this formula is proportional to the Fredholm determinant of 
the linear integral operator $\tilde{I} +\widetilde{V}$. The detailed 
description of this integral operator is given in Appendix A.

The results of the paper \cite{KS1} allow one to express the 
Fredholm  determinant and factor $\hb_{12}(u,v)$  in terms of the 
solution of an operator-valued RHP $(0)$ with a certain jump matrix 
$\h G(\lambda)$. In fact, the operator $\hb_{12}$ in (\ref{Bdetrep})
 is the corresponding 
entry of the operator coefficient $\hb$ of the asymptotic expansion
\eq{Basyexpan}. The logarithmic derivatives of $\det\freop$ 
with respect to distance $x$ and time $t$ also can be written in terms
of the coefficients $\hb$ and $\h c$ from \eq{Basyexpan}:
\be{Blogdir1}
\begin{array}{l}
{\dis \partial_x\log\det\freop=
i\tr\hb_{11}, }\nona{20}
{\dis \partial_t\log\det\freop=
i\tr(\hat c_{22}-\hat c_{11}).}
\end{array}
\ee

The second logarithmic derivatives depend on the `matrix elements'
$\hb_{12}$ and $\hb_{21}$ only:

\be{Blogdir2}
\begin{array}{l}
{\dis
\partial_x\partial_x\log\det\freop=
-\tr\bigl(\hb_{12}\hb_{21}\bigr),}\nona{20}

{\dis
\partial_t\partial_x\log\det\freop=
i\tr(\partial_x \hb_{12}\cdot \hb_{21}
-\partial_x \hb_{21}\cdot \hb_{12}).}
\end{array}
\ee
Finally these two operators satisfy a non-Abelian generalization of the
classical Nonlinear Schr\"o\-din\-ger equation
\be{Bopeq}
\begin{array}{c}
{\dis -i\partial_t\hb_{12}
=-\partial_x^2\hb_{12}+2\hb_{12}\hb_{21}\hb_{12},}\non
{\dis i\partial_t\hb_{21}
=-\partial_x^2\hb_{21}+2\hb_{21}\hb_{12}\hb_{21}.}
\end{array}
\ee
Recall that one should understand the products of operators
$\hb_{12}$ and $\hb_{21}$ in the r.h.s. of \eq{Blogdir2} and
\eq{Bopeq} in the sense \eq{Bprod}

Thus, we have described $\cal B$ in terms of the solution of the
operator-valued RHP (\ref{defnormcond} - \ref{defjumpcond})
(the explicit expression for the jump matrix $\h G(\lambda)$
 will be  given in the 
next section). Equations \eq{Blogdir1} define
${\cal B}$ up to a constant factor. Equations \eq{Blogdir2},
\eq{Bopeq} will help  us to analyse the higher corrections
to the leading asymptotics of $\hb_{12}$ and $\hb_{21}$.

Let us turn now to the auxiliary quantum operators---dual fields.
These  operators  $\psi(\lambda)$, $\phi_{D}(\lambda)$ and
$\phi_{A}(\lambda)$ (originally the last two fields were denoted as
$\phi_{D_1}(\lambda)$ and $\phi_{A_2}(\lambda)$) were introduced in
\cite{KKS1} in order to remove two-body scattering and to reduce the
model to free fermionic one.  As it was mentioned already, they
act in an auxiliary Fock space. Each of these  fields can be
presented as a sum of creation and annihilation parts
\be{Bdualfields}
\begin{array}{rcl}
\phi_{A}(\lambda)&=&q_{A}(\lambda)+p_{D}(\lambda),\\
\phi_{D}(\lambda)&=&q_{D}(\lambda)+p_{A}(\lambda),\\
\psi(\lambda)&=&q_\psi(\lambda)+p_\psi(\lambda).
\end{array}
\ee
Here $p(\lambda)$ denotes the annihilation parts of the dual fields:
$p(\lambda)\ketd=0$; $q(\lambda)$ denotes the creation parts of the dual fields:
$\brad q(\lambda)=0$.

The only nonzero commutation relations are
\be{Bcommutators}
\begin{array}{l}
{}[p_{A}(\lambda),q_\psi(\mu)]=
[p_\psi(\lambda),q_{A}(\mu)]=\ln h(\mu,\lambda),\num
{}[p_{D}(\lambda),q_\psi(\mu)]=
[p_\psi(\lambda),q_{D}(\mu)]=\ln h(\lambda,\mu),\num
{}[p_\psi(\lambda),q_\psi(\mu)]=\ln [h(\lambda,\mu)h(\mu,\lambda)],
\quad\mbox{where}\quad
  {\dis h(\lambda,\mu)=\frac{\lambda-\mu+ic}{ic}.}
\end{array}
\ee
Recall that $c$ is the coupling constant in \eq{IHamilton}.
It follows immediately from \eq{Bcommutators} that the dual fields
belong to an Abelian sub-algebra
\be{BAbel} [\psi(\lambda),\psi(\mu)]=
[\psi(\lambda),\phi_a(\mu)]=
[\phi_b(\lambda),\phi_a(\mu)]=0,
\ee
where $a,b=A,D$. Due to the property \eq{BAbel}, the Fredholm
determinant $\det\freop$ is well-defined.

The Fredholm determinant $\det\freop$ and the factor $\hb_{12}$ in the
representation \eq{Bdetrep} functionally depend on the dual fields.
Hence $\cal B$ is an operator in auxiliary Fock space (for explicit
formul\ae~see \eq{A1detrep}--\eq{A1kernel1}). From the 
equations \eq{Blogdir1}, \eq{Blogdir2} one can conclude, that the 
solution of the corresponding operator-valued RHP, as well as the 
jump matrix, also should depend on these auxiliary operators. We 
shall see in the next section that the jump matrix $\h G(\lambda)$ 
does depend on dual fields, therefore it is an operator in the
auxilary Fock space as well.

However, we would like to emphasize that the operator nature of the RHP
$(0)$ \it is not related \rm to the fact that $\hx(\lambda)$ and
$\h G(\lambda)$  depend on dual fields. Even if we replace all dual
fields by some complex functions, the RHP $(0)$ still remains the
operator-valued one, since the entries of $\hx(\lambda)$ and
$\h G(\lambda)$ are integral kernels.

The calculation of the asymptotics of the correlation function
\eq{Itempcorrel} consists of two stages. At the first stage one has to
solve the operator-valued RHP $(0)$ in order to find explicit
expression for the operator $\cal B$. The second stage consists of
the averaging with respect to auxiliary vacuum vectors. The role of
dual fields is significantly different at these two stages. While the
averaging is based essentially on the operator properties of dual
fields, at the first stage the analytical properties of the last ones
are much more important. While solving the RHP $(0)$ one can  consider
the dual fields as some complex functions, which are holomorphic in
a neighborhood of the real axis. Indeed, all these auxiliary operators
commute with each other, and their matrix elements are holomorphic
functions if their arguments belong to the strip $|\Im\lambda|<c/2$.

However such a treatment of dual fields leads us to a problem, which
we had touched briefly in the Introduction. The matter is that the
operator-valued RHP considered in the present paper can be solved
only asymptotically for the large time $t$ and long distance $x$.
Thus, we obtain only asymptotic solution for RHP, and, hence, we find
only the asymptotics of the operator $\cal B$. Strictly speaking, one
has to be sure that vacuum expectation value of the asymptotics is
equal to the asymptotics of the vacuum expectation value. Otherwise
corrections to the asymptotic expression for $\cal B$ may give
non-vanishing contribution into $\brad{\cal B}\ketd$.

Below (section \ref{MA}) we shall see that there exists a set of different asymptotic
expansions of the operator $\cal B$, which are equivalent, if we
treat the dual fields as complex functions. All of them provide us
with the same result for the vacuum expectation value if and only if
we take into consideration the complete asymptotic series. However,
if we restrict our selves with the leading term of asymptotics and
finite set of corrections, then these asymptotic series give
different results for $\brad{\cal B}\ketd$. In  section \ref{MA}  
we will suggest an algorithm of choosing the  asymptotic expansion 
which  is compatible with the avaraging over
the dual fields.

In summary we can formulate our point of view on the
dual field problem as follows. We consider the
dual fields as the complex functions, which
are holomorphic in a finite neighborhood of the real axis. Moreover, 
since the matrix elements of $\exp{\phi_{a}(\lambda)}$ and  $\exp{\psi(\lambda)}$
are rational 
functions of $\lambda$, we shall consider
$\exp{\phi_{a}(\lambda)}$ and  $\exp{\psi(\lambda)}$  as the rational functions
of $\lambda$, bounded in the strip $|\Im{\lambda}| < c/2$. 
This permits us to solve the operator-valued RHP for the large time 
and long distance separation by deforming the original contour
to the contours in the complex plane.  At the same time we do not forget 
about the quantum operator nature of these auxiliary operators.  
Therefore we do not restrict our selves with the leading term of the 
asymptotics for the operator $\cal B$. We draw our special attention 
to the corrections and their behavior after averaging with respect to 
the auxiliary vacuum.  We shall consider all these questions in more 
detail in sections \ref{ID}, \ref{MA}, where we will 
demonstrate how one can modify the procedure of solving the RHP in 
order to obtain correct result for the asymptotics of the vacuum mean value.

\section{The  operator-valued RHP\label{F}}

The operator-valued RHP describing the temperature correlation
function of local fields \eq{Itempcorrel} was formulated in
\cite{KS1}. We have presented already the general form of this RHP in 
the previous section. Here we give the explicit expression for the
jump matrix and explain all necessary definitions and notations.

According to \cite{KS1}, the operator-valued RHP in question
consists of finding the operator $\hx(\lambda)$ possessing the properties,
\ba{Fovrhp}
&&{\dis
1^a.~\hx(\lambda)\to\h I,
\quad \lambda\to\infty,}\nona{20}
&&{\dis
2^a.~\hx(\lambda)\quad\mbox{is analytical function of $\lambda$ if}~
\lambda\notin R,}\num 
&&{\dis
3^a.~\hx_-(\lambda)=\hx_+(\lambda)\h G(\lambda),\qquad
\lambda\in R,}\nonumber
\ea
where the operator-valued entries of the jump matrix $\h G(\lambda)$ 
are defined by the equations,
\be{Fjumpmat}
\begin{array}{rcl}
{\dis
\h G_{11}(\lambda|u,v)}&=& {\dis \delta(u-v)}{\dis
-Z(v,\lambda)\delta(u-\lambda)\vartheta(\lambda)
e^{\phi_D(\lambda)};}\num
 {\dis
\h G_{12}(\lambda|u,v)}&=& {\dis
2\pi i (\vartheta(\lambda)-1)
\delta(u-\lambda)\delta(v-\lambda)
e^{\psi(\lambda)+\tau(\lambda)};}\num
 {\dis
\h G_{21}(\lambda|u,v)}&=&
{\dis -\frac{i}{2\pi}Z(u,\lambda)Z(v,\lambda)\vartheta(\lambda)
e^{\phi_A(\lambda)+\phi_D(\lambda)
-\psi(\lambda)-\tau(\lambda)};}\num
{\dis
\h G_{22}(\lambda|u,v)}&=&
 {\dis
\delta(u-v)
-Z(u,\lambda)\delta(v-\lambda)\vartheta(\lambda)
e^{\phi_A(\lambda)}}.
\end{array}
\ee
Here $\phi_A(\lambda)$, $\phi_D(\lambda)$ and $\psi(\lambda)$ are the
dual fields \eq{Bdualfields}. Recall that starting from this section
we consider $\exp{\phi_{A}(\lambda)}$,
$\exp{\phi_{D}(\lambda)}$, and $\exp{\psi(\lambda)}$ as the 
classical functions which are rational and bounded in  a finite strip 
( $|\Im{\lambda}| < c/2$) near the real axis. Moreover, we will 
assume (see the arguments given in the end of this section ) 
the following symmetric properties of the functions $\psi(\lambda)$ 
and $\phi(\lambda) \equiv \phi_{A}(\lambda) -\phi_{D}(\lambda)$ :

\begin{equation}\label{symmetry1}
\psi(\lambda) = \bar \psi(\bar \lambda),
\end{equation}
and
\begin{equation}\label{symmetry2}
\phi(\lambda) = -\bar \phi(\bar \lambda),
\end{equation}

The function $Z(\lambda,\mu)$ is equal to
\be{FZ}
Z(\lambda,\mu)=
\frac{e^{-\phi_D(\lambda)}}{h(\mu,\lambda)}+
\frac{e^{-\phi_A(\lambda)}}{h(\lambda,\mu)},
\ee
where $h(\lambda,\mu)=(\lambda-\mu+ic)/ic$ was introduced in
\eq{Bcommutators}. The Fermi weight $\vartheta(\lambda)$ was
defined in \eq{IFermi}, and it is rapidly decreasing function at
$\lambda\to\infty$. The only object depending on the time $t$ and
the distance $x$ is the function $\tau(\lambda)\equiv
\tau(\lambda; x,t)$ defined by the equation :
\be{Ftau1}
\tau(\lambda)=it\lambda^2-ix\lambda,
\ee
and representing the dispersion law of the underlying classical model.

In the present paper we study the large time and the long distance
asymptotic behavior of the solution of the RHP $(a)$. More
precisely we consider the case when $t\to\infty$, $x\to\infty$,
while their ratio remains fixed:
$$
x/2t\equiv \lambda_0= O(1).
$$

Therefore
it is convenient to use new independent variables $t$ and $\lambda_0$
instead of $t$ and $x$. The function $\tau(\lambda)$ then turns into
\be{Ftau2}
\tau(\lambda)=it(\lambda-\lambda_0)^2-it\lambda_0^2.
\ee

Since the jump matrix $\h G(\lambda)$ \eq{Fjumpmat} depends on
delta-functions whose arguments contain 
parameter $\lambda$, one has to understand the jump condition $3^a$ in a weak
sense. In order to avoid the complications caused by the presence of
the distributions depending on the complex parameter,
we make the following regularization of the jump matrix \eq{Fjumpmat} .

Let
\be{Fregular}
\begin{array}{l}
{\dis
\delta_\epsilon(u-\lambda)=
\frac{1}{2\sqrt{\pi\epsilon}}e^{-\frac{(u-\lambda)^2}{4\epsilon}},}\non
{\dis
\Ne(\lambda)=\stint Z^2(u,\lambda)\delta_\epsilon
(u-\lambda)\,du,}
\end{array}
\ee
and
\be{Fdefvectors}                                 
\hspace{-1cm}\begin{array}{lr}
{\dis
\odr\equiv|1,u\rangle=
\frac{\delta_\epsilon(u-\lambda)Z(u,\lambda)}
{\sqrt{\Ne(\lambda)Z(\lambda,\lambda)}},}
&{\dis
\odl\equiv\langle1,v|=
\sqrt{\frac{Z(\lambda,\lambda)}
{\Ne(\lambda)}}Z(v,\lambda) ,}\numa{30}
{\dis
\dvr\equiv|2,u\rangle=\
\sqrt{\frac{Z(\lambda,\lambda)}
{\Ne(\lambda)}}Z(u,\lambda) ,}
&{\dis
\dvl\equiv\langle2,v|=\frac
{\delta_\epsilon(v-\lambda)Z(v,\lambda)}
{\sqrt{\Ne(\lambda)Z(\lambda,\lambda)}}.}
\end{array}
\ee
Obviously
\be{Fnormvectors}
\odl1\rangle\equiv\stint\langle1,u|1,u\rangle\,du=\dvl2\rangle
\equiv\stint\langle2,u|2,u\rangle\,du=1.
\ee
Also, in virtue of the symmetry equation (\ref{symmetry2}), one can  see
that $\Ne(\lambda)$ has no zeros on the real line.

We define the  entries of the regularized jump matrix  by the equations,
\be{Fjumpmatreg}
\begin{array}{rcl}
{\dis
\h G_{11}(\lambda)}&=& {\dis \hi
-\vartheta(\lambda)Z(\lambda,\lambda)
e^{\phi_D(\lambda)}\odr\odl;}\num
{\dis
\h G_{12}(\lambda)}&=& {\dis
2\pi i (\vartheta(\lambda)-1)Z(\lambda,\lambda)
e^{\psi(\lambda)+\tau(\lambda)}\odr\dvl;}\num
 {\dis
\h G_{21}(\lambda)}&=&
{\dis -\frac{i}{2\pi}\vartheta(\lambda)Z(\lambda,\lambda)
e^{\phi_A(\lambda)+\phi_D(\lambda)
-\psi(\lambda)-\tau(\lambda)}\dvr\odl;}\num
{\dis
\h G_{22}(\lambda)}&=&
 {\dis
\hi-\vartheta(\lambda)Z(\lambda,\lambda)
e^{\phi_A(\lambda)}\dvr\dvl},
\end{array}
\ee
where $\hi \equiv \delta(u-v)$. It is easy to see that in the limit
$\epsilon\to 0$ the jump matrix \eq{Fjumpmatreg} turns into
\eq{Fjumpmat}.

The RHP $(a)$ with the regularized jump matrix corresponds to a new
regularized operator ${\cal B}$. In particular, the Fredholm 
determinant $\det\freop$ should be replaced by $\det
\Bigl(\tilde I+\widetilde V_\epsilon\Bigr)$ (see appendix B).
New regularized kernel $\widetilde V_\epsilon$, as well as
original kernel $\widetilde V$, is not singular, and the limit           
$\widetilde V_\epsilon\to\widetilde V,\quad\epsilon\to 0$ is
well defined. It is remarkable that all relationships
\eq{Blogdir1}--\eq{Bopeq} remain valid for the  regularized
objects for {\it finite} values of $\epsilon$ (not only in
the limit $\epsilon \to 0$ !). In the rest of the paper
we will only work with the regularized operator
$\widetilde V_\epsilon$, and we will denote it as 
$\widetilde V$  omitting the subscript $\epsilon$.
We will also assume that the large time limit we
study commutes with the limit $\epsilon \to 0$. As an
indirect justification of this conjecture, we take
another special property of our regularization which
will be revealed later:
the first leading terms which we evalute
for the regularized Fredholm determinant {\it do not
depend on $\epsilon$}. 

As in  \cite{KS1}, under the assumptions made in the previous
section concerning the general theory of the operator-valued
RHPs we are dealing with,  solvability of the RHP 
(\ref{Fovrhp}, \ref{Fjumpmatreg}) implies  uniqness of
its solution. Moreover, using the representation property
(\ref{det}) and equation (\ref{detcauchy}) we conclude that
the solution $\hx(\lambda)$ of the  RHP $(a)$ with the jump
matrix (\ref{Fjumpmatreg}) satisfies the equation,
\begin{equation}\label{det1}
\det \hx(\lambda) = 1 \quad \mbox{for all} \quad \lambda \, .
\end{equation}

\vskip .2in

The very possibility  of solving the RHP
 (\ref{Fovrhp}, \ref{Fjumpmatreg})
 asymptoticly as $t \to \infty$ is based on the presence of the 
rapidly oscillating exponents
 $e^{\pm\tau(\lambda)}$ in the jump matrix.
In the next section we start the asymptotic analysis
of the solution of the RHP (\ref{Fovrhp}, \ref{Fjumpmatreg})
using an operator-valued generalization of the
nonlinear steepest descend method suggested in \cite{DZ} for the oscillatory
matrix-valued Riemann-Hilbert problems.

We conclude this section by one important note.

The model QNLS has two different phases, corresponding to the 
positive and negative chemical potential $h$ in the Hamiltonian
\eq{IHamilton}. In particular, the ground state of the model for
$h<0$ coincides with the bare Fock vacuum, while for $h>0$ the
ground state is the Dirac sea \cite{LL}. At finite temperature the 
difference is not so essential, but nevertheless one could expect
that the asymptotics of the correlation function is different for
these two cases. Since this asymptotics is defined by the solution of 
the RHP $(a)$,  the properties of the jump matrix 
\eq{Fjumpmatreg} should be different for $h>0$ and $h<0$. This fact
does take place in the free fermionic limit \cite{IIKV} and it is
related to the zeros of the diagonal elements of the jump matrix.
Out off free fermionic point one should consider zeros of the 
determinants of $\h G_{11}$ and $\h G_{22}$. Since both of these
operators are the sum of identity operator and one-dimensional
projector, it is easy to see that
\be{FdetG11}
\begin{array}{l}
{\dis
\det \h G_{11}(\lambda)=1-\vartheta(\lambda)
Z(\lambda,\lambda)e^{\phi_D(\lambda)}
=\frac{
e^{\frac{\varepsilon(\lambda)}{T}}-
e^{-\phi(\lambda)}}
{e^{\frac{\varepsilon(\lambda)}{T}}+1},}\non
{\dis
\det \h G_{22}(\lambda)=1-\vartheta(\lambda)
Z(\lambda,\lambda)e^{\phi_A(\lambda)}
=\frac{
e^{\frac{\varepsilon(\lambda)}{T}}-
e^{\phi(\lambda)}}
{e^{\frac{\varepsilon(\lambda)}{T}}+1},}
\end{array}
\ee

In the free fermionic point (coupling constant $c$ goes to infinity)
one can put dual fields in these formul\ae~equal to zero, since due
to commutation relations \eq{Bcommutators}, the vacuum expectation
value is trivial in this limit. As we had mentioned in the 
Introduction,  the solution of the  Yang--Yang equation \eq{IYY}  
$\varepsilon(\lambda)$ is positive, if $h<0$, and   
$\varepsilon(\lambda)$ has two real roots, if $h>0$. Therefore the 
determinants of $\h G_{11}$ and $\h G_{22}$ possess just the same
properties.

Out off free fermionic point, in order to answer the question, 
whether the determinants of $\hat G_{11}$ and $\hat G_{22}$ have 
zeros at the real axis, one need to know the properties of the
function $\exp\{\varepsilon(\lambda)/T\}-\exp\{\pm\phi(\lambda)\}$. 
These properties must reflect
by a natural way the properties of the operator $\exp\{\varepsilon
(\lambda)/T\}-\exp\{\pm\phi(\lambda)\}$. 
It was shown in [1] that dual fields can be expressed in terms of
canonical Bose fields. Using these representations one can prove,
in particular, that $\psi^\dagger(\bar\lambda)=\psi(\lambda)$ and
$\phi^\dagger(\bar\lambda)=-\phi(\lambda)$. Therefore we demand the
corresponding classical functions to possess similar properties
\eq{symmetry1}, \eq{symmetry2}. 
The operator $\exp\{\pm \phi(\lambda)\}$ obviously
is a unitary operator with the spectrum, belonging to the unite 
circle. There are serious reasons to assume that all the points of 
the unite circle belong to the spectrum of this operator (although 
this fact have not been checked rigorously). Therefore, if $h<0$, and 
hence $\varepsilon(\lambda)>0$, we conclude that the spectrum of the 
operator $\exp\{\varepsilon(\lambda)/T\}-\exp\{\pm\phi_A(\lambda)\}$ 
does not contain zero:
\be{spectrneg}
0\notin\sigma\left(e^{\varepsilon(\lambda)/T}-e^{\pm 
\phi(\lambda)}\right),\qquad h<0.
\ee
On the contrary for $h>0$ the function $\varepsilon(\lambda)$ has
two real roots, hence
\be{spectrpos}
0\in\sigma\left(e^{\varepsilon(\lambda)/T}-e^{\pm 
\phi(\lambda)}\right),\qquad h>0.
\ee
Thus, taking into account \eq{spectrneg}, \eq{spectrpos}, we shall
demand the classical function 
$\phi(\lambda)=\phi_A(\lambda)-\phi_D(\lambda)$ to be such, that the 
determinants of the operators $\hat G_{11}$ and $\hat G_{22}$ have no 
real roots for $h<0$, and they both have two real roots for $h>0$.

\section{Negative chemical potential\label{NCP}}
In this section we begin the asymptotic analysis of the RHP 
$(a)$ with the jump matrix ( \ref{Fjumpmatreg}). We refer the reader to the
works \cite{DZ2} and \cite{DIZ1}, which present the
nonlinear steepest descent method for the classical
NLS, to see that each of our basic operator constructions has its
matrix counterpart. In particular, in this and the next
sections we perform the first step of the method, i.e.
the transformation of the jump matrix to
the proper up-and-low-triangular forms followed by
the deformation of the real line to the steepest
descent contours with respect of the exponent
$\tau(\lambda)$. All the considerations of this section
are very close to those of \cite{FIK}. 

We start with the case of
negative chemical potential. Let as before $\lambda_{0}$ denote
the saddle point of the exponent $\tau(\lambda)$, i.e. (see
\eq{Ftau1} - \eq{Ftau2}),
\be{NCPsaddle}
\lambda_{0} = \frac {x}{2t}\, ,
\ee
and consider the following substitution
\be{NCPsub}
\hx(\lambda)=\h\Phi(\lambda)\Rho(\lambda),
\ee
where $\h\Phi(\lambda)$ is new unknown operator-valued matrix, and
$\Rho(\lambda)$ is diagonal matrix
\be{NCPRho}
\Rho=\left(\begin{array}{cc}
{\dis\h\varrho(\lambda)}&0\num
0&{\dis\bigl(\h\varrho^T(\lambda)\bigr)^{-1}}
\end{array}\right).
\ee
Here $\h\varrho(\lambda)$ is the solution of the RHP
\ba{NCPscalrhp}
&&{\dis
1^b.~\h\varrho(\lambda)\to\hi \equiv \delta(u-v),
\quad \lambda\to\infty,}\nona{20}
&&{\dis
2^b.~\h\varrho(\lambda)\quad\mbox{is analytical function of $\lambda$ if}~
\lambda\notin R,}\num
&&{\dis
3^b.~\h\varrho_-(\lambda)=\h\varrho_+(\lambda)
\left( \theta(\lambda_0-\lambda)\h G_{11}
+\theta(\lambda-\lambda_0)
\bigl(\h G_{22}^T\bigr)^{-1}\right),\quad
\lambda\in R,}\nonumber
\ea
and its kernel $\h\varrho(\lambda| u,v)$  is integrable at the point $\lambda=\lambda_0$
(cf. the function $\delta(\lambda)$ in \cite{DZ2}, \cite{DIZ1}).
The RHP $(b)$
looks like a scalar RHP, however one should remember that
$\h\varrho(\lambda)\equiv \h\varrho(\lambda|u,v)$ and
$\h G_{jk}(\lambda)\equiv \h G_{jk}(\lambda|u,v)$ are integral operators.
The symbols $\h\varrho^T(\lambda)$  as well as $\h G_{22}^T(\lambda)$
mean the transposition of these operators:
$\h\varrho^T(\lambda|u,v)=\h\varrho(\lambda|v,u)$ and
$\h G_{22}^T(\lambda|u,v)=\h G_{22}(\lambda|v,u)$ respectively.
$\theta(\lambda)$ is the step function
\be{NCPstep}
\theta(\lambda)=\left\{\begin{array}{l}
1,\qquad \lambda>0,\\
0,\qquad \lambda<0.
\end{array}\right.
\ee
Since the jump operator in \eq{NCPscalrhp} is discontinuous at
$\lambda=\lambda_0$ we have demanded an additional condition of
integrability for the solution of the RHP $(b)$ (cf. the standard conditions 
\cite{GK} of the
theory of factorization of matrix functions). Recall also, that
we consider the case, when $\det\h G_{11}$ and $\det\h G_{22}$ are
not equal to zero at the real axis, hence the jump operator
$\theta(\lambda_0-\lambda)\h G_{11}(\lambda)
+\theta(\lambda-\lambda_0) \Bigl(\h G_{22}^T(\lambda)\Bigr)^{-1}$ is
well defined.

Thus we obtain a new RHP for $\h\Phi(\lambda)$
\ba{NCPrhpPhi}
&&{\dis
1^c.~\h\Phi(\lambda)\to\h I,
\quad \lambda\to\infty,}\nona{20}
&&{\dis
2^c.~\h\Phi(\lambda)\quad\mbox{is analytical function of $\lambda$ if}~
\lambda\notin R,}\num
&&{\dis
3^c.~\h\Phi_-(\lambda)=\h\Phi_+(\lambda)
\h G_\Phi(\lambda),\qquad\lambda\in R,}\nonumber
\ea
where
\be{NCPGPhi}
\h G_\Phi(\lambda)=\Rho_+(\lambda)\h G(\lambda)\Rho^{-1}_-(\lambda).
\ee
The coefficients $\hb$ and $\h c$ of the asymptotic expansion of the
original operator $\hx(\lambda)$  (the RHP $(a)$) can be easily
expressed in terms of asymptotic expansions of $\h\Phi(\lambda)$ and
$\h\varrho(\lambda)$:
\be{asPhi}
\h\Phi=\h I+\frac{\h\Phi_0}{\lambda}+\frac{\h\Phi_1}{\lambda^2}
+\dots,
\ee
\be{asrho}
\h\varrho=\hi+\frac{\h\varrho_0}{\lambda}+\frac{\h\varrho_1}{\lambda^2}
+\dots.
\ee
In particular, it is easy to see that
\be{NCPlogdirnew1}
\begin{array}{l}
{\dis\hb_{11}=(\h\Phi_0)_{11}+\h\varrho_0,}\num
{\dis\h c_{22}-\h c_{11}=
(\h\Phi_1)_{22}-(\h\Phi_1)_{11}-(\h\Phi_0)_{22}\h\varrho_0^T-
(\h\Phi_0)_{11}\h\varrho_0 -\h\varrho_1 -\h\varrho_1^T
+\left(\h\varrho_0^T\right)^2.}
\end{array}
\ee

Consider the function $\Delta(\lambda)=\det\h\varrho(\lambda)$. This
is not an operator
but  a complex function. The asymptotic expansion of this
function at $\lambda\to\infty$ plays an important role. It is convinient
to write down this expansion in the exponential form, 
\be{NCPasDelta}
\Delta=
\exp\left\{\frac{\Delta_0}{\lambda}+\frac{\Delta_1}{\lambda^2}
+\dots\right\}.
\ee

In order to find the first logarithmic derivatives of the Fredholm
determinant we need to know only the traces of the operators
$\hb_{11}$ and $\h c_{22}-\h c_{11}$ (see \eq{Blogdir1}). Therefore
taking the trace of \eq{NCPlogdirnew1}  we arrive at
\be{NCPlogdirnew}
\begin{array}{l}
{\dis\tr\hb_{11}=\tr(\h\Phi_0)_{11}+\Delta_0,}\num
{\dis\tr(\h c_{22}-\h c_{11})=
\tr\left[(\h\Phi_1)_{22}-(\h\Phi_1)_{11}-
(\h\Phi_0)_{22}\h\varrho_0^T-(\h\Phi_0)_{11}\h\varrho_0\right]
-2\Delta_1.}
\end{array}
\ee

Consider now the jump matrix $\h G_\Phi$.
Using the equations,

\be{NCPtransp}
(\odr\odl)^T=\dvr\dvl,\qquad
(|k\rangle \langle j|)^T = |k\rangle \langle j|, \quad k\neq j,
\ee

and
\be{NCPprod}
(|k\rangle \langle j|)(|j\rangle \langle l|) = |k\rangle \langle l|,\quad
k,j,l = 1,2,
\ee
one can easily check that
\be{NCPquasidet}
\begin{array}{l}
{\dis
\h G_{22}-\h G_{21}\bigl(\h G_{11}\bigr)^{-1}
\h G_{12}=\bigl(\h G_{11}^T\bigr)^{-1},}\num
{\dis
\h G_{11}-\h G_{12}\bigl(\h G_{22}\bigr)^{-1}
\h G_{21}=\bigl(\h G_{22}^T\bigr)^{-1}.}
\end{array}
\ee
Due to these properties the jump matrix $\h G_\Phi$ has the form
\be{NCPGPhi1}
\h G_{\Phi}=\left(
\begin{array}{cc}
{\dis \hi}&{\dis \h Pe^\tau}\num
{\dis \h Qe^{-\tau}}&{\dis \hi+\h Q \h P}
\end{array}\right),\qquad \mbox{for}\quad\lambda<\lambda_0
\ee
where
\be{NCPQ1a}
\h Qe^{-\tau}=\bigl(\h\varrho_+^T\bigr)^{-1}\h G_{21}
(\h G_{11})^{-1}\h\varrho_+^{-1},
\ee
\be{NCPP1}
\h Pe^{\tau}=\h\varrho_-(\h G_{11})^{-1}\h G_{12}
\h\varrho_-^T,
\ee
and
\be{NCPGPhi1a}
\h G_{\Phi}=\left(
\begin{array}{cc}
{\dis \hi+\h{\tilde P}\h{\tilde Q}}&{\dis \h{\tilde P}e^\tau}\num
{\dis \h{\tilde Q}e^{-\tau}}&{\dis \hi}
\end{array}\right),\qquad \mbox{for}\quad\lambda>\lambda_0
\ee
where
\be{NCPQ2}
\h{\tilde Q}e^{-\tau}=\bigl(\h\varrho_-^T\bigr)^{-1}(\h G_{22})^{-1}
\h G_{21}\h\varrho_-^{-1},
\ee
\be{NCPP2a}
\h{\tilde P}e^{\tau}=\h\varrho_+\tilde G_{12}(\tilde G_{22})^{-1}
\h\varrho_+^T.
\ee
Therefore one can factorize the jump matrix
\be{NCPfactor}
\begin{array}{l}
{\dis
\h G_\Phi=\h M_+\h M_-, \qquad\mbox{for}\quad\lambda<\lambda_0,
}\num
{\dis
\h G_\Phi=\h N_+\h N_-, \qquad\mbox{for}\quad\lambda>\lambda_0,}
\end{array}
\ee
where
\be{NCPMplminus}
\h M_+=\left(\begin{array}{cc}
{\dis \hi}&{\dis 0}\num
{\dis \h Qe^{-\tau}}&{\dis \hi}
\end{array}\right), \qquad
\h M_-=\left(\begin{array}{cc}
{\dis \hi}&{\dis \h Pe^\tau}\num
{\dis 0}&{\dis \hi}
\end{array}\right),
\ee
\be{NCPNplminus}
\h N_+=\left(\begin{array}{cc}
{\dis \hi}&{\dis \h{\tilde P}e^\tau}\num
{\dis 0}&{\dis \hi}
\end{array}\right),\qquad
\h N_-=\left(\begin{array}{cc}
{\dis \hi}&{\dis 0}\num
{\dis \h{\tilde Q}e^{-\tau}}&{\dis \hi}
\end{array}\right).
\ee
Notice that $\h Q$ and $\h{\tilde P}$ can be analytically continued
into a neighborhood of the real axis in the upper half-plane.
Similarly, $\h{\tilde Q}$ and $\h P$ can be continued into a strip
in the lower half-plane.

\begin{figure}[t]
\begin{center}
\begin{picture}(370,120)
\put(0,60){\vector(1,0){335}}
\put(290,64){$\Im\lambda=0$}
{\thicklines
\put(160,100){\line(1,0){160}}
\put(10,100){\vector(1,0){10}}
\put(300,100){\vector(1,0){10}}
\put(0,100){\line(1,0){80}}
\put(120,60){\line(1,1){40}}
\put(120,60){\line(-1,1){40}}
\put(160,20){\line(1,0){160}}
\put(10,20){\vector(1,0){10}}
\put(300,20){\vector(1,0){10}}
\put(0,20){\line(1,0){80}}
\put(120,60){\line(-1,-1){40}}
\put(120,60){\line(1,-1){40}}
}
\put(290,104){$C_1$}
\put(30,104){$C_2$}
\put(30,4){$C_3$}
\put(290,4){$C_4$}
\put(265,83){$\Omega_1$}
\put(180,70){$\h U=\h\Phi\h N_+$}
\put(5,83){$\Omega_2$}
\put(35,70){$\h U=\h\Phi\h M_+$}
\put(5,30){$\Omega_3$}
\put(35,43){$\h U=\h\Phi\h M_-^{-1}$}
\put(265,30){$\Omega_4$}
\put(180,43){$\h U=\h\Phi\h N_-^{-1}$}
\put(100,110){$\h U=\h\Phi$}
\put(100,10){$\h U=\h\Phi$}
\put(115,70){$\lambda_0$}
\end{picture}
\end{center}
\caption{Contour for new RH problem}\label{contour}
\end{figure}
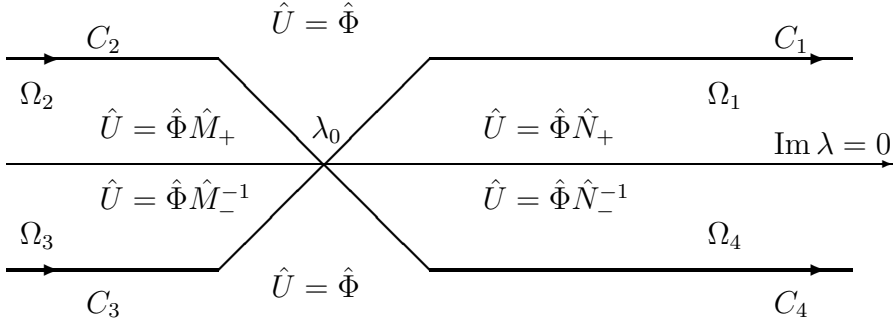

Consider new contour, which is shown on the Fig.\ref{contour}
The contour consists of four branches $C_1$, $C_2$, $C_3$ and $C_4$,
having the origin in the saddle point $\lambda_0$ (arrows show the
positive direction).  The branches $C_j$ and the real axis are
the boundaries of the four half-infinite `wedges' $\Omega_j$,
($j=1,\dots,4$). Matrices $N_\pm$ can be analytically continued into
the wedges $\Omega_1$ and $\Omega_4$ respectively, matrices $M_\pm$
--- into the wedges $\Omega_2$ and $\Omega_3$.

Define the matrix $\h U$ as follows:
\be{NCPdefU}
\begin{array}{l}
{\dis
\h U=\h\Phi,\qquad \lambda\notin
\Omega_1\cup\Omega_2\cup \Omega_3\cup \Omega_4,}\num
{\dis
\h U=\h\Phi\h N_+,\qquad \lambda\in\Omega_1,}\num
{\dis
\h U=\h\Phi\h M_+,\qquad \lambda\in\Omega_2,}\num
{\dis
\h U=\h\Phi\h M_-^{-1},\qquad \lambda\in\Omega_3,}\num
{\dis
\h U=\h\Phi\h N_-^{-1},\qquad \lambda\in\Omega_4,}
\end{array}
\ee
(see Fig.\ref{contour}). Then the matrix $\h U$ has no jump at the real axis,
however it has jumps at the new contour $C=C_1\cup C_2\cup C_3
\cup C_4$. The RHP for $\h U$ can be formulated now as

\ba{NCPrhpU}
&&{\dis
1^d.~\h U(\lambda)\to\h I
\quad \lambda\to\infty,}\nona{20}
&&{\dis
2^d.~\h U(\lambda)\quad\mbox{is analytical function of $\lambda$ if}~
\lambda\notin C,}\non
&&{\dis
3^d.~\h U_-=\h U_+\h N_+,\qquad \lambda\in C_1,}\num
&&{\dis\hphantom{3^d.~}
\h U_-=\h U_+\h M_+,\qquad \lambda\in C_2,}\num
&&{\dis\hphantom{3^d.~}
\h U_-=\h U_+\h M_-,\qquad \lambda\in C_3,}\num
&&{\dis\hphantom{3^d.~}
\h U_-=\h U_+\h N_-,\qquad \lambda\in C_4.}\label{NCPrhpU1}
\ea
where $\h U_\mp$ are boundary values of $\h U$ from the right and
left of the contour $C$ respectively.

It is easy to see that all jump matrices in
\eq{NCPrhpU}--\eq{NCPrhpU1} have the form
\be{NCPform}
\h I+o(e^{- t^{2\delta}}),\qquad \mbox{for}\quad
|\lambda-\lambda_0|>t^{-1/2+\delta},
\ee
where $\delta$ is an arbitrary small positive number.
Thus, we can put $\h N_\pm=\h M_\pm\approx\h I$ up to exponentially
decreasing corrections for all $\lambda \in C$ except small vicinity of
the saddle point. Formally appealing to the integral equation
\eq{sie}, written for the contour $C$ and the operator-valued function 
$\h U_{+}$ (cf. the {\it nonformal} arguments of \cite{DZ2} for the usual
matrix case), we conclude  that
\be{NCPsolU}
\h U\approx \h I, \qquad\Rightarrow\qquad
\h \Phi\approx \h I, \qquad\Rightarrow\qquad
\hx\approx \Rho.
\ee
and hence the coefficients of the asymptotic expansion $\h\Phi_0$
and $\h\Phi_1$ are asymptoticly  equal to zero:
\be{NCPPhi01}
\h\Phi_0,\, \h\Phi_1 \approx 0.
\ee

Thus, similar to \cite{FIK},  the traces of the operators $\h b_{11}$ and
$\h c_{22}-\h c_{11}$ asymptoticly depend only on the coefficients of the
$\lambda$ -  expansion \eq{NCPasDelta} of the function $\Delta(\lambda)$
(see \eq{NCPlogdirnew}):
\be{NCPtraces}
\begin{array}{l}
{\dis\tr\hb_{11}=\Delta_0 + o(1),}\num
{\dis\tr(\h c_{22}-\h c_{11})=-2\Delta_1 + o(1).}
\end{array}
\ee
On the other hand, one can easily find $\Delta(\lambda)$ since the
RHP $(b)$ for the operator $\h\varrho(\lambda)$ implies the following scalar
RHP for its determinant $\Delta(\lambda)$ (cf. \eq{detnormcond} -
\eq{detjumpcond}):
\ba{NCPscalrhpD}
&&{\dis
1^e.~\Delta(\lambda)\to 1,
\quad \lambda\to\infty,}\nona{20}
&&{\dis
2^e.~\Delta(\lambda)\quad\mbox{is analytical function of $\lambda$ if}~
\lambda\notin R,}\num
&&{\dis
3^e.~\Delta_-(\lambda)=\Delta_+(\lambda)
\det\left( \theta(\lambda_0-\lambda)\h G_{11}
+\theta(\lambda-\lambda_0)
\bigl(\h G_{22}^T\bigr)^{-1}\right),\quad
\lambda\in R,}\nonumber
\ea
Using the fact that both of the operators $\h G_{11}$ and $\h G_{22}$
are a sum of the identity operator $\hi$ and an one-dimensional
projector we find
\be{NCPG22invT}
\bigl(\h G_{22}^T\bigr)^{-1}=
\hi+\frac{Z(\lambda,\lambda)\vartheta(\lambda)
e^{\phi_A(\lambda)}}{ 1-
Z(\lambda,\lambda)\vartheta(\lambda)e^{\phi_A(\lambda)}}
\odr\odl,
\ee
and (cf. \eq{det})
\be{NCPdetjump}
\det\left(
\theta(\lambda_0-\lambda)\h G_{11}(\lambda)
+\theta(\lambda-\lambda_0)
\Bigl(\h G_{22}^T(\lambda)\Bigr)^{-1}\right)
=\left(1-\vartheta(\lambda)
\Bigl(1+e^{\phi(\lambda)\sign(\lambda-\lambda_0)}
\Bigr)\right)^{\sign(\lambda_0-\lambda)},
\ee
where $\sign(\lambda)=\theta(\lambda)-\theta(-\lambda)$ is the
sign function, and $\phi(\lambda)=\phi_A(\lambda)-\phi_D(\lambda)$.
Thus from \eq{detcauchy} it follows that the solution of the RHP $(e)$ 
is given by the explicit formula
\be{NCPsolDelta}
\Delta=\exp\left\{-\frac{1}{2\pi i}
\stint\frac{d\mu}{\mu-\lambda}\sign(\lambda_0-\mu)
\ln\left(1-\vartheta(\mu)
\Bigl(1+e^{\phi(\mu)\sign(\mu-\lambda_0)}
\Bigr)\right)\right\},
\ee
and, hence,
\be{NCPsolDelta0}
\Delta_0=\frac{1}{2\pi i}
\stint\,d\mu\sign(\lambda_0-\mu)
\ln\left(1-\vartheta(\mu)
\Bigl(1+e^{\phi(\mu)\sign(\mu-\lambda_0)}
\Bigr)\right),
\ee
\be{NCPsolDelta1}
\Delta_1=\frac{1}{2\pi i}
\stint\,\mu d\mu\sign(\lambda_0-\mu)
\ln\left(1-\vartheta(\mu)
\Bigl(1+e^{\phi(\mu)\sign(\mu-\lambda_0)}
\Bigr)\right).
\ee
The logarithmic derivatives of the Fredholm determinant with respect
to the variables $t$ and $\lambda_0$ are equal to
\be{NCPlogdirtl}
\begin{array}{l}
{\dis
\frac{1}{2t}\partial_{\lambda_0}\ln\det\freop
=i\tr\hb_{11}=i\Delta_0 + o(1),}\non
{\dis
\left(\partial_t-\frac{\lambda_0}{t}\partial_{\lambda_0}
\right)\ln\det\freop
=i\tr(\h c_{22}-\h c_{11})=-2i\Delta_1 + o(1).}
\end{array}
\ee
Integrating these equations with respect to $t$ and $\lambda_0$
we arrive at
\be{NCPasdet2}
\ln\det(\tilde I+\tilde V)=\frac{1}{2\pi}
\stint\,d\mu|x-2\mu t|
\ln\left(1-\vartheta(\mu)
\Bigl(1+e^{\phi(\mu)\sign(\mu-\lambda_0)}
\Bigr)\right) + {\cal O} (\ln t),
\ee
where $x=2t\lambda_0, \quad t\to \infty$. It is worth
observing that the leading term in  the
r.h.s. of \eq{NCPasdet2} does not depend on the
regularization parameter $\epsilon$. It is also should be
mentioned that 
equation \eq{NCPasdet2} for the asymptotics of the Fredholm
determinant exactly coincides with the one
 obtained in \cite{S1} by independent and more direct methods.
This fact partially justifies our, mostly formal, manipulations
with the operator-valued Riemann-Hilbert problems.

\section{Positive chemical potential\label{PCP}}

In the case of positive chemical potential the method of
factorization, considered in the section \ref{NCP} is not applicable
directly to the jump matrix \eq{Fjumpmatreg}. Indeed, now
the determinants of the operators $\h G_{11}(\lambda)$ and
$\h G_{22}(\lambda)$  have zeros on the real axis, and the
scalar RHP $(e)$ for the function $\Delta(\lambda) = \det 
\h\varrho(\lambda)$ is ill posed.
The factorization \eq{NCPfactor} also does not exist, since the
operators $\h Q$, $\h P$, $\h{\tilde P}$ and $\h{\tilde Q}$ have
singularities on the real axis (see \eq{NCPQ1a} - \eq{NCPP2a}).

In this situation one can use the approach similar to the one proposed
for the free fermionic limit of the problem in \cite{IIKV}. First, we have
to specify the position of the
determinants roots relative to the saddle point $\lambda_0$.
Denote the roots of the function,
$$
\det \hat G_{11}(\lambda) = 1 - \vartheta(\lambda) Z(\lambda, \lambda)e^{\phi_{D}(\lambda)}
=1 - \vartheta(\lambda) (1 + e^{-\phi(\lambda)}),
$$
$$
\phi(\lambda) = \phi_{A}(\lambda) - \phi_{D}(\lambda),
$$ 
as $\Lambda_j$, $j=1,2$, $\Lambda_1<\Lambda_2$.
We shall concentrate our attention on  the case,
$$
\Lambda_1<\Lambda_2<\lambda_0,
$$
assuming simultaniously
that the roots of $\det \hat G_{22}(\lambda)$ lie to the left of 
$\lambda_{0}$ as well (see the arguments in the end of the section 
\ref{F}). All the other possible cases of the roots position can be 
considered in a similar fashion.  Moreover, similar to the free 
fermion case \cite{IIKV}, the difference in the way how the roots are 
distributed along the real line does not affect the final answer for 
the asymptotics.  What is important is the assumption that non of the 
roots is close to the saddle point.  

Secondly, observe that the integral operator $\tilde I+\widetilde
V$  can be  continued into some vicinity of the real axis,
hence the integration contour can be slightly deformed in this
vicinity. The Fredholm determinant evidently does not depend on such
deformation.

\begin{figure}[h]
\begin{center}
\begin{picture}(320,80)
\put(40,40){\line(1,0){45}}
\put(50,40){\vector(1,0){10}}
\put(105,40){\oval(40,40)[t]}
\put(125,40){\line(1,0){80}}
\put(104,60){\vector(1,0){4}}
\put(160,40){\vector(1,0){10}}
\put(104,38){$\scriptscriptstyle{\bullet}$}
\put(108,25){$\Lambda_1$}
\put(225,40){\oval(40,40)[b]}
\put(245,40){\line(1,0){60}}
\put(224,20){\vector(1,0){4}}
\put(290,40){\vector(1,0){10}}
\put(225,38){$\scriptscriptstyle{\bullet}$}
\put(283,45){$\lambda_0$}
\put(274,39){$\scriptscriptstyle{\bullet}$}
\put(228,45){$\Lambda_2$}
\put(40,47){$\Gamma$}
\end{picture}
\end{center}
\caption{Deformation of the contour}\label{deform}
\end{figure}
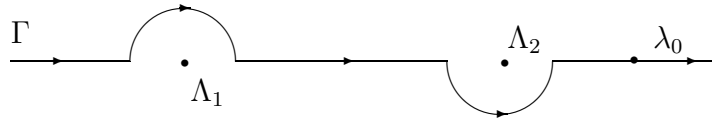

The deformed contour $\Gamma$ is shown on the Fig.\ref{deform}.
We will see below that this particular choice of the contour deformation
will guarantee the solvability of the deformed scalar
problem \eq{NCPscalrhpD} for the function $\Delta(\lambda)$. 

\vskip .2in
As in the case of negative chemical potential we make the
substitution
\be{PCPsub}
\hx(\lambda)=\h\Phi(\lambda)\Rho(\lambda),
\ee
where as before $\Rho(\lambda)$ is diagonal matrix
\be{PCPRho}
\Rho=\left(\begin{array}{cc}
{\dis\h\varrho(\lambda)}&0\num
0&{\dis\bigl(\h\varrho^T(\lambda)\bigr)^{-1}}
\end{array}\right),
\ee
and $\h\varrho(\lambda)$ satisfies the RHP
\ba{PCPscalrhp}
&&{\dis
1^f.~\h\varrho(\lambda)\to\hi,
\quad \lambda\to\infty,}\nona{20}
&&{\dis
2^f.~\h\varrho(\lambda)\quad\mbox{is analytical function of $\lambda$ if}~
\lambda\notin \Gamma,}\num
&&{\dis
3^f.~\h\varrho_-(\lambda)=\h\varrho_+(\lambda)
\left( \theta(\lambda_0-\lambda)\h G_{11}
+\theta(\lambda-\lambda_0)
\bigl(\h G_{22}^T\bigr)^{-1}\right),\quad
\lambda\in \Gamma,}\nonumber
\ea
The difference  between the Riemann--Hilbert problem
$(b)$, considered in the previous section, and RHP  $(f)$ is that
the last one is formulated on the jump contour $\Gamma$ instead of the
real axis. Similary, both the operator-valued RHP for
$\hat \Phi(\lambda)$ and the scalar RHP for $\Delta(\lambda) = \det
\h\varrho(\lambda) $ 
are set now on the contour $\Gamma$. Due to the choice made for
the contour $\Gamma$, the function 
$$
\arg \det \left(\theta(\lambda_{0} -\lambda)\hat G_{11}(\lambda) +
\theta(\lambda -\lambda_{0})(\hat G^{T}_{22})^{-1}(\lambda)
\right)
$$ 
can be  defined in such way that it is continuous for all
$\lambda \neq \lambda_{0}$ and approaches $0$ as 
$\lambda \to \pm \infty$. In other words, the $\Delta$-RHP has
no index and hence is well posed. Using again the general
formula \eq{detcauchy} (with the real line replaced by $\Gamma$), 
we have that 
\be{PCPsolDelta}
\Delta=\exp\left\{-\frac{1}{2\pi i}
\int_{\Gamma}\frac{d\mu}{\mu-\lambda}\sign(\lambda_0-\Re \mu)
\ln\left(1-\vartheta(\mu)
\Bigl(1+e^{\phi(\mu)\sign(\Re \mu-\lambda_0)}
\Bigr)\right)\right\},
\ee
(cf. \eq{NCPsolDelta}).

The jump matrix $\hat G_{\Phi}$ is given by  the same equations \eq{NCPGPhi1}-
\eq{NCPP2a} as before. The operators $\hat Q$ and $\hat P$ have singularities
at $\Lambda_{1,2}$, and hence the obstructions to the analytic continuation
of the operator matrices $\hat M_{+}$ and  $\hat M_{-}$ 
occure (the singularities of $\hat N_{\pm}$ are not relevant
since we have assumed that the zeros of $\det \hat G_{22}$ lie to the
left of $\lambda_{0}$). Therefore, before factorizing the $\hat \Phi$-
jump matrix let us make one more substitution (cf. \cite{IIKV}):

\be{PCPsub1}
\hP(\lambda)=(\lambda\hI+\hPP)\hPo(\lambda)
\left(\begin{array}{cc}
(\lambda-\Lambda_2)\hi &0\\
0 &(\lambda-\Lambda_1)\hi
\end{array}\right)^{-1}.
\ee
Here $\hPP$ is an operator  satisfying the following conditions:
\be{PCPproj}
\begin{array}{l}
\dis (\Lambda_2\hI+\hPP)\hPo(\Lambda_2)\left(
\begin{array}{c}1\\0\end{array}\right)=0,\non
\dis (\Lambda_1\hI+\hPP)\hPo(\Lambda_1)\left(
\begin{array}{c}0\\1\end{array}\right)=0.
\end{array}
\ee
The RHP for the operator $\hPo(\lambda)$ has the form
\ba{PCPrhpPhio}
&&{\dis
1^g.~\hPo(\lambda)\to\h I,
\quad \lambda\to\infty,}\nona{20}
&&{\dis
2^g.~\hPo(\lambda)\quad\mbox{is analytical function of $\lambda$ if}~
\lambda\notin R,}\num
&&{\dis
3^g.~\hPo_-(\lambda)=\hPo_+(\lambda)
\h G_s(\lambda),\qquad\lambda\in R,}\nonumber
\ea
where the jump matrices $\h G_s$ and $\hG_\Phi$ \eq{NCPGPhi} are
related by
\be{PCPjump}
\hG_s(\lambda)=
\left(\begin{array}{cc}
(\lambda-\Lambda_2)\hi &0\\
0 &(\lambda-\Lambda_1)\cdot\hi
\end{array}\right)^{-1}\hG_\Phi(\lambda)
\left(\begin{array}{cc}
(\lambda-\Lambda_2)\hi &0\\
0 &(\lambda-\Lambda_1)\hi
\end{array}\right).
\ee
Of course, now one should choose for $\hG_\Phi$ the operator
$\h\varrho$ satisfying RHP $(f)$.

Since the roots $\Lambda_1$ and $\Lambda_2$ as well as the 
roots of $\det \hat G_{22}(\lambda)$ all are smaller then
$\lambda_0$, the factorization of the matrix $\hG_s$ in the domain
to the right from the saddle point produces the matrices
$\hat N_{\pm}$ whose analyticity in the relevant
half planes is obvious. In the domain $\lambda<\lambda_0$
the jump matrix $\hG_s$ has the form

\be{PCPSGs}
\hG_s(\lambda)=\left(\begin{array}{lr}
\hi& \h P^{(1)}e^\tau\\
\h Q^{(1)}e^{-\tau}&\hi+\h Q^{(1)}\h P^{(1)}
\end{array}\right).
\ee
Here
\be{PCPSp1q1}
\h P^{(1)}(\lambda)=\h P(\lambda)\left(\frac{\lambda-\Lambda_1}
{\lambda-\Lambda_2}\right),\qquad
\h Q^{(1)}(\lambda)=\h Q(\lambda)\left(\frac{\lambda-\Lambda_2}
{\lambda-\Lambda_1}\right),
\ee
and operators $\h Q$ and $\h P$ are given by \eq{NCPQ1a} and
\eq{NCPP1}. Thus, the jump matrix $\hG_s$ can be factorized
in the domain $\lambda<\lambda_0$ as
\be{PCPfactor}
\h G_s=\h M_+\h M_-,
\ee
where
\be{PCPMplminus}
\h M_+=\left(\begin{array}{cc}
{\dis \hi}&{\dis 0}\num
{\dis \h Q^{(1)}e^{-\tau}}&{\dis \hi}
\end{array}\right), \qquad
\h M_-=\left(\begin{array}{cc}
{\dis \hi}&{\dis \h P^{(1)}e^\tau}\num
{\dis 0}&{\dis \hi}
\end{array}\right),
\ee
Using explicit formul\ae~ \eq{Fjumpmatreg} for the operators $\h G_{jk}$
one can write
\be{PCPDQ1}
Q^{(1)}(\lambda)=q^{(1)}(\lambda)\left(\h\varrho_+^T\right)^{-1}(\lambda)\, 
|2\rangle\langle1|\, \h\varrho_+^{-1}(\lambda),
\ee
\be{PCPDP1}
P^{(1)}(\lambda)=p^{(1)}(\lambda)\h\varrho_-(\lambda)\, 
|1\rangle\langle2|\, \h\varrho_-^T(\lambda),
\ee
where
\be{PCPDp1}
p^{(1)}(\lambda)=\frac{2\pi iZ(\lambda,\lambda)
(\vartheta(\lambda)-1)e^{\psi(\lambda)}}
{1-Z(\lambda,\lambda)\vartheta(\lambda)e^{\phi_D(\lambda)}}
\left(\frac{\lambda-\Lambda_1}{\lambda-\Lambda_2}\right),
\ee
\be{PCPDq1}
q^{(1)}(\lambda)=\frac{\vartheta(\lambda)Z(\lambda,\lambda)
e^{\phi_A(\lambda)+\phi_D(\lambda)-\psi(\lambda)}}
{2\pi i\left(1-Z(\lambda,\lambda)\vartheta(\lambda)
e^{\phi_D(\lambda)}\right)}
\left(\frac{\lambda-\Lambda_2}{\lambda-\Lambda_1}\right).
\ee
Note that the denominator in these formul\ae~is just the determinant
of the operator $\hG_{11}$:
\be{PCPdetG11}
\det\hG_{11}(\lambda)=1-Z(\lambda,\lambda)\vartheta(\lambda)
e^{\phi_D(\lambda)},
\ee
and it has zeros when $\lambda=\Lambda_{1,2}$. However, it is easy to
see that $p^{(1)}(\lambda)$ has no pole at $\lambda=\Lambda_1$ as
well as $q^{(1)}(\lambda)$ has no pole at $\lambda=\Lambda_2$. Thus
the function $p^{(1)}(\lambda)$ (and, hence, the matrix $\h M_-$) can
be analytically continued into some vicinity of the real axis to the
lower half-plane.  Similarly the matrix $\h M_+$ can be  continued into
a vicinity of the real axis to the upper half-plane. Hence, we again
arrive at the contour Fig.\ref{contour}. 

The following considerations are  absolutely identical to the
considerations of the previous section. Introducing the operator
$\hat U (\lambda)$ by  equations \eq{NCPdefU} ( with  
$\hat \Phi$ replaced by $\hPo$), we again obtain that
all the jump matrices associated with the function $\hat U (\lambda)$
are exponentially small for the large value of $t$ on the
contour Fig.\ref{contour} away from $\lambda_{0}$. Therefore, we have again that
\be{PCPhPo}
\hat U\approx \h I.
\ee
However, in distinction of the previous case, in order to find the
original solution $\hx$ we need to take into account the contribution
of the operator $\hPP$ \eq{PCPsub1}. 

The system of equations
\eq{PCPproj} gives
\be{PCPP1}
\begin{array}{l}
{\dis
\hPP_{11}=-\Lambda_1\hi+\Lambda_{12}
\hPo_{11}(\Lambda_2)\hD,}\non
{\dis
\hPP_{12}=\Lambda_{21}
\hPo_{11}(\Lambda_2)\hD~\hPo_{12}(\Lambda_1)
\left(\hPo_{22}
(\Lambda_1)\right)^{-1},}\non
{\dis
\hPP_{21}=\Lambda_{12}\hPo_{21}(\Lambda_2)\hD,}\non
{\dis
\hPP_{22}=-\Lambda_{1}\hi+\Lambda_{21}
\hPo_{21}(\Lambda_2)\hD~\hPo_{12}(\Lambda_1)
\left(\hPo_{22}(\Lambda_1)\right)^{-1}.}
\end{array}
\ee
where $\Lambda_{jk}=\Lambda_j-\Lambda_k$ and
\be{PCPquasidet}
\hD=\left[\hPo_{11}(\Lambda_2)-\hPo_{12}(\Lambda_1)
\left(\hPo_{22}
(\Lambda_1)\right)^{-1}\hPo_{21}(\Lambda_2)\right]^{-1}.
\ee

The solution $\hPo(\lambda)$ of the RHP $(g)$ is asymptoticly equal to
$\h M_+^{-1}$ and $\h M_-$ in the domains $\Omega_2$ and $\Omega_3$
respectively.  Hence,
\be{PCPSL1}
\hPo(\Lambda_1)\approx \h M_-(\Lambda_1)=\left(
\begin{array}{cc}
\hi&\h P^{(1)}(\Lambda_1)e^{\tau(\Lambda_1)}\\
0&\hi
\end{array}\right),
\ee
\be{PCPSL2}
\hPo(\Lambda_2)\approx \h M_+^{-1}(\Lambda_2)=\left(
\begin{array}{cc}
\hi&0\\
-\h Q^{(1)}(\Lambda_2)e^{-\tau(\Lambda_2)}&\hi
\end{array}\right).
\ee
Thus, we obtain
\be{PCPShi}
\begin{array}{l}
{\dis
\hPo_{11}(\Lambda_1),\,  \hPo_{11}(\Lambda_2),\, 
 \hPo_{22}(\Lambda_1),\,  \hPo_{22}(\Lambda_2)\approx \hi,}\non
{\dis
\hPo_{21}(\Lambda_1),\,  \hPo_{12}(\Lambda_2)\approx 0,}\non
{\dis
\hPo_{21}(\Lambda_2)\approx -\h Q^{(1)}(\Lambda_2)e^{-\tau(\Lambda_2)},}\non
{\dis
\hPo_{12}(\Lambda_1)\approx \h P^{(1)}(\Lambda_1)e^{\tau(\Lambda_1)}.}
\end{array}
\ee
Therefore, equations \eq{PCPP1} can be replaced by the equations,
\be{PCPP2}
\begin{array}{l}
{\dis
\hPP_{11}\approx -\Lambda_1\hi+\Lambda_{12}\hD,\qquad
\hPP_{12}\approx \Lambda_{21}\hD~\hPo_{12}(\Lambda_1),}\non
{\dis
\hPP_{21}\approx \Lambda_{12}\hPo_{21}(\Lambda_2)\hD,\qquad
\hPP_{22}\approx -\Lambda_{1}\hi+\Lambda_{21}
\hPo_{21}(\Lambda_2)\hD~\hPo_{12}(\Lambda_1),}
\end{array}
\ee
and
\be{PCPquasidet1}
\hD\approx \left[\hi-\hPo_{12}(\Lambda_1)\hPo_{21}(\Lambda_2)\right]^{-1}.
\ee

Define new vectors
\be{Idef}
\begin{array}{ccc}
{\dis \ODi=
\frac{\delta_\epsilon(u-\Lambda_i)Z(u,\Lambda_i)}
{\sqrt{\Ne(\Lambda_i)Z(\Lambda_i,\Lambda_i)}},}
&\qquad&
{\dis \DODi=
\sqrt{\frac{Z(\Lambda_i,\Lambda_i)}
{\Ne(\Lambda_i)}}Z(v,\Lambda_i) ,}\non
{\dis \DVi=
\sqrt{\frac{Z(\Lambda_i,\Lambda_i)}
{\Ne(\Lambda_i)}}Z(u,\Lambda_i) ,}
&\qquad&
{\dis \DDVi=
\frac{\delta_\epsilon(v-\Lambda_i)Z(v,\Lambda_i)}
{\sqrt{\Ne(\Lambda_i)Z(\Lambda_i,\Lambda_i)}},}
\end{array}
\ee
where $i=1,2$, and introduce two quantities
\be{Ialpha}
\alpha=\DODd\varrho_+^{-1}(\Lambda_2)\varrho_-(\Lambda_1)\ODo,
\ee
\be{Ibeta}
\beta=\DDVo\varrho_-^{T}(\Lambda_1)\left(\varrho_+^T(\Lambda_2)
\right)^{-1}\DVd.
\ee
Then it is easy to see that
\be{PCPhPo1221}
\begin{array}{l}
{\dis
\hPo_{21}(\Lambda_2)\approx -q^{(1)}(\Lambda_2)e^{-\tau
(\Lambda_2)}\left(\h\varrho_+^T(\Lambda_2)\right)^{-1}
\DVd\hspace{-1mm}\DODd\h\varrho_+^{-1}(\Lambda_2)
,}\non
{\dis
\hPo_{12}(\Lambda_1)\approx p^{(1)}(\Lambda_1)e^{\tau(\Lambda_1)}
\h\varrho_-(\Lambda_1)
\ODo\hspace{-1mm}\DDVo
\h\varrho_-^T(\Lambda_{1}),}
\end{array}
\ee
and hence
\be{PCPIdet}
\hD\approx \hi-\frac{\beta p^{(1)}(\Lambda_1)q^{(1)}(\Lambda_2)
e^{\tau(\Lambda_1)-\tau(\Lambda_2)}}
{1+\alpha\beta p^{(1)}(\Lambda_1)q^{(1)}(\Lambda_2)
e^{\tau(\Lambda_1)-\tau(\Lambda_2)}}
\varrho_-(\Lambda_1)\ODo\hspace{-1mm}\DODd\varrho_+^{-1}(\Lambda_2),
\ee
\be{PCPIP11}
\hPP_{11}\approx -\Lambda_2\hi+
\frac{\Lambda_{21} \beta p^{(1)}(\Lambda_1)q^{(1)}(\Lambda_2)
e^{\tau(\Lambda_1)-\tau(\Lambda_2)}}
{1+\alpha\beta p^{(1)}(\Lambda_1)q^{(1)}(\Lambda_2)
e^{\tau(\Lambda_1)-\tau(\Lambda_2)}}
\varrho_-(\Lambda_1)\ODo\hspace{-1mm}\DODd\varrho_+^{-1}(\Lambda_2),
\ee
\be{PCPIP12}
\hPP_{12}\approx \frac{\Lambda_{21} p^{(1)}(\Lambda_1)e^{\tau(\Lambda_1)}}
{1+\alpha\beta p^{(1)}(\Lambda_1)q^{(1)}(\Lambda_2)
e^{\tau(\Lambda_1)-\tau(\Lambda_2)}}
\varrho_-(\Lambda_1)\ODo\hspace{-1mm}\DDVo\varrho_-^{T}(\Lambda_1).
\ee

Now, let us imtroduce the coefficients of the  asymptotic expansion
\be{SPphi0}
\hPo(\lambda)=\hI+\frac1\lambda\hB^{(0)}+\frac1{\lambda^2}\h C^{(0)}
\dots,
\ee
Since  $\hPo\approx\h I$ for complex $\lambda$, we 
have that $\hB^{(0)}, \h C^{(0)}\approx 0$.
Using expansion \eq{SPphi0} and expansion \eq{asrho}
for the function $\h\varrho$, we obtain the following asymptotic
representations for the operators $\hat b_{kj}$ and $\hat c_{kj}$,
\be{PCPSPb11}
\hb_{11}=\hPP_{11}+\h\varrho_0+\Lambda_2\hi +  o(1),
\ee
\be{PCPSPc11}
\h c_{22}-\h c_{11}=
(\Lambda_1\hi+\hPP_{22})(\Lambda_1\hi-\h\varrho_0^T)-
(\Lambda_2\hi+\hPP_{11})(\Lambda_2\hi+\h\varrho_0)-
\h\varrho_1-\h\varrho_1^T+\left(\h\varrho_0^T\right)^2
+ o(1),
\ee
and
\be{PCPSPb12}
\hb_{12}=\hPP_{12} + o(1).
\ee
The logarithmic derivatives of the Fredholm determinant with respect
to $x$ and $t$ (i.e. $\lambda_0$ and $t$) are equal to the traces of the
operators \eq{PCPSPb11} and \eq{PCPSPc11} respectively.  Consider more detailed
the first one. From \eq{PCPIP11} and \eq{PCPSPb11} we have
\be{PCPSPexplb11}
\hb_{11} = \h\varrho_0+\frac{\beta\Lambda_{21}
p^{(1)}(\Lambda_1)q^{(1)}(\Lambda_2) e^{\tau(\Lambda_1)-\tau(\Lambda_2)}}
{1+\alpha\beta p^{(1)}(\Lambda_1)Q^{(1)}(\Lambda_2)
e^{\tau(\Lambda_1)-\tau(\Lambda_2)}}
\h\varrho_-(\Lambda_1)\ODo\hspace{-1mm}
\DODd\h\varrho_+^{-1}(\Lambda_2)
+o(1)
\ee
hence
\ba{PCPSPtrb11}
&&{\dis\hspace{-1cm}
\tr\hb_{11}=\tr\h\varrho_0+\frac{\alpha\beta\Lambda_{21}
p^{(1)}(\Lambda_1)q^{(1)}(\Lambda_2) e^{\tau(\Lambda_1)-\tau(\Lambda_2)}}
{1+\alpha\beta p^{(1)}(\Lambda_1)q^{(1)}(\Lambda_2)
e^{\tau(\Lambda_1)-\tau(\Lambda_2)}} + o(1)}\non
&&{\dis\hspace{1cm}
=\tr\h\varrho_0-i\frac{\partial}{\partial x}\ln
\left(1+\alpha\beta
p^{(1)}(\Lambda_1)q^{(1)}(\Lambda_2) e^{\tau(\Lambda_1)-\tau(\Lambda_2)}
\right)
+o(1).}
\ea
The trace of the operator $\h\varrho_0$ as before is equal to
the first coefficient of the asymptotic expansion of its determinant
$\Delta_0$ (see \eq{NCPasDelta}), which due to \eq{PCPsolDelta} is
equal to
\be{PCPsolDelta0}
\Delta_0=\frac{1}{2\pi i}
\int_\Gamma\,d\mu\sign(\lambda_0-\Re \mu)
\ln\left(1-\vartheta(\mu)
\Bigl(1+e^{\phi(\mu)\sign(\Re \mu - \lambda_0)}
\Bigr)\right).
\ee
In comparison with \eq{NCPsolDelta0} the difference is that one
should integrate now with respect to deformed contour $\Gamma$ instead
of the real axis.

Similarly one can find the trace of the operator $\h c_{22}-\h c_{11}$.
We do not present here the details of these  calculations; they
are quite strightforward, although rather long. The resulting formula for the
leading term of the asymptotics of the Fredholm determinant is
\ba{PCPasdet2}
&&{\dis\hspace{-15mm}
\det(\tilde I+\tilde V)=
\left(1+\alpha\beta
p^{(1)}(\Lambda_1)q^{(1)}(\Lambda_2) e^{\tau(\Lambda_1)-\tau(\Lambda_2)}
\right)}\non
&&{\dis\hspace{-8mm}
\times \exp\left\{
\frac{1}{2\pi} \int_\Gamma\,d\mu|x-2\mu t|
\ln\left(1-\vartheta(\mu)
\Bigl(1+e^{\phi(\mu)\sign(\Re \mu-\lambda_0)}
\Bigr)\right)\right\}.}
\ea
The symbol $|x-2\mu t|$ is understood according to the equation,
$$
|x-2\mu t| \equiv (x-2\mu t)\sign(\Re \mu-\lambda_0).
$$

We would like to draw attention of the reader to the difference between
the cases of positive and  negative chemical potential. In the last case
the determinants of the operators $\hG_{11}$ and $\hG_{22}$
\be{PCPdethG111}
\det\hG_{11(22)}(\lambda)=1-\vartheta(\lambda)
\Bigl(1+e^{-(+)\phi(\lambda)}\Bigr)
\ee
have no zeros at the real axis, and the zero-index condition,
\be{zeroindex}
\arg \det \left(\theta(\lambda_{0} -\lambda)\hat G_{11}(\lambda) +
\theta(\lambda -\lambda_{0})(\hat G^{T}_{22})^{-1}(\lambda)
\right) \to 0, \quad |\lambda| \to \infty,
\ee
takes place on the real line.
Thus the integral in \eq{NCPasdet2}
is well defined. In the case of  positive chemical potential
$\det\hG_{11}$  have real roots, and therefore one should choose
the integration contour more accurately in order to satisfy \eq{zeroindex}.
 We have seen, that this means
that the real axis have to be replaced by the contour $\Gamma$.

The second important difference, is that the leading term of the
solution of the RHP $(a)$ is diagonal in the case of negative
chemical potential \eq{NCPsolU}. Thus the coefficient $\hb_{12}$
is asymptoticly equal to zero (in the next
section we will see that actually $\hb_{12}=
{\cal O}(t^{-1/2})$). This means that in the case of negative chemical potential
the accuracy we reached so far  is not enough for
evaluation of the full combination $\int\,dudv\hb_{12}\cdot\det\freop$,
which describes the correlation function. On the contrary in the case
of positive chemical potential we obtain non-zero value for $\hb_{12}$
already in the leading term of the asymptotics,
\be{PCPSPRHsolb12}
\hb_{12}=\frac{\Lambda_{21} p^{(1)}(\Lambda_1)e^{\tau(\Lambda_1)}}
{1+\alpha\beta p^{(1)}(\Lambda_1)q^{(1)}(\Lambda_2)
e^{\tau(\Lambda_1)-\tau(\Lambda_2)}}
\h\varrho_-(\Lambda_1)\ODo\hspace{-1mm}\DDVo\h\varrho_-^{T}(\Lambda_1)
\ee
$$
+ o(1).
$$
Observe, that the pre-exponent factor in \eq{PCPasdet2} cancels the
denominator in \eq{PCPSPRHsolb12}, so we obtain
\ba{PCPasdet3}
&&{\dis\hspace{-17mm}
{\cal B}=
\Lambda_{21} p^{(1)}(\Lambda_1)e^{\tau(\Lambda_1)}\stint\,dudv
\h\varrho_-(\Lambda_1)\ODo\hspace{-1mm}\DDVo\h\varrho_-^{T}(\Lambda_1)
}\non
&&{\dis\hspace{-8mm}
\times \exp\left\{
\frac{1}{2\pi} \int_\Gamma\,d\mu|x-2\mu t|
\ln\left(1-\vartheta(\mu)
\Bigl(1+e^{\phi(\mu)\sign(\Re \mu-\lambda_0)}
\Bigr)\right)\right\}.}
\ea

As we have mentioned already, the different asymptotics for
negative and positive chemical potential has deep physical origin.
This difference becomes essentially important in the low-temperature
limit.

\section{The localized RHP\label{L} }

The estimates of the previous sections are not enough for
comprehensive analysis of the correlation function \eq{Itempcorrel}.
In particular, we saw that for the case of negative chemical
potential the leading term of the asymptotics of the
RHP $(a)$ solution is diagonal, therefore we can not find the
factor $\hb_{12}$, which is necessary for study of the correlation
function. In order to improve the estimates obtained in the
previous sections one needs
to perform the second step of the nonlinear steepest descent method,
i.e. to consider
the RHP in the vicinity of the saddle point---so called localized
Riemann--Hilbert problem.  Setting $\h U\approx\h I$ in \eq{NCPsolU}
we neglect the contribution of the vicinity
$|\lambda-\lambda_0|<t^{-1/2}$. Now we consider this
problem in more detail. The difference between positive and
negative chemical potential now is not very essential, therefore
for simplicity we concentrate our attention at the negative chemical
potential only.

Since the vicinity of the saddle point $|\lambda-\lambda_0|<t^{-1/2}$
is small for large $t$ one could replace jump matrices
$\h N_\pm(\lambda)$ and $\h M_\pm(\lambda)$ by their values in the
point $\lambda_0$: $\h N_\pm(\lambda_0)$ and $\h M_\pm(\lambda_0)$.
However, $\h N_\pm(\lambda)$ and $\h M_\pm(\lambda)$ are well
defined  only in the vicinity of $\lambda_0$, but not exactly in the
saddle point. Indeed, all of them depend on $\h\varrho$ (see
\eq{NCPQ1a}, \eq{NCPP1}, \eq{NCPQ2}, \eq{NCPP2a}, \eq{NCPMplminus}
and \eq{NCPNplminus}). In turn, $\h\varrho$ satisfies RHP $(b)$
with jump operator having a discontinuity in the saddle point.  Recall the
jump condition for the boundary values of $\h\varrho$:
\be{LscalRHrho2}
\h\varrho_-=\h\varrho_+\h{\cal D}\qquad
\lambda\in R.
\ee
where
\be{LR} \h{\cal D}=\theta(\lambda_0-\lambda)\h G_{11}
+\theta(\lambda-\lambda_0)
\bigl(\h G_{22}^T\bigr)^{-1}.
\ee
It is easy to see that
\be{limR}
\h{\cal D}_l\equiv\lim_{\lambda\to\lambda_0-0}\h{\cal D}(\lambda)\ne
\h{\cal D}_r\equiv\lim_{\lambda\to\lambda_0+0}\h{\cal D}(\lambda).
\ee
Thus, the solution $\h\varrho$ of the RHP $(b)$ has a branch point at
$\lambda=\lambda_0$.

The explicit  expression for jump operator $\h{\cal D}$ is given by
\eq{Fjumpmatreg}, \eq{LR}. It has the following structure
\be{LR11}
\h{\cal D}=\theta(\lambda_0-\lambda)(\hi+f_1(\lambda)\h\omega)+
\theta(\lambda-\lambda_0)(\hi+f_2(\lambda)\h\omega).
\ee
Here $f_1(\lambda)$ and $f_2(\lambda)$ are equal to
\be{Lf12}
\begin{array}{l}
{\dis
f_1(\lambda)=-\vartheta(\lambda)Z(\lambda,\lambda)
e^{\phi_D(\lambda)},}\numa{35}
{\dis f_2(\lambda)=\frac{\vartheta(\lambda)Z(\lambda,\lambda)
e^{\phi_A(\lambda)}}
{1-Z(\lambda,\lambda)\vartheta(\lambda)e^{\phi_A(\lambda)}},}
\end{array}
\ee
and
\be{Lomega}
\h\omega(\lambda)\equiv \h\omega(\lambda|u,v)=\odr\odl.
\ee
The operator $\h\omega$ will play an im\-por\-tant role.  Obviously
$\h\omega$ is one-dimensional projector with the trace equal to $1$,
hence,
\be{Lprop1}
\h\omega^n=\h\omega.
\ee
Defining an operator-valued function
$(\lambda-\lambda_0)^{\xi\h\omega}$, where $\xi$ is a number, as a formal
Taylor series with respect to $\h\omega$ we obtain
\be{Ldeffunct1}
(\lambda-\lambda_0)^{\xi\h\omega}=\hi +\h\omega
\left((\lambda-\lambda_0)^{\xi}-1\right)
\ee
We shall consider formula \eq{Ldeffunct1} as non-formal
de\-fi\-ni\-ti\-on of the ope\-ra\-tor-valued function
$(\lambda-\lambda_0)^{\xi\h\omega}$. We will also assume that
the branch of the function $(\lambda-\lambda_0)^{\xi}$ is
fixed by the condition,
\be{fixbranch}
-\pi < \arg (\lambda-\lambda_0) < \pi.
\ee 

 It is easy to check the following
properties
\be{Lprop3}
\left((\lambda-\lambda_0)^{\xi\h\omega}\right)^{-1}
=(\lambda-\lambda_0)^{-\xi\h\omega}
=\hi +\h\omega\left((\lambda-\lambda_0)^
{-\xi}-1\right),
\ee
\be{Lprop4}
(\lambda-\lambda_0)^{\xi\h\omega}_+
(\lambda-\lambda_0)^{-\xi\h\omega}_-
=\hi +\theta(\lambda_0-\lambda)\h\omega
\left(e^{2\pi i\xi}-1\right)
=\theta(\lambda_0-\lambda)e^{2\pi i\xi\h\omega}+
\theta(\lambda-\lambda_0)\hi,
\ee

Put
\be{Lsubforrho}
\h\varrho(\lambda)=\h\varrho^{(c)}(\lambda)
(\lambda-\lambda_0)^{is\h\omega_0},
\ee
where 
$\h\omega_0=\h\omega(\lambda_0)$ and $s$ is a complex parameter. Then
jump condition for $\h\varrho^{(c)}$ takes the form
\be{Ljump2}
\h\varrho^{(c)}_-=\h\varrho^{(c)}_+\h{\cal D}_0,
\ee
where
\be{LR13}
\h{\cal D}_0(\lambda)=(\lambda-\lambda_0)_+^{is\h\omega_0}
\h{\cal D}(\lambda)(\lambda-\lambda_0)_-^{-is\h\omega_0}.
\ee
We want equation \eq{Lsubforrho} to represent the
singularity of $\h\varrho(\lambda)$
near the saddle point $\lambda_{0}$, i.e. we want 
the $\pm$-values of the factor $\h\varrho^{(c)}(\lambda)$ to be continuous
at $\lambda_{0}$. To ensure this property we demand that $\h{\cal D}_0(\lambda)$ 
has no jump at $\lambda_0$ (cf. the analysis of the usual RHPs near the
points of the discontinuities of the jump coefficients \cite{GK}).
Observe that the limit values of $\h{\cal D}(\lambda)$ in the saddle
point  $\h{\cal D}_l$ and $\h{\cal D}_r$ commute
with $(\lambda-\lambda_0)^{is\h\omega_0}$. Thus,  using
\eq{Lprop4} we have
\be{Llrlimits}
\begin{array}{l} {\dis
\lim_{\lambda\to\lambda_0-0}\h{\cal D}_0(\lambda)= e^{-2\pi
s\h\omega_0}(\hi+f_1(\lambda_0)\h\omega_0),}\num
{\dis
\lim_{\lambda\to\lambda_0+0}\h{\cal D}_0(\lambda)=
\hi+f_2(\lambda_0)\h\omega_0.}
\end{array}
\ee
Making right-hand sides of \eq{Llrlimits} equal to each other we find
\be{Ls1}
s=\frac{1}{2\pi}\ln
\frac{1+f_1(\lambda_0)}{1+f_2(\lambda_0)},
\ee
or, using \eq{Lf12}
\be{Ls2}
s=\frac{1}{2\pi}\ln\left[\left(
1-\vartheta_0Z_0e^{\phi_D(\lambda_0)}\right)
\left(1-\vartheta_0Z_0e^{\phi_A(\lambda_0)}\right)\right],
\ee
where
\be{Ltheta0}
\vartheta_0=\vartheta(\lambda_0),\qquad
Z_0=Z(\lambda_0,\lambda_0).
\ee

Thus, we have found the behavior of $\h\varrho$ in the saddle point.
It is given by \eq{Lsubforrho}, where $\h\varrho^{(c)}$ has
the limits as $\lambda$ approaches $\lambda_0$ from the
upper and from the lower half planes, operator $\h\omega$ is defined in
\eq{Lomega}, $s$ is given by \eq{Ls2}.

\begin{figure}[t]
\begin{center}
\begin{picture}(270,140)
\put(100,60){\vector(1,0){160}}
\put(215,64){$\Im\lambda=0$}
{\thicklines
\put(120,100){\line(1,-1){80}}
\put(120,20){\line(1,1){80}}
\put(120,100){\vector(1,-1){13}}
\put(160,60){\vector(1,-1){35}}
\put(120,20){\vector(1,1){13}}
\put(160,60){\vector(1,1){35}}
}
\put(205,100){$\h N_+$}
\put(100,100){$\h M_+$}
\put(205,20){$\h N_-$}
\put(100,20){$\h M_-$}
\put(155,70){$\lambda_0$}
\end{picture}
\end{center}
\caption{Jump contour and jump matrices in the vicinity of the
saddle point}\label{vicin}
\end{figure}
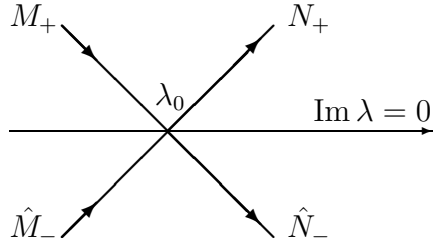

Let us turn back now to the RHP $(d)$ \eq{NCPrhpU} for the operator
$\h U$. We consider this problem in the vicinity of $\lambda_0$.
The corresponding jump contour  is shown on the Fig.\ref{vicin}.
It consists of four branches.
One can  continue the contour on the Fig.\ref{vicin} from the
vicinity $|\lambda- \lambda_0|<t^{-1/2}$ to infinity and
consider new RHP with jump conditions on two infinite axes. Such a
replacement, of course, gives a non-zero contribution in the
solution, but this contribution is exponentially small due to the
presence of factors $e^{\pm\tau}$ in the jump matrices (cf.
the classical analysis of the oscillatory integrals).

Let us introduce a notation $\hW$, which will denote matrices $M_+$,
$M_-$, $N_+$ or $N_-$ for the corresponding  branches of
the contour.  Due to the fact that the functions $\h\varrho^{(c)}_{\pm}(\lambda)$
 are  continuous at
$\lambda_0$, we replace $\hW$ by $\hW_0=\{\h M_\pm^{(0)},\quad \h
N_\pm^{(0)}\}$, where
\be{LMp0}
\h M_+^{(0)}=\left(\begin{array}{ccc}
{\dis \hi}&{}&{\dis 0}\num
{\dis \bigl(\h\varrho^{(c)T}_0\bigr)^{-1}
(\lambda-\lambda_0)^{-is\h\omega_0^T}
\cdot\h Q^{(0)}\cdot(\lambda-\lambda_0)^{-is\h\omega_0}
\bigl(\h\varrho^{(c)}_0\bigr)^{-1}e^{-\tau}}
&{}&{\dis\hi}
\end{array}\right),
\ee
\vskip0.5cm
\be{LNm0}
\h N_-^{(0)}=\left(\begin{array}{ccc}
{\dis\hi}&{}&{\dis 0}\num
{\dis \bigl(\h\varrho^{(c)T}_0\bigr)^{-1}
(\lambda-\lambda_0)^{-is\h\omega_0^T}
\cdot\h{\tilde Q}{}^{(0)}\cdot(\lambda-\lambda_0)^{-is\h\omega_0}
\bigl(\h\varrho^{(c)}_0\bigr)^{-1}e^{-\tau}}
&{}&{\dis\hi}
\end{array}\right).
\ee
\vskip0.5cm
\be{LMm0}
\h M_-^{(0)}=\left(\begin{array}{ccc}
{\dis\hi}&{}&{\dis \h\varrho^{(c)}_0
(\lambda-\lambda_0)^{is\h\omega_0}\cdot\h P^{(0)}\cdot
(\lambda-\lambda_0)^{is\h\omega_0^T}\h\varrho^{(c)T}_0e^\tau}\num
{\dis 0}&{}&{\dis\hi}
\end{array}\right),
\ee
\vskip0.5cm
\be{LNp0}
\h N_+^{(0)}=\left(\begin{array}{ccc}
{\dis\hi}&{}&{\dis \h\varrho^{(c)}_0
(\lambda-\lambda_0)^{is\h\omega_0}
\cdot\h{\tilde P}{}^{(0)}\cdot(\lambda-\lambda_0)^{is\h\omega_0^T}
\h\varrho^{(c)T}_0e^\tau}\num
{\dis 0}&{}&{\dis\hi}
\end{array}\right).
\ee
Here $\h\varrho^{(c)}_0=\h\varrho^{(c)}(\lambda_0)$, and
\be{LQ0}
\h Q^{(0)}e^{-\tau(\lambda_{0})}=\hG_{21}(\lambda_0)
(\hG_{11}(\lambda_0))^{-1},\quad
\h P^{(0)}e^{\tau(\lambda_{0})}=(\hG_{11}(\lambda_0))^{-1}\hG_{12}(\lambda_0)
\ee
\be{LtQ0}
\h{\tilde Q}{}^{(0)}e^{-\tau(\lambda_{0})}=(\hG_{22}(\lambda_0))^{-1}
\hG_{21}(\lambda_0),\quad
\h{\tilde P}{}^{(0)}e^{\tau(\lambda_{0})}=\hG_{12}(\lambda_0)
(\hG_{22}(\lambda_0))^{-1}.
\ee
Strictly speaking, one should add the subscript `$_+$' to $\h\varrho^{(c)}_0$
which appears in the matrices $\h M_+^{(0)}$, $\h N_+^{(0)}$ and the 
subscript `$_-$' to $\h\varrho^{(c)}_0$ which
appears in the matrices $\h M_-^{(0)}$, $\h N_-^{(0)}$.
Unless the distinction is important, we usually will 
omit the subscripts to avoid the overcomplication of the notations.

Based on the experience with the classical NLS (see \cite{DZ2} and \cite{DIZ1}),
we expect that 
the replacement
${\cal W}\to{\cal W}_0$ will enable us to capture the terms
of order $t^{-1/2}$ in the asymptotic solution of
the original operator-valued RHP. These terms, as it
has already been explained in the introduction,
are of the most importance for the comprehensive
asymptotic analysis of the correlation function
\eq{Itempcorrel}.

Replacing ${\cal W}\to{\cal W}_0$, we approximate the
exact solution $\h U (\lambda)$ by the operator-valued
function $\h U^{(0)} (\lambda)$ which solves the following localized RHP
(cf. again \cite{DZ2} and \cite{DIZ1}),
\ba{LRH21}
&&{\dis
1^h.~\h U^{(0)}(\lambda)\to\h I
\quad \lambda\to\infty,}\nona{20}
&&{\dis
2^h.~\h U^{(0)}(\lambda)\quad\mbox{is analytical function of $\lambda$ if}~
\lambda\notin C,}\num
&&{\dis
3^h.~\h U^{(0)}_-=\h U^{(0)}_+\hW_0,\qquad \lambda\in C,}\nonumber
\ea
where $C$ is the contour depicted in Fig.\ref{vicin}. In this and the next two
sections we will construct an explicit solution of this problem.

\vskip .2in
Making the substitution
\be{LPsi}
\h U^{(0)}=\h\Psi\h S,
\ee
where
\be{LhS}
\h S=\left(\begin{array}{cc}
{\dis (\lambda-\lambda_0)^{-is\h\omega_0}
\bigl(\h\varrho^{(c)}_0\bigr)^{-1}e^{-\tau(\lambda)/2}}&{\dis 0}\num
{\dis 0}&{\dis (\lambda-\lambda_0)^{is\h\omega_0^T}
\h\varrho^{(c)T}_0e^{\tau(\lambda)/2}}
\end{array}\right)
\ee
we obtain new RHP for the operator $\h\Psi$:
\ba{LjumpPsi}
&&{\dis
1^i.~\h \Psi(\lambda)\to\h S^{-1}
\quad \lambda\to\infty,}\nona{20}
&&{\dis
2^i.~\h \Psi(\lambda)\quad\mbox{is analytical function of $\lambda$ if}~
\lambda\notin C\cup R,}\num
&&{\dis
3^{i_1}.~\h\Psi_-=\h\Psi_+\hW_\Psi,\qquad \lambda\in C,}
\ea
where
\be{LWPsi}
\hW_\Psi=\h S\hW_0\h S^{-1}=\{\h M_{\pm,\Psi}^{(0)},\quad
\h N_{\pm,\Psi}^{(0)}\},
\ee
and
\be{LMNPsi1}
\h M_{+,\Psi}^{(0)}=\left(\begin{array}{cc}
\hi&0\\ \h Q_0&\hi\end{array}\right),\qquad
\h N_{-,\Psi}^{(0)}=\left(\begin{array}{cc}
\hi&0\\ \h{\tilde Q}{}^{(0)}&\hi\end{array}\right),
\ee
\be{LMNPsi2}
\h M_{-,\Psi}^{(0)}=\left(\begin{array}{cc}
\hi&\h P_0\\0&\hi\end{array}\right),\qquad
\h N_{+,\Psi}^{(0)}=\left(\begin{array}{cc}
\hi&\h{\tilde P}{}^{(0)}\\0&\hi\end{array}\right).
\ee
Since the matrix $\h S$ depends on $(\lambda-\lambda_0)^{is\h\omega_0}$
and $\h\varrho_0^{(c)}$, we have an additional jumps on the real axis
\be{LjumpK}
3^{i_2}.~\h\Psi_-=\h\Psi_+\h K,\qquad \lambda\in R
\ee
where
\be{LK11}
\h K \equiv \h K_+=\left(\begin{array}{cc}
{\dis \hi+f_1(\lambda_0)\h\omega_0}&{\dis 0}\num
{\dis 0}&{\dis \bigl(\hi+f_1(\lambda_0)\h\omega_0^T
\bigr)^{-1}}
\end{array}\right),\qquad \lambda<\lambda_0,
\ee
\be{LK2}
\h K \equiv \h K_-=\left(\begin{array}{cc}
{\dis \hi+f_2(\lambda_0)\h\omega_0}&{\dis 0}\num
{\dis 0}&{\dis \bigl(\hi+f_2(\lambda_0)\h\omega_0^T
\bigr)^{-1}}
\end{array}\right),\qquad \lambda>\lambda_0,
\ee
For the functions $f_{1,2}(\lambda)$ see \eq{Lf12}.

It is convenient to make the following change of the variable $\lambda$,
$$
\lambda \to z : \quad 
\lambda-\lambda_0=ze^{i\frac{\pi}{4}},
$$
so that,
\be{LhS1}
\h S=\left(\begin{array}{cc}
{\dis
z^{-is\h\omega_0}e^{\pi s\h\omega_0/4}
\bigl(\h\varrho^{(c)}_0\bigr)^{-1}
e^{\frac{tz^2}{2}+\frac{it\lambda_0^2}2}}&{\dis 0}\num
{\dis 0}&{\dis z^{is\h\omega_0^T}
e^{-\pi s\h\omega_0^T/4} \h\varrho^{(c)T}_0
e^{-\frac{tz^2}{2}-\frac{it\lambda_0^2}2}}
\end{array}\right)
\ee
Note that matrices $\hW_\Psi$ and $\h K$ do not depend on $\lambda$,
hence they do not change.

Let $\h\kappa$ be the first coefficient of the asymptotic expansion of
$\h U^{(0)}$:
\be{LasyU}
\h U^{(0)}=\h I+ \frac{\h\kappa}{\lambda}+\dots.
\ee
Then the asymptotics of $\h\Psi$ is given by the expression
\be{LasyPsi}
\h\Psi=(\h I +\frac{\h\kappa}{z}e^{-i\frac{\pi}{4}}+\dots)\h S^{-1}.
\ee

Due to the  property that all the jump matrices do not depend on $z$ we
conclude that $(\partial_z\h\Psi)\h\Psi^{-1}$ is holomorphic in the
finite complex plane. The asymptotics of this `logarithmic derivative'
$(\partial_z\h\Psi)\h\Psi^{-1}$ can be found from equality
\be{LasyPsiPsi}
(\partial_z\h\Psi)\h\Psi^{-1}=(\partial_z\h U^{(0)})\h U^{(0){-1}}-
\h U^{(0)}\h S^{-1}\partial_z\h S \h U^{(0)-1}.
\ee
It is easy to see that
\be{LSS}
\h S^{-1}\partial_z\h S =\left(\begin{array}{cc}
{\dis tz\hi-\frac{is}{z}\h\omega_0}&{\dis 0}\num
{\dis 0}&{\dis -tz\hi+\frac{is}{z}\h\omega_0^T}
\end{array}\right),
\ee
therefore
\be{LasyPsi1}
(\partial_z\h\Psi)\h\Psi^{-1}\to -tz\h\sigma_3
+{\cal O}(1),\qquad z\to\infty.
\ee
Due to the Liouville theorem $(\partial_z\h\Psi)\h\Psi^{-1}$ is a
first order polynomial. Using  \eq{LasyPsi} we find
\be{LPsi22}
(\partial_z\h\Psi)\h\Psi^{-1}=-tz\h\sigma_3
-t[\h\kappa,\h\sigma_3]e^{-i\frac{\pi}{4}},
\ee
and, hence, $\h\Psi$ satisfies ordinary differential equation
\be{Ldifeq}
\partial_z\h\Psi=-t(z\h\sigma_3+
[\h\kappa,\h\sigma_3]e^{-i\frac{\pi}{4}})\h\Psi.
\ee

The arguments which led us to the differential equation \eq{Ldifeq}
are identical to the arguments used in the theory of classical
integrable systems. Similar to the pure matrix case \cite{DZ2}, \cite{DIZ1},
\cite{I1},
the first order `matrix' differential equation for $\h\Psi$ implies the
second order differential equation for components $\h\Psi_{jk}$.
After one more replacement, $\xi=z\sqrt{2t}$, we obtain, for example, for
$\Psi_{11}$:
\be{Lsecdifeq}
\frac{d^2}{d\xi^2}\h\Psi_{11}+\left(\frac{1}{2}-
\frac{\xi^2}{4}\right)\h\Psi_{11}+\h\nu\h\Psi_{11}=0.
\ee
Here $\h\nu=-2it\h\kappa_{12}\h\kappa_{21}$.  This equation looks
like the parabolic cylinder equation, except that $\h\nu$ {\it is an
operator}.  In the next section we shall give rigorous
definition of solution of equation \eq{Lsecdifeq}. At the moment  one
can consider the parabolic cylinder function (PCF) $D_{\h\nu}(\xi)$
with operator-valued index $\h\nu$ as a formal Tailor series with
respect to $\h\nu$. This series satisfies recurrence, which is valid
for usual PCF:
\be{Lprop11}
\begin{array}{l} {\dis
\frac{d}{d\xi}D_{\nu}(\xi)+\frac{\xi}{2}D_{\nu}(\xi) 
-\h\nu D_{\nu-1}(\xi)=0,}\num
{\dis
\frac{d}{d\xi}D_{\nu}(\xi)-\frac{\xi}{2}D_{\nu}(\xi)
 + D_{\nu+1}(\xi)=0,}
\end{array}
\ee
Using these  properties, one can check directly that the following
operator-valued matrix
\be{LsolutionPsi}
\h\Psi=\left(\begin{array}{cc}
{\dis D_{\h\nu}(\xi)}&
{\dis \sqrt{2t}e^{i\frac{\pi}{4}}\h\kappa_{12}
D_{\h{\tilde\nu}-\hi}(i\xi)}\num
{\dis \sqrt{2t}e^{-i\frac{\pi}{4}}\h\kappa_{21}
D_{\h\nu-\hi}(\xi)}&
{\dis D_{\h{\tilde\nu}}(i\xi)}
\end{array}\right)\h L.
\ee
satisfies equation \eq{Ldifeq}. Here
\be{Lnu}
\h\nu=-2it\h\kappa_{12}\h\kappa_{21},\qquad
\h{\tilde\nu}=2it\h\kappa_{21}\h\kappa_{12},
\ee
and $\h L$ is piecewise constant operator-valued matrix.

\section{The parabolic cylinder functions with an operator index
\label{A}}
To complete the  solution of the model RHP $(h)$, or
equivalently the RHP $(i)$ (see \eq{LjumpPsi} - \eq{LMNPsi2}),
the quantaties $\hat \kappa_{jl}$ and 
$\hat L$ involved in \eq{LsolutionPsi} are needed
to be determined explicitly in terms of the jump
matrix  $\hW_\Psi$. Also, it follows from the comparision
of  expansion  \eq{LasyU} for $\h U^{(0)}$ 
and expansion \eq{Basyexpan} for original operator $\hx$ that
\be{kappab}
\h\kappa_{12}=\hb_{12}\quad \mbox{and}\quad  \h\kappa_{21}=\hb_{21}.
\ee
Thus, in order
to obtain the asymptotics of the correlation function we need to find
just these operators.

To achieve the objectives formulated above, we have to make
the PCF
$D_{\h\nu}(\xi)$ with the operator-valued index $\h\nu$ a
non-formal object. In fact, we need to control its asymptotic
behavior as $\xi \to \infty$ everywhere on the complex plane $\xi$.
In other words, similar to the classical PCF, we want $D_{\h\nu}(\xi)$ 
to be a genuine special function. To this end we need to know more
about some special features of the operators $\hat \kappa_{jl}$ and 
$\h\nu$ which follow from the setting of the RHP$(i)$. In this 
and the next sections we show that the structure of the
jump matrices of the model RHP$(i)$ suggests a very special
anzats for these operators which allows eventually to reduce
the operator-valued problem $(i)$ to a matrix one. This in turn 
will give us the possibility to obtain the explicit expressions 
for $\hat \kappa_{jl}$.  

\vskip .2in
In the previous section we have already replaced variable $\lambda$ by
variable $\xi$:
\be{Axi}
\lambda-\lambda_0=\frac{\xi}{\sqrt{2t}}e^{i\frac{\pi}{4}}.
\ee
Further we shall work with variable $\xi$ only. Therefore we
present here jump contours in the complex plane of $\xi$.
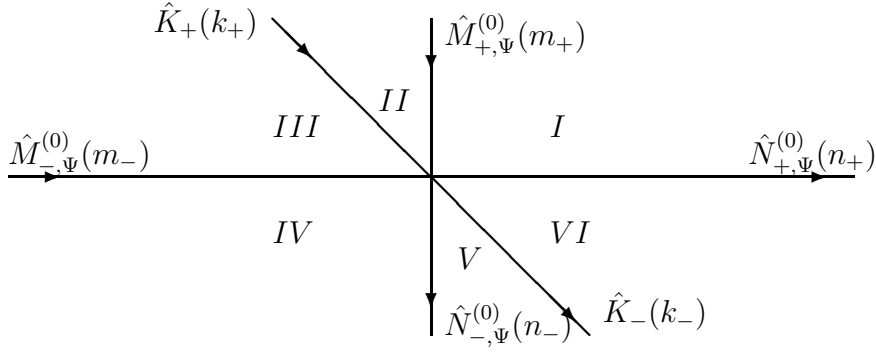
\begin{figure}[t]
\begin{center}
\begin{picture}(360,140)
{\thicklines
\put(20,70){\line(1,0){320}}
\put(180,10){\line(0,1){120}}
\put(120,130){\line(1,-1){120}}
\put(30,70){\vector(1,0){10}}
\put(320,70){\vector(1,0){10}}
\put(180,120){\vector(0,-1){10}}
\put(180,30){\vector(0,-1){10}}
\put(125,125){\vector(1,-1){10}}
\put(225,25){\vector(1,-1){10}}
}
\put(300,75){$\h N^{(0)}_{+,\Psi}(n_+)$}
\put(185,120){$\h M^{(0)}_{+,\Psi}(m_+)$}
\put(185,10){$\h N^{(0)}_{-,\Psi}(n_-)$}
\put(20,75){$\h M^{(0)}_{-,\Psi}(m_-)$}
\put(75,125){$\h K_+(k_+)$}
\put(245,15){$\h K_-(k_-)$}
\put(225,85){$I$}
\put(225,45){$VI$}
\put(160,95){$II$}
\put(120,85){$III$}
\put(120,45){$IV$}
\put(190,35){$V$}
\end{picture}
\end{center}
\caption{Jump contour and jump matrices in the ``$\xi$"-plane}
\label{rotcon}
\end{figure}
The complex $\xi$-plane consists of six sectors $I,\dots,VI$. The
boundaries of the sectors $C_{i-j},\quad i,j=I,\dots,VI$ are
shown on the  Fig.\ref{rotcon}. They are jump contours of the matrix
$\h\Psi$.  The corresponding  operator-valued jump matrices also
placed on the Fig.\ref{rotcon}. In the parenthesis we have placed scalar
analogs of jump matrices, which will be discussed later.

The piecewise constant matrix $\h
L$ and operators $\h\kappa_{12}$ and $\h\kappa_{21}$ are uniquely
defined by the condition that
 matrix operator
$\h\Psi$ \eq{LsolutionPsi} satisfies the jump equations indicated in
\eq{LjumpPsi}-\eq{LMNPsi2} and
possesses asymptotics \eq{LasyPsi}. These restrictions  should
provide us the explicit expressions for  $\h\kappa_{jk}$ and $\h L$.

In what follows we shall consider  vectors $\odr$, $\dvr$, $\odl$
and $\dvl$ only in the point $\lambda=\lambda_0$. Therefore, we
introduce new objects
\be{Adefnewvectors}
\begin{array}{ll}
{\dis
\OD=\frac{\delta_\epsilon(u-\lambda_0)Z(u,\lambda_0)}
{\sqrt{\Ne(\lambda_0)Z_0}},}
\qquad&
{\dis
\DOD=\sqrt{\frac{Z_0}
{\Ne(\lambda_0)}}Z(v,\lambda_0) ,}\non
{\dis
\DV=\sqrt{\frac{Z_0}{\Ne(\lambda_0)}}Z(v,\lambda_0) ,}
\qquad&
{\dis
\DDV=\frac{\delta_\epsilon(v-\lambda_0)Z(v,\lambda_0)}
{\sqrt{\Ne(\lambda_0)Z_0}}.}
\end{array}
\ee
Recall once more that $Z_0=Z(\lambda_0,\lambda_0)$. Denoting as before
$\h\omega(\lambda_0)=\h\omega_0$, we have
\be{Aomega}
\h\omega_0=\OD\sss\DOD,\qquad
\h\omega_0^T=\DV\sss\DDV.
\ee
Using the Taylor series definition, we obtain for an arbitrary holomorphic
at the origin function $f(x)$:
\be{uf1}
\begin{array}{l}
{\dis f\left(x\OD\sss\DOD\right)=f(0)\left(\hi-\OD\sss\DOD\right)
+\OD\sss\DOD f(x),}\num
{\dis f\left(x\DV\sss\DDV\right)=f(0)\left(\hi-\DV\sss\DDV\right)
+\DV\sss\DDV f(x),}
\end{array}
\ee
\be{uf2}
\begin{array}{l}
{\dis \DV\sss\DOD f\left(x\OD\sss\DOD\right)=f(x)\DV\sss\DOD,}\num
{\dis \OD\sss\DDV f\left(x\DV\sss\DDV\right)=f(x)\OD\sss\DDV.}
\end{array}
\ee
Note also the following generalization of \eq{uf1}, 
\be{uf1g}
\begin{array}{l}
{\dis f\left(a\hi + x\OD\sss\DOD\right)=f(a)\left(\hi-\OD\sss\DOD\right)
+\OD\sss\DOD f(x+a),}\num
{\dis f\left(a\hi + x\DV\sss\DDV\right)=f(a)\left(\hi-\DV\sss\DDV\right)
+\DV\sss\DDV f(x+a),}
\end{array}
\ee
which takes place for a function $f(x)$ holomorphic at the point $x = a$.
 
Now, we are ready to describe our anzats. The structure of jump
matrices $\h N^{(0)}_{\pm,\Psi}$ and $\h M^{(0)}_{\pm,\Psi}$ allows
us to look for the operators $\h\kappa_{12}$ and $\h\kappa_{21}$
in the form of one-dimensional projectors. Therefore we set
\be{Ahkappa12pm}
\h\kappa_{12}=\kappa_{12}^\pm\rca\pm\OD\sss\DDV\rcta\pm,
\ee
\be{Ahkappa21pm}
\h\kappa_{21}=\kappa_{21}^\pm\bigl(\rcta\pm\bigr)^{-1}
\DV\sss\DOD\bigl(\rca\pm\bigr)^{-1}.
\ee
Here $\kappa_{12}^\pm$ and $\kappa_{21}^\pm$ are constants, which
should be defined. One should choose in both formul\ae~the sign
plus for sectors $I,II,VI$, but the sign minus for sectors
$III,IV,V$. However, operators $\h\kappa_{jk}$, being the
coefficients of the asymptotic expansion of the operator $\h U$, must
be the same in any sector of complex $\xi$-plane (since, by
definition, they do not depend on $\xi$). Therefore, the values of
$\kappa_{jk}^+$ and $\kappa_{jk}^-$ on the contour $C_{II-III}\cup
C_{V-VI}$ should be related with the jump of the operator $\rc$ on
the same contour, in order to provide the unique value of the
operators $\h\kappa_{jk}$.  Thus, we demand
\be{Ademend}
\kappa_{12}^+\rca{+}\OD\sss\DDV\rcta{+}=
\kappa_{12}^-\rca{-}\OD\sss\DDV\rcta{-}.
\ee
Using \eq{Ljump2}--\eq{Llrlimits} we find
\be{Acorollary}
\begin{array}{l}
{\dis\bigl(\rca{-}\bigr)^{-1}\rca{+}=\left(\hi+f_2(\lambda_0)
\OD\sss\DOD\right)^{-1},}\num
{\dis\biggl(\bigl(\rca{+}\bigr)^{-1}\rca{-}\biggr)^T=
\hi+f_2(\lambda_0)
\DV\sss\DDV.}
\end{array}
\ee
This leads us to the jump condition for $\kappa_{12}$
\be{Akappa12pmr}
\kappa_{12}^-=\kappa_{12}^+(1+f_2(\lambda_0))^{-2}.
\ee
Similar consideration gives
\be{Akappa21pmr}
\kappa_{21}^-=\kappa_{21}^+(1+f_2(\lambda_0))^{2}.
\ee

Substituting \eq{Ahkappa12pm}, \eq{Ahkappa21pm} into expression
\eq{Lnu} for $\h\nu$ we obtain
\be{Ahnu}
\h\nu=\nu\rc\OD\sss\DOD\bigl(\rc\bigr)^{-1},
\ee
and
\be{Athnu}
\h{\tilde\nu}=-\nu\bigl(\rct\bigr)^{-1}\DV\sss\DDV\rct
=-\h\nu^T,
\ee
where
\be{Anu}
\nu=-2it\kappa_{12}^+\kappa_{21}^+
=-2it\kappa_{12}^- \kappa_{21}^-.
\ee
It is worth mentioned that due to \eq{Akappa12pmr} and
\eq{Akappa21pmr} the number $\nu$ (not only the 
operator $\h\nu$ !) is the same in all sectors of
$\xi$-plane.

Using explicit expressions for the operators $\h\nu$ and
$\h{\tilde\nu}$ \eq{Ahnu}, \eq{Athnu} and formul\ae~\eq{uf1},
\eq{uf1g}  one can
define `operator-indexed' PCFs as

\be{SUFPCF}
D_{\h\nu}(\xi)=\rc\left[D_0(\xi)\left(\hi-\OD\sss\DOD\right)
+\OD\sss\DOD D_\nu(\xi)\right]\bigl(\rc\bigr)^{-1}.
\ee

\be{SUFPC1}
D_{\h\nu-\hi}(\xi)=\rc\left[D_{-1}(\xi)\left(\hi-\OD\sss\DOD\right)
+\OD\sss\DOD D_{\nu-1}(\xi)\right]\bigl(\rc\bigr)^{-1}.
\ee

\be{SUFPCF2}
D_{\h{\tilde\nu}}(i\xi)=\bigl(\rct\bigr)^{-1}\left[D_0(i\xi)\left(\hi-\DV\sss\DDV\right)
+\DV\sss\DDV D_{-\nu}(i\xi)\right]\rct.
\ee

\be{SUFPCF3}
D_{\h{\tilde\nu}- \hi}(i\xi)=\bigl(\rct\bigr)^{-1}\left[D_{-1}(i\xi)\left(\hi-\DV\sss\DDV\right)
+\DV\sss\DDV D_{-\nu - 1}(i\xi)\right]\rct.
\ee

\vskip .2in
\noindent
We would like to mention that $\nu$ in these formulas is a number, and
therefore $D_{\nu}(\xi)$ is the usual (not operator-indexed!) PCF. It
follows from \eq{Acorollary} that
\be{commute1}
\rca{-} \OD\sss\DOD\bigl(\rca{-}\bigr)^{-1} =
\rca{+} \OD\sss\DOD\bigl(\rca{+}\bigr)^{-1},
\ee
and
\be{commute2}
\bigl(\rcta{-}\bigr)^{-1}\DV\sss\DDV\rcta{-}=
\bigl(\rcta{+}\bigr)^{-1}\DV\sss\DDV\rcta{+}.
\ee
Therefore, the r.h.s.s of \eq{SUFPCF}- \eq{SUFPCF3} do not have
jumps on the boundaries $C_{II - III}$ and  $C_{V - VI}$.

One can check once more, that functions, defined by \eq{SUFPCF}- \eq{SUFPCF3}
satisfy recurrence \eq{Lprop11}.

\section{Reduction to the matrix RHP\label{R}}

Define two mappings $\hAop M$ (`parallel mapping') and $\hAoo M$
(`orthogonal mapping') of $2\times2$ matrices into algebra of
`hat'-operators. The  first mapping is defined for arbitrary
matrix $M$:
\be{def2}
\hAop M=\left(\begin{array}{cc}
M_{11}\OD\sss\DOD&M_{12}\OD\sss\DDV\num
M_{21}\DV\sss\DOD&M_{22}\DV\sss\DDV
\end{array}\right),
\ee
The second mapping is defined only for diagonal $2\times 2$ matrices:
\be{def3}
\hAoo M=
\left(\begin{array}{cc}
\left(\hi-\OD\sss\DOD\right)M_{11} & 0\num
0 & \left(\hi-\DV\sss\DDV\right)M_{22}
\end{array}\right).
\ee
Similar mappings were introduced by Korepin, and they 
have already been used in \cite{IKW}. Similar to \cite{IKW} it
is strightforward to check that
\be{uf3}
M \mapsto \hAo M \equiv \hAoo I+\hAop M,
\ee
where $I$ is the unite matrix, is a representation of $Gl(2,C)$ (cf. 
\eq{repr}, \eq{vecmod1} in the introduction) .
We would like to emphasize that in distinction of the representation,
considered in \cite{IKW}, {\it the mappings $\hAop M$ and $\hAoo M$
preserve the analytical structure of a matrix function $M(\lambda)$}
 since the projectors
$\OD\sss\DOD$, $\OD\sss\DDV$ etc. do not depend on $\lambda$.

The mappings $\hAop M$ and $\hAoo M$  are not
representations. In particular $\hAop I\ne\h I$ and $\hAoo I\ne\h I$.
Nevertheless these mappings possess very important properties:
\be{property123}
\begin{array}{l}
{\dis\hAop M\hAoo N=\hAoo N\hAop M=0,}\num
{\dis\hAop M\hAop N=\hAop {MN},}\num
{\dis\hAoo M\hAoo N=\hAoo {MN}.}
\end{array}
\ee
These properties allows one to reduce the operator-valued RHP
$(i)$ for $\h\Psi$ to two matrix RHP for parallel and orthogonal
parts of its pre-image.

Let
\be{def4}
\h{\cal R}^{(0)}=\left(\begin{array}{cc}
{\dis\rc}&{\dis 0}\num
{\dis 0}&{\dis\bigl(\rct\bigr)^{-1}}
\end{array}\right).
\ee
Since $\rc$ has two values $\rca{+}$ and $\rca{-}$, the
operator-valued matrix $\h{\cal R}^{(0)}$ also has two 
values:$\h{\cal R}^{(0)}_{+}$ 
in sectors $I, II, VI$, and $\h{\cal R}^{(0)}_{-}$  in 
sectors $III, IV, V$. The jump condition is
\be{MjumpV}
\h{\cal R}^{(0)}_-=\h{\cal R}^{(0)}_+\h K_-,
\ee
where $\h K_-$ is defined in \eq{LK2}.

Let us introduce also two scalar matrices $\beta^\bot$ and $\beta^\pa$:
\be{betao}
\beta^\bot=\left(\begin{array}{cc}
D_0(\xi)&0\num
0 & D_0(i\xi)
\end{array}\right),
\ee
\be{beta}
\beta^\pa=\left(\begin{array}{cc}
D_\nu(\xi)&\kappa_{12}\sqrt{2t}e^{i\frac{\pi}{4}}D_{-\nu-1}(i\xi)\num
\kappa_{21}\sqrt{2t}e^{-i\frac{\pi}{4}}D_{\nu-1}(\xi)
&D_{-\nu}(i\xi)
\end{array}\right).
\ee
The matrix $\beta^\pa$ depends on $\kappa_{12}$ and $\kappa_{21}$, which
have two values in different sectors of the complex plane. Hence,
$\beta^\pa$ also has two values:  $\beta^\pa_+$ in sectors $I,
II, VI$, and $\beta^\pa_-$ in sectors $III, IV, V$. The jump
condition is (cf. \eq{Akappa12pmr}, \eq{Akappa21pmr}) 
\be{MPjumpbeta}
\beta^\pa_-=(k_-)^{-1}\beta^\pa_+\cdot k_-,
\ee
where
\be{MPkm}
k_-=\left(\begin{array}{cc}
1+f_2(\lambda_0)&0\num
0&\bigl(1+f_2(\lambda_0)\bigr)^{-1}
\end{array}\right).
\ee

Let matrix-operator $\h L$ has the structure
\be{hL}
\h L=\h{\cal R}^{(0)}\left(\hAoo{\ell^\bot}+\hAop{\ell^\pa}\right),
\ee
where
\be{ello}
\ell^\bot=\left(\begin{array}{cc}
\ell^\bot_{11}&0\num
0 & \ell^\bot_{22}
\end{array}\right),
\ee
and
\be{ell}
\ell^\pa=\left(\begin{array}{cc}
\ell_{11}&\ell_{12}\num
\ell_{21} & \ell_{22}
\end{array}\right).
\ee
Then combining Eqs. \eq{SUFPCF}-\eq{SUFPCF3}, \eq{betao}--\eq{ell}
and using the properties of mappings \eq{property123} we obtain
\be{repPsi2}
\h\Psi=\h{\cal R}^{(0)}\biggl(\hAoo{\beta^\bot\ell^\bot}
+\hAop{\beta^\pa\ell^\pa}\biggr).
\ee

It is easy to check, that the matrix $\h S^{-1}$ (see \eq{LhS1}) also
has the structure:
\be{reps}
\h S^{-1}=\h{\cal R}^{(0)}\left(\hAoo{s^\bot}+\hAop{s^\pa}\right), 
\ee 
where 
\be{so} 
s^\bot=\left(\begin{array}{cc}
{\dis e^{-\frac{\xi^2}{4}-\frac{it\lambda_0^2}{2}}}&
{\dis 0}\num
{\dis 0}&
{\dis e^{\frac{\xi^2}{4}+\frac{it\lambda_0^2}{2}}}
\end{array}\right),
\ee
\be{s}
s^\pa=\left(\begin{array}{cc}
{\dis e^{-\frac{\xi^2}{4}}\xi^{is}(2t)^{-\frac{is}{2}}
e^{-\frac{\pi s}{4}-\frac{it\lambda_0^2}{2}}}&\hspace{-1cm}
{\dis 0}\num
{\dis 0}&\hspace{-1cm}
{\dis e^{\frac{\xi^2}{4}}\xi^{-is}(2t)^{\frac{is}{2}}
e^{\frac{\pi s}{4}+\frac{it\lambda_0^2}{2}}}
\end{array}\right).
\ee
Here we have used formul\ae~\eq{Aomega} for $\h\omega_0$ and
$\h\omega^T_0$.

Consider now the mappings for jump matrices.
The jump conditions for the operator $\h\Psi$ at the contours
$C_{I-VI}$, $C_{I-II}$, $C_{III-IV}$ and $C_{IV-V}$ are given by
the matrix $\h{\cal W}_\Psi=\{\h N^{(0)}_{\pm,\Psi},\h
M^{(0)}_{\pm,\Psi}\}$ (see Eqs.\eq{LjumpPsi}--\eq{LMNPsi2} and
Fig.\ref{rotcon}).  Formul\ae~for $\h Q^{(0)}$, $\h P^{(0)}$,
$\h{\tilde Q}{}^{(0)}$ and $\h{\tilde P}{}^{(0)}$ are given in
\eq{LQ0}--\eq{LtQ0}. Using explicit expressions for $\h G_{jk}$
(see \eq{Fjumpmatreg}) we find
\be{JMq0}
\h Q^{(0)}=q^{(0)}\DV\sss\DOD,\qquad
\h{\tilde Q}{}^{(0)}={\tilde q}{}^{(0)}\DV\sss\DOD,
\ee
\be{JMp0}
\h P^{(0)}=p^{(0)}\OD\sss\DDV,\qquad
\h{\tilde P}{}^{(0)}={\tilde p}{}^{(0)}\OD\sss\DDV,
\ee
where
\be{Jmqo}
q^{(0)}=\frac{Z_0\vartheta_0
e^{\phi_D(\lambda_0)+\phi_A(\lambda_0)-\psi(\lambda_0)}}
{2\pi i(1-Z_0\vartheta_0e^{\phi_D(\lambda_0)})},\qquad
p^{(0)}=\frac{2\pi i(\vartheta_0-1)Z_0e^{\psi(\lambda_0)}}
{1-Z_0\vartheta_0e^{\phi_D(\lambda_0)}},
\ee
\be{Jmpo}
{\tilde q}{}^{(0)}=\frac{Z_0\vartheta_0
e^{\phi_D(\lambda_0)+\phi_A(\lambda_0)-\psi(\lambda_0)}}
{2\pi i(1-Z_0\vartheta_0e^{\phi_A(\lambda_0)})},
\qquad
{\tilde p}{}^{(0)}=\frac{2\pi i(\vartheta_0-1)Z_0e^{\psi(\lambda_0)}}
{1-Z_0\vartheta_0e^{\phi_A(\lambda_0)}}.
\ee
Obviously, all the jump matrices $\h{\cal W}_\Psi$ can
be written in the form
\be{JmhW}
\h {\cal W}_\Psi=\hAoo I+\hAop {w_\Psi}.
\ee
Here
\be{Jmw}
w_\Psi=\{n_\pm,m_\pm\},
\ee
and
\be{Jmnmpm}
m_+=\left(\begin{array}{cc}
1&0\num
q^{(0)}&1
\end{array}\right),\qquad
n_-=\left(\begin{array}{cc}
1&0\num
{\tilde q}{}^{(0)}&1
\end{array}\right),
\ee
\be{Jmnmmp}
m_-=\left(\begin{array}{cc}
1&p^{(0)}\num
0&1
\end{array}\right),\qquad
n_+=\left(\begin{array}{cc}
1&{\tilde p}{}^{(0)}\num
0&1
\end{array}\right).
\ee
The jump condition at the contours $C_{II-III}$ and $C_{V-VI}$ are
given by matrix $\h K_\pm$, which also can be presented as
\be{JmhK}
\h K_\pm=\hAoo I+\hAop {k_\pm},
\ee
where
\be{Jmkp}
k_+=\left(\begin{array}{cc}
1+f_1(\lambda_0)&0\num
0&\bigl(1+f_1(\lambda_0)\bigr)^{-1}
\end{array}\right),
\ee
\be{Jmkm}
k_-=\left(\begin{array}{cc}
1+f_2(\lambda_0)&0\num
0&\bigl(1+f_2(\lambda_0)\bigr)^{-1}
\end{array}\right).
\ee
The jump matrices $m_\pm$, $n_\pm$ and $k_\pm$ are placed on the
Fig.\ref{rotcon} in the parenthesis.

Thus all operators entering the RHP $(i)$ and its solution are
written as the $\hAo$- mappings of $2\times 2$ matrices.
Let us summarize the preliminary results.

Using the mappings $\hAoo \cdot$ and $\hAop \cdot$ we have presented the
operator-valued matrix $\h\Psi$ as a sum of orthogonal and
parallel parts (see \eq{repPsi2}):
\be{SRHrepPsi2}
\h\Psi=\h{\cal R}^{(0)}\biggl(\hAoo{\beta^\bot\ell^\bot}
+\hAop{\beta^\pa\ell^\pa}\biggr).
\ee
We have similar representation for the operator $\h S^{-1}$
\be{SRHreps}
\h S^{-1}=\h{\cal R}^{(0)}\left(\hAoo{s^\bot}+\hAop{s^\pa}\right).
\ee
This gives us the asymptotic conditions for orthogonal and parallel
parts of $\h\Psi$. Indeed, since $\h\Psi\to\h S^{-1}$, when
$\xi$ goes to infinity, then the pre-images of the parallel and
orthogonal parts of $\h\Psi$ should go to the pre-images of the parallel
and orthogonal parts of $\h S^{-1}$ respectively:
\be{MPasybetao}
\beta^\bot\ell^\bot\stackrel{\xi\to\infty}{\longrightarrow}
s^\bot,
\ee
\be{MPasybeta}
\beta^\pa\ell^\pa\stackrel{\xi\to\infty}{\longrightarrow}
\left(\begin{array}{cc}
{\dis s_{11}^\pa}&{\dis
\kappa_{12}\sqrt{2t}e^{-i\frac{\pi}{4}}{\xi}^{-1}s_{22}^\pa}\num
{\dis \kappa_{21}\sqrt{2t}e^{-i\frac{\pi}{4}}{\xi}^{-1}s_{11}^\pa}
&{\dis s_{22}^\pa}
\end{array}\right).
\ee
Finally, all the jump matrices also are written in terms of
orthogonal and parallel parts:
\be{SRHmhW}
\hW_\Psi=\hAoo
I+\hAop {w_\Psi},
\ee
\be{SRHmhK}
\h K_\pm=\hAoo I+\hAop {k_\pm}.
\ee
The parallel and orthogonal parts of $\h\Psi$ should separately
satisfy jump conditions. Thus, we have reduced the operator-valued
RHP to two matrix ones. Namely, we need now to find piecewise
constant matrices $\ell^\bot$, $\ell^\pa$ and constants
$\kappa_{12}^\pm$ and $\kappa_{21}^\pm$ such that:
\begin{enumerate}
\item
both of the matrices $\beta^\bot\ell^\bot$ and
$\beta^\pa\ell^\pa$ possess the asymptotics prescribed;
\item
both of the matrices $\beta^\bot\ell^\bot$ and
$\beta^\pa\ell^\pa$ possess the jump conditions prescribed.
\end{enumerate}

We start with the asymptotic condition for the orthogonal part.
One should find matrix $\ell^\bot$ such that
\be{SRHbotpart}
\beta^\bot\ell^\bot\stackrel{\xi\to\infty}{\longrightarrow}
\left(\begin{array}{cc}
{\dis e^{-\frac{\xi^2}{4}-\frac{it\lambda_0^2}{2}}}&
{\dis 0}\num
{\dis 0}&
{\dis e^{\frac{\xi^2}{4}+\frac{it\lambda_0^2}{2}}}
\end{array}\right).
\ee
Here $\beta^\bot$ is given by \eq{betao}:
\be{SRHbetao}
\beta^\bot=\left(\begin{array}{cc}
D_0(\xi)&0\num
0 & D_0(i\xi)
\end{array}\right).
\ee
Since $D_0(\xi)=\exp(-\xi^2/4)$, we conclude that
\be{SRHsolell0}
\ell^\bot=\exp\{-\frac{it\lambda_0^2}{2}\sigma_3\}.
\ee

Let us discuss now the jump conditions for orthogonal part.
Observe that orthogonal parts of all jump matrices are equal to
$\hAoo I$. Thus, matrix $\beta^\bot\ell^\bot$ has no jumps on the
contours $C_{I-VI}$, $C_{I-II}$, $C_{III-IV}$ and $C_{IV-V}$. It
also has no jumps at the contours  $C_{II-III}$ and $C_{V-VI}$ in
spite of orthogonal part of $\h\Psi$ depends on matrix $\h{\cal R}^{(0)}$,
which may have jump just at the contours  $C_{II-III}$
and $C_{V-VI}$. Let us check the  corresponding jump condition. We
have (see \eq{Acorollary})
\be{SRHjumprhoo}
\rca{-}=\rca{+}\left(\hi-\OD\sss\DOD+(1+f_2(\lambda_0))
\OD\sss\DOD\right),
\ee
(see \eq{Llrlimits}). Multiplying this equality by
$\hi-\OD\sss\DOD$ from the right, we obtain
\be{SRHjumprhoo1}
\rca{-}\left(\hi-\OD\sss\DOD\right)
=\rca{+}\left(\hi-\OD\sss\DOD\right).
\ee
Thus, the orthogonal part of $\rc$ (and, hence, orthogonal
part of $\h{\cal R}^{(0)}$) has no jump at the contours  $C_{II-III}$ and
$C_{V-VI}$.

The orthogonal part of $\h\Psi$ is found:
\be{SRHPsio}
\h\Psi^\bot=\h{\cal R}^{(0)}\hAoo{\beta^\bot\ell^\bot},
\ee
where $\beta^\bot$ and $\ell^\bot$ are given by \eq{SRHbetao}
and \eq{SRHsolell0}.

Consider now the parallel part of $\h\Psi$. In order
to simplify some formul\ae, we make replacement
\be{SRHellpa}
\ell^\pa=\tilde\ell\exp\left(\left\{-\frac{is}{2}\ln(2t)
-\frac{\pi s }{4}-
\frac{it\lambda_0^2}{2}\right\}\sigma_3\right).
\ee

First, one should find piecewise constant matrix $\tilde\ell$ such that
matrix $\beta^\pa\tilde\ell$ possesses the following asymptotics
\be{SRHasyp}
\beta^\pa\tilde\ell\stackrel{\xi\to\infty}{\longrightarrow}
\left(\begin{array}{cc}
{\dis \xi^{is}e^{-\frac{\xi^2}{4}}}&{\dis
\kappa_{12}\sqrt{2t}e^{-i\frac{\pi}{4}}\xi^{-is-1}
e^{\frac{\xi^2}{4}}}\num
{\dis \kappa_{21}\sqrt{2t}e^{-i\frac{\pi}{4}}
\xi^{is-1}e^{-\frac{\xi^2}{4}}}
&{\dis \xi^{-is}e^{\frac{\xi^2}{4}}}
\end{array}\right).
\ee
Recall that $\beta^\pa$ is equal to
\be{SRHbeta}
\beta^\pa=\left(\begin{array}{cc}
D_\nu(\xi)&\kappa_{12}\sqrt{2t}e^{i\frac{\pi}{4}}D_{-\nu-1}(i\xi)\num
\kappa_{21}\sqrt{2t}e^{-i\frac{\pi}{4}}D_{\nu-1}(\xi)
&D_{-\nu}(i\xi)
\end{array}\right).
\ee

The asymptotics of PCF $D_\nu(\xi)$ strongly depends on argument
of $\xi$. In general form it may be written as
\be{SRHasyPCF}
D_\nu(\xi)\stackrel{|\xi|\to\infty}{\longrightarrow}
\xi^\nu e^{-\frac{\xi^2}{4}}+c(\nu)
\xi^{-\nu-1} e^{\frac{\xi^2}{4}},
\ee
where constant $c(\nu)$ does not depends on $|\xi|$, but only
on $\arg\xi$ \cite{BE}. It easy to see that the asymptotics of
\eq{SRHasyp} can coincide with \eq{SRHbeta} only if
\be{SRHnu}
\nu=is.
\ee
Then we need to choose matrix $\tilde\ell$ in order to provide
the correct asymptotics behavior in every sectors of the complex
$\xi$-plane. We drop out the detail  of these calculations
and present here the  matrix $\tilde\ell$ in each sector:
\be{one}
\tilde\ell=\left(\begin{array}{lcr}
{\dis 1}&{}&{\dis \sqrt{\frac{\pi}{t}}\frac{e^{-\frac{i\pi}{4}}}
{\kappa_{21}\Gamma(\nu)}}\num
{\dis 0}&{}&{\dis e^{\frac{i\pi\nu}{2}}}
\end{array}\right),
\qquad\mbox{in the sector $I$},
\ee
\vspace{0.5cm}
\be{two}
\tilde\ell=\left(\begin{array}{lcr}
{\dis e^{2i\pi\nu}}&{}&
{\dis \sqrt{\frac{\pi}{t}}\frac{e^{-\frac{i\pi}{4}}}
{\kappa_{21}\Gamma(\nu)}}\num
{\dis \sqrt{\frac{\pi}{t}}\frac{e^{\frac{3i\pi\nu}{2}
+\frac{i\pi}{4}}}
{\kappa_{12}\Gamma(-\nu)}}&{}&
{\dis e^{\frac{i\pi\nu}{2}}}
\end{array}\right),
\qquad\mbox{in the sector $II$},
\ee
\vspace{0.5cm}
\be{three}
\tilde\ell=\left(\begin{array}{lcr}
{\dis 1}&{}&
{\dis \sqrt{\frac{\pi}{t}}\frac{e^{2i\pi\nu-\frac{i\pi}{4}}}
{\kappa_{21}\Gamma(\nu)}}\num
{\dis \sqrt{\frac{\pi}{t}}\frac{e^{-\frac{i\pi\nu}{2}
+\frac{i\pi}{4}}}
{\kappa_{12}\Gamma(-\nu)}}&{}&
{\dis e^{\frac{5i\pi\nu}{2}}}
\end{array}\right),
\qquad\mbox{in the sector $III$},
\ee
\vspace{0.5cm}
\be{four}
\tilde\ell=\left(\begin{array}{lcc}
{\dis 1}&{}&
{\dis 0}\num
{\dis \sqrt{\frac{\pi}{t}}\frac{e^{-\frac{i\pi\nu}{2}
+\frac{i\pi}{4}}}
{\kappa_{12}\Gamma(-\nu)}}&{}&
{\dis e^{\frac{i\pi\nu}{2}}}
\end{array}\right),
\qquad\mbox{in the sector $IV$},
\ee
\vspace{0.5cm}
\be{fivesix}
\tilde\ell=\left(\begin{array}{lcc}
{\dis 1}&{}&
{\dis 0}\num
{\dis 0}&{}&
{\dis e^{\frac{i\pi\nu}{2}}}
\end{array}\right),
\qquad\mbox{in the sectors $V$ and $VI$}.
\ee

Let us illustrate how one can obtain these results.  Consider
sectors $V$ and $VI$ where $-\frac{\pi}{2}\leq \arg\xi \leq 0$. For these
values of argument the asymptotics of the PCF is extremely simple:
\be{SRHasy1}
D_\nu(\xi)\to
\xi^\nu e^{-\frac{\xi^2}{4}},
\ee
\be{SRHasy1i}
D_{-\nu}(i\xi)\to
e^{-\frac{i\pi\nu}{2}}\xi^{-\nu} e^{\frac{\xi^2}{4}}.
\ee
Putting  $\tilde\ell_{12}=\tilde\ell_{21}=0$,  $\tilde\ell_{11}=1$
and $\tilde\ell_{22}=e^{\frac{i\pi\nu}{2}}$ we satisfy the asymptotic
condition. It is worth mentioned that matrix $\tilde\ell$ has no
jump at the contour $C_{V-VI}$, which is in completely agreement with
the jump condition \eq{SRHjumpell56} below. Thus, we arrive at \eq{fivesix}.

Consider now sector $I$. Here the asymptotics of $D_\nu(\xi)$ is
just the same, but $D_{-\nu}(i\xi)$ behaves as follows
\be{SRHasy2i}
D_{-\nu}(i\xi)\to
e^{-\frac{i\pi\nu}{2}}\left( \xi^{-\nu} e^{\frac{\xi^2}{4}}-
\frac{\sqrt{2\pi}}{\Gamma(\nu)}e^{-i\frac{\pi}{2}}
\xi^{\nu-1} e^{-\frac{\xi^2}{4}}\right) .
\ee
Putting again $\tilde\ell_{21}=0$ and $\tilde\ell_{11}=1$, we
obtain, that the first column of $\beta^\pa\tilde\ell$
possesses correct asymptotics. But the second column has an additional
term in comparison with sectors $V$ and $VI$. In order to compensate
this term, one should put
$\tilde\ell_{22}=e^{\frac{i\pi\nu}{2}}$
and $\tilde\ell_{12} =\sqrt{\frac{\pi}{t}}\frac{e^{-\frac{i\pi}{4}}}
{\kappa_{21}\Gamma(\nu)}$. We obtain formula \eq{one}.

Similar considerations allow to find all other formul\ae~
\eq{two}--\eq{four} for matrix $\tilde\ell$.
 \vskip .2in

Finally, we should check, that all the jump conditions are valid.
After replacement $\ell^\pa\to\tilde\ell$ the jump conditions for
$\beta^\pa\tilde\ell$ have changed. Let us formulate them
explicitly.

The diagonal matrices $k_\pm$ do not change under
replacement \eq{SRHellpa}. However,  the matrices $m_\pm$ and
$n_\pm$ should be replaced by $m'_\pm$ and $n'_\pm$ respectively:
\be{Jmnmpmprim}
m'_+=\left(\begin{array}{cc}
1&0\num
q'^{(0)}&1
\end{array}\right),\qquad
n'_-=\left(\begin{array}{cc}
1&0\num
{{\tilde q}'}{}^{(0)}&1
\end{array}\right),
\ee
\be{Jmnmmpprim}
m'_-=\left(\begin{array}{cc}
1&p'^{(0)}\num
0&1
\end{array}\right),\qquad
n'_+=\left(\begin{array}{cc}
1&{{\tilde p}'}{}^{(0)}\num
0&1
\end{array}\right),
\ee
where
\be{Jmqprim}
q'^{(0)}=q^{(0)}(2t)^\nu e^{-\frac{i\pi\nu}{2}+it\lambda_0^2},\qquad
{\tilde q'}{}^{(0)}={\tilde q}{}^{(0)}(2t)^\nu
e^{-\frac{i\pi\nu}{2}+it\lambda_0^2},
\ee
\be{Jmpprim}
p'^{(0)}=p^{(0)}(2t)^{-\nu} e^{\frac{i\pi\nu}{2}-it\lambda_0^2},\qquad
{{\tilde p}'}{}^{(0)}={\tilde p}{}^{(0)}(2t)^{-\nu}
e^{\frac{i\pi\nu}{2}-it\lambda_0^2}.
\ee
Thus, the matrix $\tilde\ell$ should possess the following
jumps:
\be{SRH16}
\tilde\ell_{VI}=\tilde\ell_{I}n'_+, \qquad
\mbox{at the contour $C_{I-VI}$},
\ee
\be{SRH12}
\tilde\ell_{II}=\tilde\ell_{I}m'_+, \qquad
\mbox{at the contour $C_{I-II}$},
\ee
\be{SRH34}
\tilde\ell_{IV}=\tilde\ell_{III}m'_-, \qquad
\mbox{at the contour $C_{III-IV}$},
\ee
\be{SRH45}
\tilde\ell_{IV}=\tilde\ell_{V}n'_-, \qquad
\mbox{at the contour $C_{IV-V}$}.
\ee
As for contours $C_{II-III}$ and $C_{V-VI}$, the situation is
slightly more complicated. Recall jump conditions for $\h{\cal R}^{(0)}$ and
$\beta^\pa$
\be{SRHjumpV}
\h{\cal R}^{(0)}_-=\h{\cal R}^{(0)}_+(\hAoo I+\hAop{k_-}),
\ee
\be{SRHjumpbeta}
\beta^\pa_-=(k_-)^{-1}\beta^\pa_+\cdot k_-.
\ee
Then for the parallel part of $\h\Psi$ we have at the contour
$C_{V-VI}$:
\ba{SRHjump56}
{\dis \h\Psi^\pa_-}&=&{\dis\h{\cal R}^{(0)}_-\hAop{\beta^\pa_-\tilde\ell_V}
=\h{\cal R}^{(0)}_+\hAop{\beta^\pa_+k_-\tilde\ell_V(k_-)^{-1}}
\hAop{k_-}}\non
&=&{\dis \h{\cal R}^{(0)}_+\hAop{\beta^\pa_+\tilde\ell_{VI}}
\hAop{k_-}}.
\ea
Hence, we find that
\be{SRHjumpell56}
\tilde\ell_{VI}=k_-\tilde\ell_{V}(k_-)^{-1},
\qquad\mbox{at the contour $C_{V-VI}$}.
\ee
Similarly
\be{SRHjumpell23}
\tilde\ell_{II}=k_-\tilde\ell_{III}(k_+)^{-1},
\qquad\mbox{at the contour $C_{II-III}$}.
\ee

One can easily  check that the jumps on the contours $C_{V-VI}$ and
$C_{II-III}$ are  valid automatically. The jumps on the other
contours give us explicit expressions for $\kappa^\pm_{12}$ and
$\kappa^\pm_{21}$. Indeed, for example, due to \eq{SRH16} we have
\be{SRHjumpell12}
(\tilde\ell_{VI})_{12}=(\tilde\ell_{I})_{12}+\tilde
p'^{(0)}(\tilde\ell_{I})_{11},
\ee
which implies
\be{SRHtaksebe}
(\tilde\ell_{I})_{12}=-{\tilde p}'^{(0)},
\ee
and, hence,
\be{SRHkappa21p}
\kappa^+_{21}=-\sqrt{\frac{\pi}{t}}\frac{e^{-\frac{i\pi}{4}}}
{{\tilde p}'^{(0)}\Gamma(\nu)}.
\ee
Then due to \eq{Anu}
\be{SRHkappa12p}
\kappa^+_{12}=\frac{i\nu}{2t\kappa^+_{21}}.
\ee
Using equalities \eq{Akappa12pmr} and \eq{Akappa21pmr} we can
find $\kappa^-_{12}$ and $\kappa^-_{21}$.

The jumps on the other contours give us just the same expressions
for $\kappa^\pm_{jk}$. One should only take into account the
identity
\be{SRHident}
{\tilde p}'{}^{(0)}q'{}^{(0)}=
{\tilde q}'{}^{(0)}p'{}^{(0)}=1-e^{2i\pi\nu}.
\ee
This identity can be checked directly, using explicit expressions
for all variables entering \eq{SRHident} (see \eq{Jmqprim}-\eq{Jmpprim},
\eq{Jmqo}-\eq{Jmpo}).

Combining all necessary formul\ae~after simple algebra we arrive at
\be{SRHhkappa12}
\h\kappa_{12}=\frac{\gamma\nu}{2\pi i}\sqrt{\frac{\pi}{t}}
\rca{-}\OD\sss\DDV\rcta{+},
\ee
\be{SRHhkappa21}
\h\kappa_{21}=-\frac{1}{\gamma}\sqrt{\frac{\pi}{t}}
\bigl(\rcta{+}\bigr)^{-1}\DV\sss\DOD\bigl(\rca{-}\bigr)^{-1}.
\ee
Here
\be{SRHc}
\gamma=2\pi Z_0(\vartheta_0-1)\Gamma(\nu)(2t)^{-\nu}
e^{\psi(\lambda_0)+\frac{i\pi\nu}{2}+\frac{3i\pi}{4}
-it\lambda_0^2},
\ee
and
\be{SRHnu1}
\nu=-\frac{1}{2\pi i}\ln\left[\left(
1-\vartheta_0Z_0e^{\phi_D(\lambda_0)}\right)
\left(1-\vartheta_0Z_0e^{\phi_A(\lambda_0)}\right)\right].
\ee
\vskip .2in

\noindent
This  gives us  the solution $\h\Psi$ of the localized RHP
$(i)$. Let us present here once more the main formul\ae. The operator
$\h\Psi$ is equal to
\be{RESsol}
\h\Psi=\left(\begin{array}{cc}
{\dis\rc}&{\dis 0}\num
{\dis 0}&{\dis\bigl(\rct\bigr)^{-1}}
\end{array}\right)
\biggl(\hAoo{\beta^\bot\ell^\bot}
+\hAop{\beta^\pa\ell^\pa}\biggr).
\ee
The definition of the mappings $\hAoo\cdot$ and $\hAop\cdot$ is given
in \eq{def2}, \eq{def3}. The matrix $\beta^\bot\ell^\bot$ is equal to
\be{RESbetaort}
\beta^\bot\ell^\bot=\exp\left\{-\left(\frac{\xi^2}{4}+
\frac{it\lambda_0^2}{2}\right)\sigma_3\right\}.
\ee
The parallel part of the matrix $\beta$ is
\be{RESbetapar}
\beta^\pa=\left(\begin{array}{cc}
D_\nu(\xi)&\kappa_{12}\sqrt{2t}e^{i\frac{\pi}{4}}D_{-\nu-1}(i\xi)\num
\kappa_{21}\sqrt{2t}e^{-i\frac{\pi}{4}}D_{\nu-1}(\xi)
&D_{-\nu}(i\xi)
\end{array}\right).
\ee
Here $D_\nu(\xi)$ is the PCF, $\nu$ is given by \eq{SRHnu1}.
The constants $\kappa_{12}$ and $\kappa_{21}$ have two values:
$\kappa_{jk}^+$ for the sectors $I$, $II$ and $VI$, and
$\kappa_{jk}^-$ for the sectors $III$, $IV$ and $V$ (see
Fig.\ref{rotcon}). The values $\kappa_{jk}^+$ are defined in
\eq{SRHkappa21p}, \eq{SRHkappa12p}. The constants $\kappa_{jk}^-$
are related with the last ones by \eq{Akappa12pmr}, \eq{Akappa21pmr}.

The matrix $\ell^\pa$ it is given by Eqs. \eq{SRHellpa},
\eq{one}--\eq{fivesix}.

For our purposes it is most important to know the operators
$\h\kappa_{12}$ and $\h\kappa_{21}$  since the equations,
\be{RESkapb}
\h\kappa_{12}=\hb_{12},\qquad \h\kappa_{21}=\hb_{21}.
\ee
These operators are given by \eq{SRHhkappa12}, \eq{SRHhkappa21}.

\vskip .2in
The case of the positive chemical potential can be delt with in
a similar fashion. Instead of the operators $\h Q^{(0)}$ and
$\h P^{(0)}$ one should consider operators
\be{RESrepl}
\h Q^{(0)}_1=\h Q^{(0)}\left(\frac{\lambda_{0}-\Lambda_2}
{\lambda_{0}-\Lambda_1}\right),\qquad
\h P^{(0)}_1=\h P^{(0)}\left(\frac{\lambda_{0}-\Lambda_1}
{\lambda_{0}-\Lambda_2}\right).
\ee
The corresponding replacement should be done for the scalars
$q^{(0)}$ and $p^{(0)}$ as well. The rest of calculations is
just the same as for the case of the negative chemical potential.
The difference, however, is that for the negative chemical potential
formul\ae~\eq{RESkapb} provide us with the leading term of the large
time asymptotic expansion of the operators $\hb_{12}$ and $\hb_{21}$,
while for the positive chemical potential the operators
$\h\kappa_{12}$ and $\h\kappa_{21}$ define only corrections to the
coefficients $\hb_{12}$ and $\hb_{21}$. Indeed, in the last case, as
we have seen, the leading term of the asymptotic expansion of
$\hb_{12}$ is given by \eq{PCPSPRHsolb12}. In comparison with this
expression, the operators \eq{SRHhkappa12}, \eq{SRHhkappa21} have the
order of $t^{-1/2}$, hence, they are only corrections to the leading
terms.  Another words in the case of positive chemical potential the
operators $\h\kappa_{12}$ and $\h\kappa_{21}$ are equal to the
coefficients of the asymptotic expansion \eq{SPphi0}
$\h B_{12}^{(0)}$ and $\h B_{21}^{(0)}$ respectively:
\be{RESkapB}
\h\kappa_{12}=\h B_{12}^{(0)},
\qquad \h\kappa_{21}=\h B_{21}^{(0)}.
\ee
 \vskip.2in

Finally, we need to say few words about the operator $\rc$, entering
Eqs. \eq{SRHhkappa12}, \eq{SRHhkappa21}. This operator is defined
by the equation,
\be{rhodef}
\rc = \h\varrho^{(c)}(\lambda_{0}) 
\ee
where the operator-valued function $\h\varrho^{(c)}(\lambda)$ in turn
is defined as a unique solution of the RHP \eq{Ljump2}-\eq{LR13}. This
RHP is equivalent to the singular integral equation (cf. \eq{sie}),
\be{sierho}
\h\varrho^{(c)}_{+}(\lambda) = \hi + \frac{1}{2\pi }
\int_{-\infty}^{+\infty}\frac{d\mu}{\mu -\lambda -i0}
\h\varrho^{(c)}_{+}(\mu)\left( \hi - \hat{\cal D}_{0}(\mu)\right).
\ee
Therefore, to determine $\rc$ one has to solve this integral equation.
In other words, the time-independent operator factor $\rc$
in our final formulae is still a transcendental object.
It is clear however, that it does not depend on the function
$\psi(\lambda)$  since the jump matrix $\hat{\cal D}_{0}(\lambda)$
only depends on the functions $\phi_A(\lambda)$, $\phi_D(\lambda)$,
and the ratio $\lambda_{0} = x/2t$ (and, of course, on the chemical
potential, coupling constant and temperature).
We shall see later that this property is sufficient in order to
calculate the asymptotics of the correlation function up to the
numerical constant (depending on $\lambda_{0}$, $h$, $c$, and $T$).

\section{Differential equations\label{DE}}

In this section we continue to consider the case of negative chemical
potential. 

The results of the previous sections  provides us the
leading terms of the large time asymptotics for the coefficients $\hb_{jk}$. 
In particular, we have found that
\be{ACFb12}
\h b_{12}\approx \frac{\tilde\gamma\nu}{i\sqrt{2\pi}}(2t)^{-\nu-1/2}
e^{\psi(\lambda_0)-it\lambda_0^2}
\rca{-}\OD\sss\DDV\rcta{+},
\ee
\be{ACFb21}
\h b_{21}\approx -\frac{\sqrt{2\pi}}{\tilde\gamma}(2t)^{\nu-1/2}
e^{-\psi(\lambda_0)+it\lambda_0^2}
\bigl(\rcta{+}\bigr)^{-1}\DV\sss\DOD\bigl(\rca{-}\bigr)^{-1}.
\ee
We have extracted explicitly the dependence on $t$ and function
$\psi(\lambda)$. The constant $\tilde\gamma$ is equal to
\be{ACFc}
\tilde\gamma=2\pi Z_0(\vartheta_0-1)\Gamma(\nu)
e^{\frac{i\pi\nu}{2}+\frac{3i\pi}{4}},
\ee
and it depends only on functions $\phi_D(\lambda)$ and
$\phi_A(\lambda)$. Recall also, that
\be{ACFnu}
\nu=-\frac{1}{2\pi i}\ln\left[\left(
1-\vartheta_0Z_0e^{\phi_D(\lambda_0)}\right)
\left(1-\vartheta_0Z_0e^{\phi_A(\lambda_0)}\right)\right].
\ee
Corrections to the asymptotic formulae \eq{ACFb12}, \eq{ACFb21}
decay faster than $t^{-1/2}$. However, in the poduct $\hb_{12}\hb_{21}$
the oscillations cancel, and after integrating equation \eq{Blogdir2} twice
these corrections might produce the non-vanishing contributions
into $\det (\tilde I + \tilde V)$. 

In order to make our asymptotic estimates more accurate, we use
differential equations for $\h b_{12}$ and $\h b_{21}$ \eq{Bopeq}:
\be{ACFdifeq1}
i\partial_t\h b_{21}+\partial_x^2\h b_{21}=
2\h b_{21}\h b_{12}\h b_{21},
\ee
\be{ACFdifeq2}
-i\partial_t\h b_{12}+\partial_x^2\h b_{12}=
2\h b_{12}\h b_{21}\h b_{12}.
\ee
We shall look for the solution of this system in the following form
\be{ACFzeta12}
\h b_{12}=(2t)^{-\nu-1/2}
e^{-it\lambda_0^2}\h\zeta(\lambda_0,t),
\ee
\be{ACFzeta21}
\h b_{21}=(2t)^{\nu-1/2}
e^{it\lambda_0^2}\h\eta(\lambda_0,t).
\ee

It is convenient to rewrite the differential equations \eq{ACFdifeq1},
\eq{ACFdifeq2} in terms of variables $\lambda_0$ and $t$. After
simple algebra we arrive at the differential equations for the
operators $\h\zeta$, $\h\eta$:
\be{RSsumasoiti1}
-4i\frac{\partial \h\zeta}{\partial t}
-\frac{4}{t}(\h\zeta\h\eta\h\zeta-i\nu\h\zeta)
=-\frac{1}{t^2}\biggl\{\Bigl[(\nu')^2
\ln^2(2t)-\nu''\ln (2t)\Bigr]\h\zeta
-2\nu'\h\zeta'\ln (2t)+\h\zeta''\biggr\},
\ee
\be{RSsumasoiti2}
4i\frac{\partial \h\eta}{\partial t}
-\frac{4}{t}(\h\eta\h\zeta\h\eta-i\nu\h\eta)
= -\frac{1}{t^2}\biggl\{\Bigl[(\nu')^2
\ln^2(2t)+\nu''\ln (2t)\Bigr]\h\eta
+2\nu'\h\eta'\ln (2t)+\h\eta''\biggr\}.
\ee
Here prime means the derivative with respect to $\lambda_0$.
It is easy to see that we can assume the asymptotic expansions
\cite{AS} obtained for the solutions of the scalar Nonlinear
Schr\"odinger equation:
\be{ACFasyb12}
\h\zeta=
\h\zeta_0(\lambda_0)+\sum_{n=1}^{\infty}\sum_{k=0}^{2n}
\h\zeta_{nk}(\lambda_0)\frac{\bigl(\ln(2t)\bigr)^k}{t^n},
\ee
\be{ACFasyb21}
\h\eta=
\h\eta_0(\lambda_0)+\sum_{n=1}^{\infty}\sum_{k=0}^{2n}
\h\eta_{nk}(\lambda_0)\frac{\bigl(\ln(2t)\bigr)^k}{t^n},
\ee
where coefficients $\h\zeta_{nk}$ and $\h\eta_{nk}$ depend on
$\lambda_0$ only. Substituting these decompositions into
\eq{RSsumasoiti1} and \eq{RSsumasoiti2} we obtain for the
leading terms $\zeta_0$ and $\eta_0$
\be{ACFrelations1}
\hspace{-2cm}\begin{array}{l}
{\dis i\nu\h\zeta_0=
\h\zeta_0
\h\eta_0 \h\zeta_0,}\num
{\dis i\nu\h\eta_0=
\h\eta_0\h\zeta_0
\h\eta_0,}
\end{array}
\ee
These relations \eq{ACFrelations1} are valid automatically, since
due to Eqs. \eq{ACFb12}, \eq{ACFb21} we have
\be{ACFag}
\h\zeta_0\h\eta_0
=i\nu\rca{-}\OD\sss\DOD\bigl(\rca{-}
\bigr)^{-1},
\ee
\be{ACFga}
\h\eta_0\h\zeta_0
=i\nu\bigl(\rcta{+}\bigr)^{-1}\DV\sss\DDV
\rcta{+}.
\ee

In turn for the first corrections $\zeta_{1k}$ and
$\eta_{1k}$ we obtain the following recurrence
\be{ACFrelations2}
\begin{array}{l}
{\dis \h\zeta_{12}\h\eta_0+
\h\zeta_0\h\eta_{12}=0,}\num
{\dis \h\zeta_{11}\h\eta_0+
\h\zeta_0\h\eta_{11}=\frac{1}{2i}
\left(\nu'\h\zeta_0
\h\eta_0\right)',}\num
{\dis \h\zeta_{10}\h\eta_0+
\h\zeta_0\h\eta_{10}=\frac{1}{2i}
\left(\nu'\h\zeta_0\h\eta_0\right)'
+\frac{1}{4}\left(\h\zeta_0\h\eta'_0
-\h\zeta'_0\h\eta_0\right)'.}
\end{array}
\ee
Similar to the scalar Nonlinear Schr\"odinger equation
 the
relations \eq{ACFrelations2} allow us to find more precise estimate
for the product $\h b_{12}\h b_{21}=-i\partial_x\h b_{11}$. Recall
that in order to obtain asymptotic expression of the Fredholm
determinant we need to find only trace of $\h b_{11}$, but not
$\h b_{11}$ it self. Obviously
\be{ACFtrce1}
\tr \h\zeta_0\h\eta_0=i\nu.
\ee
With the trace of the expression
$\left(\h\zeta_0\h\eta'_0-\h\zeta'_0\h\eta_0\right)'$---the
situation is more complicated, since we do not know explicit
expression for $\rc$. On the other hand it is possible to extract the
dependence on the function $\psi(\lambda)$.  Using
formul\ae~\eq{ACFb12} and \eq{ACFb21}, we find
\be{ACFtrace2}
\h\zeta_0\h\eta'_0
-\h\zeta'_0\h\eta_0
=-2i\h\nu\psi'(\lambda_0)+ \h C_0,
\ee
where $\h C_0$ does not depend on the function $\psi(\lambda)$.
Thus we have
\be{ACFb11prim}
\tr\h b'_{11}=-\nu+\frac{i}{2t}(\nu'\nu)'(\ln(2t)+1)-
\frac{i}{2t}(\nu\psi'(\lambda_0))'+\frac{1}{t}\tr\h{C}_0+
{\cal O}(\ln^2t/t^{2}).
\ee
{\sl Note}.~~One could expect that due to the \eq{ACFasyb12}, 
\eq{ACFasyb21} the last equation is valid up to the terms of order 
$\ln^4t/t^{2}$. These corrections do enter the expressions 
for $\hb_{12}$ and $\hb_{21}$, however they vanish (as well as the terms
$\ln^3t/t^{2}$) for the combination $\tr\hb_{12}\hb_{21}$. 
                            
Finally one should use the identity
\be{ACFlogdirx}
\frac{1}{2t}{\partial_{\lambda_0}}
\ln\det(\tilde I+\widetilde V)=i\tr \h b_{11},
\ee
and after integrating of \eq{ACFb11prim} with respect to $\lambda_0$
we obtain improved formula for Fredholm determinant. Of course, the
integration constant may depend on $t$, therefore we need to use
also the relations
\be{ACFlogdirt}
\left({\partial_t}-\frac{\lambda_0}{t}\partial_{\lambda_0}\right)
\ln\det(\tilde I+\widetilde V)
=i\tr\left( \h c_{22}
-\h c_{11}\right),
\ee
and
\be{ACFdirb11t}
\partial_t\h b_{11}=\frac{1}{2t}
\left(\partial_{\lambda_0}\h b_{12}\cdot\h b_{21}
-\h b_{12}\cdot
\partial_{\lambda_0}\h b_{21}\right).
\ee
These equations allow us to find integrating constant up to some
function, which might depend on $\lambda_{0}$ since we
did not specify ${\hat C}_{0}$ in \eq{ACFtrace2}. 
We skip the details; they are very similar to the
free fermionic case spelled out in \cite{IIKV}. Thus, we have for the
Fredholm determinant the following estimate,
\ba{ACFasdet1}
&&\hspace{-6mm}\ln\det(\tilde
I+\tV)={\dis \frac{1}{2\pi}\stint\biggl(|x-2\lambda t|+
i\sign(\lambda_0-\lambda)\frac{d\psi(\lambda)}{d\lambda}
\biggr)}\non
&&\hspace{2cm}\times{\dis
\ln\Bigl\{1-\vartheta(\lambda)
\biggl(1+e^{\phi(\lambda)\sign(\lambda-\lambda_0)}\biggr)\Bigr\}
\,d\lambda}\non
&&\hspace{2cm}{\dis -\frac{\nu^2}{2}\ln(2t)+C_0+
{\cal O}(\ln^2t/t).}
\ea
Here $C_0$ includes integrating  constant and other terms
depending on functions $\phi_A$, $\phi_D$, the ratio $\lambda_{0}$
and the non-dynamical physical  parameters
$h$, $c$, and $T$.

In order to calculate the correlation function of local fields we
need to find the combination
\be{REScomb}
{\cal B}=\stint\,dudv\hb_{12}(u,v)\cdot\det\freop.
\ee
The formul\ae~we obtained for the Fredholm determinant and the operator
$\hb_{12}$ allow us to find the asymptotic expression for this object
\ba{ACFresult}
&&\hspace{-11mm}
{\cal B}={\dis
C(\phi_{D},\phi_{A})(2t)^{-\frac{(\nu+1)^2}{2}}
e^{\psi(\lambda_0)-it\lambda_0^2}}\non
&&\hspace{-6mm}{\dis\times\exp\left\{
\frac{1}{2\pi}\stint\biggl(|x-2\lambda t|
+i\sign(\lambda_0-\lambda)\frac{d\psi(\lambda)}{d\lambda}
\biggr)\right.}\non
&&\hspace{-6mm}{\dis\left.\vphantom{\stint}\times
\ln\Bigl\{1-\vartheta(\lambda)
\biggl(1+e^{\phi(\lambda)\sign(\lambda-\lambda_0)}\biggr)\Bigr\}
\,d\lambda\right\}\Bigl(1+{\cal O}({\ln^2t}/{t})\Bigr).}
\ea
Here we have extracted all dependence on $t$ and function $\psi
(\lambda)$. The factor $C(\phi_{D},\phi_{A})$ depends no 
$\lambda_{0}$, $h$, $c$, and $T$, and it  does not depend
on $\psi(\lambda)$; parameter $\nu$ is equal to
\be{ACFnu1}
\nu=-\frac{1}{2\pi i}\ln\left[\left(
1-\vartheta_0Z_0e^{\phi_{D}(\lambda_0)}\right)
\left(1-\vartheta_0Z_0e^{\phi_{A}(\lambda_0)}\right)\right].
\ee
and the function $\phi(\lambda)$ is defined by the equation,
\be{phidef11}
\phi(\lambda) = \phi_{A}(\lambda) - \phi_D(\lambda).
\ee

In the case of positive chemical potential the 
calculations are quite similar. It is easy to show that the
operators $\h B_{12}^{(0)}$ and $\h B_{21}^{(0)}$ \eq{SPphi0}
satisfy the same system of the differential equations \eq{ACFdifeq1},
\eq{ACFdifeq2} as the operators $\hb_{12}$ and $\hb_{21}$.
The asymptotic behavior of $\h B_{12}^{(0)}$ and $\h B_{21}^{(0)}$
was found in the previous section---Eq. \eq{RESkapB}. Thus one
can use the same approach in order to obtain more precise estimates.
We do not present here the details since in the section 11
we will develop a modified method allowing to obtain the most
important, i.e. $\psi'(\lambda)$ -  correction
to the leading term of the asymptotics directly, without using the
differential equations.

Finally, we need to say several words about the regularization
introduced in the section \ref{F}. The only object in
\eq{ACFresult}, which depends on $\epsilon$, is the factor
$C(\phi_D,\phi_A)$, since it depends on $\rc$ and $\OD\sss\DDV$.
The exponential and power law factors do not depend on the
regularization. We shall see later that $C(\phi_D,\phi_A)$ is only
responsible for a numerical factor in the asymptotics of the correlation
function \eq{Itempcorrel}

\section{Modified approach. Preliminary discussions\label{ID}}

Assuming the dual fields $\psi(\lambda)$, $\phi_A(\lambda)$ and
$\phi_D(\lambda)$ to be classical functions, we  found the
asymptotic expression for $\cal B$, which is the combination
of the Fredholm determinant $\det\freop$ and factor $\hb_{12}$
\eq{Bdetrep}. This asymptotic expression has the form
\be{IDcalB}
{\cal B}={\cal F}_c(\phi_A,\phi_D,\psi)
t^{{\cal F}_p(\phi_A,\phi_D)}
e^{t{\cal F}_e(\phi_A,\phi_D)}\left(1+{\cal O}(\ln^2t/t)\right),
\ee
where ${\cal F}_c$, ${\cal F}_p$ and ${\cal F}_e$ are functionals of
the dual fields.  Their representations are given in
\eq{ACFresult}. The functionals ${\cal F}_e\Bigl(\phi_A,\phi_D\Bigr)$
and ${\cal F}_p\Bigl(\phi_A,\phi_D\Bigr)$, defining the exponential
and power laws respectively, are found explicitly
\be{IDYe}
t{\cal F}_e(\phi_A,\phi_D)=-it\lambda_0^2+
\frac{1}{2\pi}\stint|x-2\lambda t|
\ln\Bigl\{1-\vartheta(\lambda)
\biggl(1+e^{\phi(\lambda)\sign(\lambda-\lambda_0)}\biggr)\Bigr\}
\,d\lambda,
\ee
\be{IDYp}
{\cal F}_p\Bigl(\phi_A,\phi_D\Bigr)=-\frac12(\nu+1)^2,
\ee
where $\nu$ is given by \eq{ACFnu1}. The functional ${\cal F}_c$
depends only on the ratio $\lambda_0=x/2t$, which is fixed,
hence it behaves as a constant when $t$ goes to infinity.
This functional consists of two parts, which play essentially different
roles in the process of averaging with respect to auxiliary vacuum.
Let us extract from ${\cal F}_c$ the part, containing the field $\psi$,
namely
\be{IDextr}
{\cal F}_c={\cal F}_{c\psi}(\phi_A,\phi_D,\psi)
{\cal F}_{c\phi}(\phi_A,\phi_D),
\ee
where
\be{IDY2}
{\cal F}_{c\psi}\Bigl(\phi_A,\phi_D,\psi\Bigr)=\psi(\lambda_0)
+\frac{i}{2\pi}\stint
\sign(\lambda_0-\lambda)\frac{d\psi(\lambda)}{d\lambda}
\ln\Bigl\{1-\vartheta(\lambda)
\biggl(1+e^{\phi(\lambda)\sign(\lambda-\lambda_0)}\biggr)\Bigr\}
\,d\lambda.
\ee
For the functional ${\cal F}_{c\phi}$, which does not depend on
$\psi$, we do not have any explicit formula; this functional depends 
on the operator $\rc$.  Thus, the asymptotic behavior of $\cal B$ is 
found up to the constant factor ${\cal F}_{c\phi}$. The similar 
factor in the free fermionic case can be determined by the use of the 
auxilary differential equations with respect to the temperature and 
chemical potential (cf.  \cite{IIKV}).  Here we do not pursue such 
goals, and we restrict ourselves by establishing the exponential and 
power asymptotic laws.  Thus, in the case of the dual fields being 
classical functions, we have complited the analysis. It should be 
also mentioned that the exponential law which we found coincide with 
the one obtained in \cite{S1} via the independed and more direct 
methods.
                                                            
However, as we have already mentioned in the introduction and  
section \ref{B}, the operator nature of the dual fields strongly 
changes the situation at the stage of calculation of the vacuum mean 
value $(0|{\cal B}|0)$. One must be sure that the corrections to the 
leading term do not contribute into it after the averaging. In this 
section we demonstrate that the result \eq{ACFresult} is not 
satisfactory from this point of view.

Recall the commutation relations between creation and annihilation
parts of the dual fields \eq{Bcommutators}
\be{IDcommutators}
\begin{array}{l}
{}[p_{A}(\lambda),q_\psi(\mu)]=
[p_\psi(\lambda),q_{A}(\mu)]=\ln h(\mu,\lambda),\num
{}[p_{D}(\lambda),q_\psi(\mu)]=
[p_\psi(\lambda),q_{D}(\mu)]=\ln h(\lambda,\mu),\num
{}[p_\psi(\lambda),q_\psi(\mu)]=\ln [h(\lambda,\mu)h(\mu,\lambda)],
\end{array}
\ee
where $h(\lambda,\mu)=(\lambda-\mu+ic)/{ic}$. It is important that
operators $p_A$, $p_D$, $q_A$ and $q_D$ commute with each other:
\be{IDpq}
[p_j(\lambda),q_k(\mu)]=0, \qquad j,k\in\{A,D\},
\ee
therefore the vacuum expectation value of any expression containing
only the dual fields $\phi_A$ and  $\phi_D$  is always trivial:
\be{IDtrivial}
(0|{\cal F}(\phi_A, \phi_D)|0)={\cal F}(0,0).
\ee
A non-trivial mean value arises if and only if the dual field $\psi$
presents. That is why we paid so much attention to
this dual field in the previous sections.

Consider the asymptotic representation \eq{IDcalB}  from the point of
view of its averaging. It is easy to see that if we  restrict 
ourselves with the exponential and power law factors, then we 
immediately obtain 
\be{IDrest1} (0|t^{{\cal F}_p(\phi_A,\phi_D)} 
e^{t{\cal F}_e(\phi_A,\phi_D)}|0)=
t^{{\cal F}_p(0,0)}e^{t{\cal F}_e(0,0)},
\ee
since the functionals ${\cal F}_p$ and ${\cal F}_e$ depend only on
the fields $\phi_A$ and $\phi_D$. Even if we take into consideration
the functional ${\cal F}_{c\phi}$, then nothing will change for the
exponential and power law factors because of the same
reason---${\cal F}_{c\phi}$ does not depend on $\psi$. We would like
to emphasize especially that although the functional ${\cal
F}_{c\phi}$ is unknown, this fact should not be considered as an essential
deficiency.  The fields $\phi_A$ and $\phi_D$ do not influence
on the exponential and power laws factors in the process of
averaging.

At the same time the presence of the functional
${\cal F}_{c\psi}\Bigl(\phi_A,\phi_D,\psi\Bigr)$
completely changes the situation. It depends on the dual field
$\psi$, and it turns out that this operator changes the correlation 
radius, as well as the pre-exponent. Even the simplest factor
$e^{\psi(\lambda_0)}$ shifts the arguments of ${\cal F}_{c\phi}$,
${\cal F}_p$ and ${\cal F}_e$ due to the evident  property \cite{KS2},  
\be{IDevprop}
e^{p_\psi(\lambda_0)}{\cal F}\Bigl(\phi_A(\mu),
\phi_D(\mu)\Bigr)|0)=
{\cal F}\Bigl(\phi_A(\mu)+h(\mu,\lambda_0),
\phi_D(\mu)+h(\lambda_0,\mu)\Bigr)|0).
\ee
The contribution of the remaining part of ${\cal F}_{c\psi}$
\eq{IDY2} is much more complicated, but it is clear that
it changes the result \eq{IDrest1}.

In the present paper we are not going to study the procedure of
averaging of the dual fields. This is done in \cite{KS2}, where
the reader can find the details of the calculation of the vacuum
expectation value. The main purpose of this paper is to present
the asymptotic expression of the operator $\cal B$ in the
form adjusted to the averaging procedure. The above considerations show
that the constant factor ${\cal F}_{c\psi}$ (i.e. the factor, which
remains fixed while $t$ goes to infinity) strongly changes the
leading term of the asymptotics after calculation of the vacuum
expectation value.  It becomes clear now that the leading term of
the asymptotics, which had been found in the sections \ref{NCP} and
\ref{PCP}, is not sufficient for the description of the correlation
function.

One can ask now, whether we can restrict ourselves with the
factor ${\cal F}_{c\psi}$? Can one guarantee that corrections
${\cal O}(\ln^2t/t)$ do not give similar contribution into the vacuum
mean value?  Let us demonstrate that some corrections do provide such
non-vanishing contribution.

In order to find corrections to the expression \eq{ACFresult}
one  can use the differential equations \eq{RSsumasoiti1} and
\eq{RSsumasoiti2}. Substituting the asymptotic expansions
\eq{ACFasyb12}, \eq{ACFasyb21} into these equations we obtain an
infinite set of the recurrent relations for the operators
$\h\zeta_{nk}$ and $\h\eta_{nk}$. The last ones in turn provide us
with complete asymptotic expansion of the Fredholm determinant and
the operator $\hb_{12}$. It is clear that corrections, which do not
depend on $\psi$, can not change the leading exponential term
$\exp\{t{\cal F}_e\}$ of the asymptotics  after their averaging.
Therefore we need to pay our attention mostly to the terms containing
the field $\psi$.

Observe that $\h\zeta_{0}(\lambda_0)$ and $\h\eta_{0}(\lambda_0)$ are
proportional to $e^{\psi(\lambda_0)}$ and $e^{-\psi(\lambda_0)}$
respectively. This suggests the following substitution,
\be{IDtilz}
\h\zeta(\lambda_0,t)=e^{\psi(\lambda_0)}\h{\tilde\zeta}(\lambda_0,t),
\qquad
\h\eta(\lambda_0,t)=e^{-\psi(\lambda_0)}\h{\tilde\eta}(\lambda_0,t).
\ee
into \eq{RSsumasoiti1}. We obtain for
$\h{\tilde\zeta}$
\ba{IDsumasoiti1}
&&{\dis\hspace{-12mm}
-4i\frac{\partial \h{\tilde\zeta}}{\partial t}
-\frac{4}{t}(\h{\tilde\zeta}\h{\tilde\eta}\h{\tilde\zeta}
-i\nu\h{\tilde\zeta})
= - \frac{1}{t^2}\biggl\{\Bigl[(\nu')^2
\ln^2(2t)-\nu''\ln (2t) - 2\nu' \psi'\ln (2t)
+\psi''+\psi'^2\Bigr]\h{\tilde\zeta}}\non
&&{\dis\hspace{35mm}
-2\Bigl(\nu'\ln (2t)-\psi'\Bigr)\h{\tilde\zeta}{}'
+\h{\tilde\zeta}{}''\biggr\}.}
\ea
Similar equation arises for the operator $\h{\tilde\eta}$. It is
easy to see that the coefficient $\h\zeta_{10}$ of the
asymptotic expansion \eq{ACFasyb12} contains the term
\be{IDzeta10}
\h{\tilde\zeta}_{10}=\psi'^2(\lambda_0)\h{\tilde\zeta}_0+\dots~,
\ee
and hence $\hb_{12}$ contains the term $\frac{1}{t}\psi'^2(\lambda_0)\h\zeta_0 $.
If $t$ goes to infinity, then this term vanishes.
However, after calculation of the vacuum expectation of
$\psi'^2/t$ together with the exponential factor $e^{t{\cal F}_e}$
it becomes proportional to the positive power of $t$. Indeed, the
action of the operator $\psi$ on functions of $\phi_A$ and
$\phi_D$ is equivalent to the differentiating \cite{KS2}, therefore we have
\be{IDexamp}
(0|\frac{1}{t}\psi'^2(\lambda_0)
e^{t{\cal F}_e(\phi_A,\phi_D)}|0)\sim
te^{t{\cal F}_e(0,0)}+\dots~.
\ee

Thus we see that the term $\psi'^2/t$, which was small before averaging
and, hence, could be considered as a correction, becomes large
after calculation of the vacuum expectation. Of course, this term
forms only pre-exponent, but does not change the correlation
radius. However, it is not difficult to see that the higher corrections
contain terms $\psi'^4/t^2$,\  $\psi'^6/t^3$ etc., which turn into
$t^2$, $t^3$ etc. after the averaging. Generally, all the terms of type
$\psi^n/t^m$, where  $n>m$, give positive powers of $t$ after their
averaging together with the exponential factor. The series of
positive powers of $t$ sums up into the exponent, and as a result the
correlation radius changes.

Thus, we conclude that in order to find correct asymptotic behavior of the
correlation function it is necessary to have some information about
the complete asymptotic expansion of the Fredholm determinant and  the
operator $\hb_{12}$. At least one needs to know the structure of
the $\psi$ - dependence of the higher corrections.
 The leading term \eq{ACFresult}
itself is not sufficient. This expression correctly describes the
asymptotics of the operator $\cal B$, but not the asymptotics of
its vacuum expectation value.

\section{Modified approach. The `shifted' saddle point\label{MA}}

We have shown in the previous section that, in fact, in the
process of calculation of the vacuum expectation the dual
field $\psi$ is effectively proportional to $t$. This prompts an
idea how to modify the method we used for the asymptotic
solution of  the operator-valued
RHP. Recall the jump matrix \eq{Fjumpmatreg}
of the original RHP $(a)$:
\be{MAjumpmatreg}
\begin{array}{rcl}
{\dis
\h G_{11}(\lambda)}&=& {\dis \hi
-\vartheta(\lambda)Z(\lambda,\lambda)
e^{\phi_D(\lambda)}\odr\odl;}\num
{\dis
\h G_{12}(\lambda)}&=& {\dis
2\pi i (\vartheta(\lambda)-1)Z(\lambda,\lambda)
e^{\psi(\lambda)+\tau(\lambda)}\odr\dvl;}\num
 {\dis
\h G_{21}(\lambda)}&=&
{\dis -\frac{i}{2\pi}\vartheta(\lambda)Z(\lambda,\lambda)
e^{\phi_A(\lambda)+\phi_D(\lambda)
-\psi(\lambda)-\tau(\lambda)}\dvr\odl;}\num
{\dis
\h G_{22}(\lambda)}&=&
 {\dis
\hi-\vartheta(\lambda)Z(\lambda,\lambda)
e^{\phi_A(\lambda)}\dvr\dvl}.
\end{array}
\ee
Observe that the field $\psi$ enters this matrix only in the
combination with the function $\tau(\lambda)$: $\psi(\lambda)
+\tau(\lambda)$. All the asymptotic analysis of the operator-valued
RHP was based on decomposition of the solution in the vicinity
of the point $\lambda_0=x/2t$.  This value is the saddle point
of the function $\tau(\lambda)$:
\be{MAsadtay}
\left.\frac{d\tau(\lambda)}{d\lambda}\right|_{\lambda=\lambda_0}=0.
\ee
As we have mentioned, one can consider $\psi$ to be effectively
proportional to $t$. Therefore instead of point $\lambda_0$ we can
consider new saddle point $\Lambda$, which is defined as
\be{MAsadtaypsi}
\left.\frac{d}{d\lambda}(\tau(\lambda)+\psi(\lambda))
\right|_{\lambda=\Lambda}=0,
\ee
or equivalently
\be{MAshiftsp}
\Lambda=\lambda_0+\frac{i}{2t}\psi'(\Lambda),
\ee
where prime means derivative with respect to the argument.

From the point of view of the classical asymptotic analysis
this `shift' of the saddle point does not make much sense. 
In fact, the difference between the original saddle point $\lambda_0$
and shifted one $\Lambda$ is of order $1/t$ and it vanishes, when
$t$ goes to infinity. The formal solution of the equation \eq{MAshiftsp}
can be given by the asymptotic series with respect to $1/t$ 
(particular case of \eq{MAperser2}):
\be{MAperser1}
\Lambda=\lambda_0+
\sum_{n=1}^{\infty}\frac1{n!}
\left(\frac{i}{2t}\right)^n\frac{d^{n-1}}
{d\lambda_0^{n-1}}\left[(\psi'(\lambda_0))^n\right].
\ee
The replacement of $\lambda_0$ by $\Lambda$ means that
we have shifted the center of the asymptotic expansion by asymptotically
small value. Using  series \eq{MAperser1} one can always
re-expand any function $f(\Lambda)$ into the asymptotic series over
$1/t$ with the center at the original point $\lambda_0$:
\be{MAperser2}
f(\Lambda)=f(\lambda_0)+
\sum_{n=1}^{\infty}\frac1{n!}
\left(\frac{i}{2t}\right)^n\frac{d^{n-1}}
{d\lambda_0^{n-1}}\left[(\psi'(\lambda_0))^nf'(\lambda_0)\right].
\ee
Here we have used the Lagrange's theorem (see, for instance,  
\cite{WW}).  As a result, the `new' asymptotics would of course 
coincide with the `old' one.

Nevertheless this approach appears to be extremely useful when one recalls
the role of the operator $\psi$ in the process of averaging. Indeed,
studying the asymptotic solution of the RHP for the large $t$, we
automatically take into account the field $\psi$.

Let us turn back to the RHP $(a)$ and consider the substitution
\eq{NCPsub}
\be{MAsub}
\hx(\lambda)=\h\Phi(\lambda)\Rho(\lambda),\qquad
\Rho=\left(\begin{array}{cc}
{\dis\h\varrho(\lambda)}&0\num
0&{\dis\bigl(\h\varrho^T(\lambda)\bigr)^{-1}}
\end{array}\right).
\ee
As before, the operator $\h\varrho(\lambda)$ is the solution of the RHP
$(b)$, however now we replace $\lambda_0$ by $\Lambda$ in the corresponding
jump matrix:
\be{MAjumpmat}
\h\varrho_-(\lambda)=\h\varrho_+(\lambda)
\left( \theta(\Lambda-\lambda)\h G_{11}
+\theta(\lambda-\Lambda)
\bigl(\h G_{22}^T\bigr)^{-1}\right),\quad
\lambda\in R.
\ee
Then one should literally repeat all the constructions of the section
\ref{NCP}, replacing everywhere $\lambda_0$ by $\Lambda$. Clearly that
eventually we shall arrive at the following results for the coefficients
of the asymptotic expansion of the determinant of $\h\varrho$:
\be{MAsolDelta0}
\Delta_0=\frac{1}{2\pi i}
\stint\,d\mu\sign(\Lambda-\mu)
\ln\left(1-\vartheta(\mu)
\Bigl(1+e^{\phi(\mu)\sign(\mu-\Lambda)}
\Bigr)\right),
\ee
\be{MAsolDelta1}
\Delta_1=\frac{1}{2\pi i}
\stint\,\mu d\mu\sign(\Lambda-\mu)
\ln\left(1-\vartheta(\mu)
\Bigl(1+e^{\phi(\mu)\sign(\mu-\Lambda)}
\Bigr)\right).
\ee
In comparison with the old results \eq{NCPsolDelta0},
\eq{NCPsolDelta1} of the section \ref{NCP}, the only difference is
that $\lambda_0$ is replaced by $\Lambda$. Since
$\Lambda-\lambda_0\sim 1/t$, it is clear that new representations for
$\Delta_j$ and the old ones are asymptotically equivalent.

It should be noted that the point $\Lambda$ must not be necesserely real,
so that the jump contour in \eq{MAjumpmat} might be slightly deformed to
the complex $\lambda$ with the natural re-definition
of the $\theta$-functions. Also, the formulae \eq{MAsolDelta0} and
\eq{MAsolDelta1} should be generally understood as (see \eq{Esign}), 
\be{MAsolDelta00}
\Delta_0=\frac{1}{2\pi i}
\int_{-\infty}^{\Lambda}\,d\mu
\ln\left(1-\vartheta(\mu)
\Bigl(1+e^{-\phi(\mu)}
\Bigr)\right)
\ee
$$
 -\frac{1}{2\pi i}
\int_{\Lambda}^{\infty}\,d\mu
\ln\left(1-\vartheta(\mu)
\Bigl(1+e^{\phi(\mu)}
\Bigr)\right),
$$
and
\be{MAsolDelta11}
\Delta_1=\frac{1}{2\pi i}
\int_{-\infty}^{\Lambda}\,\mu d\mu
\ln\left(1-\vartheta(\mu)
\Bigl(1+e^{-\phi(\mu)}
\Bigr)\right)
\ee
$$
 -\frac{1}{2\pi i}
\int_{\Lambda}^{\infty}\,\mu d\mu
\ln\left(1-\vartheta(\mu)
\Bigl(1+e^{\phi(\mu)}
\Bigr)\right),
$$
with the integrations go possibly in the complex
plane. In all the following formulae we will always assume the
same interpretation of the integrals involving the sign-functions.

In order to obtain the Fredholm determinant of the operator $\tilde
I+\widetilde V$ one need to integrate \eq{NCPsolDelta0} and
\eq{NCPsolDelta1} with respect to $x$ and $t$ respectively. Using
\be{MAnewder}
\partial_x\to\frac1{2t_s}
\partial_\Lambda,\qquad
\partial_t\to\partial_t-
\frac\Lambda{t_s}\partial_\Lambda,\qquad\mbox{where}\qquad
t_s=t-\frac i2\psi''(\Lambda),
\ee
we have
\ba{MAasdet2}
{\dis\hspace{-1cm}
\ln\det(\tilde I+\tilde V)}&\approx &{\dis\frac{1}{2\pi}
\stint\,d\mu\Bigl(x-2\mu t+i\psi'(\mu)\Bigr)\sign(\Lambda-\mu)
}\non
&&{\dis\hspace{2cm} \times
\ln\left(1-\vartheta(\mu)
\Bigl(1+e^{\phi(\mu)\sign(\mu-\Lambda)}
\Bigr)\right).}
\ea

It is worth mentioning that in the framework of our old approach we
obtained the term $\psi'(\lambda)$ in the integrand \eq{MAasdet2}
only after studying the localized RHP and differential equations.
As we have mentioned already in the previous section, it is this term
which is the most important for the calculation of the vacuum expectation
value, and it is remarkable that the new approach allows us to obtain
this term already at the first stage of the asymptotic analysis of the
RHP. On the other hand, one should not be surprised by this fact. The
new treatment of the saddle point \eq{MAsadtaypsi} in fact means that
we consider function $\psi$ as it would be proportional to $t$. Therefor,
it is quite natural that in addition to the term $x-2\mu t$
we have obtained $i\psi'$. One can also notice that
\be{MAnotice}
x-2\lambda t+i\psi'(\lambda)=i\bigl(\tau(\lambda)+
\psi(\lambda)\bigr)',
\ee
and thus, the integrand in the Eq. \eq{MAasdet2} reflects the fact
that the jump matrix in the original RHP $(a)$ depends only on the
combination $\tau(\lambda)+\psi(\lambda)$. Thus, we see that the method
proposed in this section permits one to trace the dependency
of the Fredholm determinant on the field $\psi$ automatically, together
with the dependency on the time $t$.

We notice again, that in reality the function $\psi$ does not depend on $t$,
and it remains fixed when $t$ goes to infinity. Hence the representations
\eq{MAasdet2} and \eq{NCPasdet2} are asymptotically equivalent. They are not
equivalent only when we recall the quantum operator nature of the field
$\psi (\lambda)$

\vskip .2in
The  localized RHP can be considered in the same manner as it had
been done previously.  We would like to mention only one moment. The
decomposition of the $\tau(\lambda)+\psi(\lambda)$ at the point
$\lambda=\Lambda$ has the form
\be{MAdecomp}
\tau(\lambda)+\psi(\lambda)=\tau(\Lambda)+\psi(\Lambda)
+it_s(\lambda-\Lambda)^2+{\cal O}(\lambda-\Lambda)^3,
\ee
where $t_s=t-i\psi''(\Lambda)/2$ was introduced in \eq{MAnewder}.
We see that variable $t_s$ plays the role of the new time. Therefore
it is quite natural to look for the asymptotic expansion of the
solution of the RHP with respect to $t_s$, but not to $t$. Up to
replacement $t\to t_s$ and $\lambda_0\to\Lambda$ the solution of the
new localized RHP formally coincides with the solution of the old
one, in particular (cf. \eq{ACFb12}, \eq{ACFb21}),
\be{MAb12}
\h b_{12}\approx \frac{\tilde\gamma\nu}{i\sqrt{2\pi}}(2t_s)^{-\nu-1/2}
e^{\psi(\Lambda)+\tau(\Lambda)}
\rca{-}\OD\sss\DDV\rcta{+},
\ee
\be{MAb21}
\h b_{21}\approx -\frac{\sqrt{2\pi}}{\tilde\gamma}(2t_s)^{\nu-1/2}
e^{-\psi(\Lambda)-\tau(\Lambda)}
\bigl(\rcta{+}\bigr)^{-1}\DV\sss\DOD\bigl(\rca{-}\bigr)^{-1},
\ee
where $\tilde\gamma$ is equal to
\be{MAc}
\tilde\gamma=2\pi
Z(\Lambda,\Lambda)(\vartheta(\Lambda)-1)\Gamma(\nu)
e^{\frac{i\pi\nu}{2}+\frac{3i\pi}{4}},
\ee
and
\be{MAnu}
\nu \equiv \nu(\Lambda) = -\frac{1}{2\pi i}\ln\left[\left(
1-\vartheta(\Lambda)Z(\Lambda,\Lambda)
e^{\phi_D(\Lambda)}\right)
\left(1-\vartheta(\Lambda)Z(\Lambda,\Lambda)
e^{\phi_A(\Lambda)}\right)\right].
\ee
Of course, now $\rc=\h\varrho^{(c)}(\Lambda)$ and vectors
$\OD\hspace{-3pt}$,  $\DV$ etc. are associated with $\Lambda$, but not
with $\lambda_0$.

Finally, using the differential equations for $\hb_{12}$ and
$\hb_{21}$ we can find the corrections to the Eqs. \eq{MAb12},
\eq{MAb21}. The most important is the dependency of these
corrections on the function $\psi$. Substituting
\be{MAzeta12}
\h b_{12}=(2t_s)^{-\nu(\Lambda)-1/2}
e^{\tau(\Lambda)+\psi(\Lambda)}\h{\tilde\zeta}(\Lambda,t_s),
\ee
\be{MAzeta21}
\h b_{21}=(2t_s)^{\nu(\Lambda)-1/2}
e^{-\tau(\Lambda)-\psi(\Lambda)}\h{\tilde\eta}(\Lambda,t_s),
\ee
into equations \eq{ACFdifeq1}, \eq{ACFdifeq1}, we obtain, for example,
for $\h{\tilde\zeta}$
\be{MAsumasoiti1}
-4i\frac{\partial \h{\tilde\zeta}}{\partial t_s}
-\frac{4}{t_s}(\h{\tilde\zeta}\h{\tilde\eta}
\h{\tilde\zeta}-i\nu\h{\tilde\zeta})
= - \frac{1}{t_s^2}\biggl\{\Bigl[(\nu')^2
\ln^2(2t)-\nu''\ln (2t)\Bigr]\h{\tilde\zeta}
-2\nu'\h{\tilde\zeta}{}'\ln (2t)+\h{\tilde\zeta}{}''
\biggr\}
+{\cal O}(\ln(t_s)t_s^{-3}),
\ee
where prime now means the derivative with respect to $\Lambda$.
Thus, up to the corrections of the $t_s^{-3}$ order we again came back to
the equation \eq{RSsumasoiti1}. However now we have already taken into account
the factor $e^{\psi(\Lambda)}$, and hence we do not need to make a substitution
similar to \eq{IDtilz}.  Therefore, the replacement of variables
$(\lambda_0,t)\to(\Lambda,t_s)$ allows us to remove the dependency
on the function $\psi$ from the terms proportional to $t_s^{-2}$. 

It is easy to see that the asymptotic expansion similar to \eq{ACFasyb12}
formally solves the Eq.\eq{MAsumasoiti1}
\be{MAasyb12}
\h{\tilde\zeta}=
\h{\tilde\zeta}_0(\Lambda)+\sum_{n=1}^{\infty}\sum_{k=0}^{2n}
\h{\tilde\zeta}_{nk}(\Lambda)
\frac{\bigl(\ln(2t_s)\bigr)^k}{t_s^n},
\ee
One can write similar expansion for the operator $\h{\tilde\eta}$ also.

In order to study the recurrence arising  for the
coefficients $\h{\tilde\zeta}_{nk}$ and $\h{\tilde\eta}_{nk}$,
one needs to know the explicit expressions for the terms ${\cal
O}(t_s^{-3})$ in \eq{MAsumasoiti1}.  It is not difficult to
find these terms, however, for our goal, it is enough to check
that the terms of the order $t_s^{-3}$ are linear with respect to
$\psi$. In addition, equation \eq{MAsumasoiti1} contains the terms of
the order $t_s^{-4}$, which are quadratic with respect to $\psi$. It is clear
that for given $n$, the coefficients $\h{\tilde\zeta}_{nk}$ and
$\h{\tilde\eta}_{nk}$ are polynomials of $\psi$ and
its derivatives of the $(n-1)$ degree or less. In other words, 
in the framework of the modified approach any large time
correction, as it is expected, has the form $\psi^m/t_s^n$, where $m<n$.
Therefore, as we have explained in the previous section, these
corrections remain small even after their averaging with respect to
the auxiliary vacuum, and hence we can drop  them out.

Summarizing the above considerations, we obtain, in 
the case of negative chemical potential, the
following asymptotics for the operator ${\cal B}$,
\ba{MAresult0}
&&\hspace{-11mm}
{\cal B}={\dis
C(\phi_{D},\phi_{A},\Lambda)(2t_s)^{-\frac{(\nu(\Lambda)+1)^2}{2}}
e^{\psi(\Lambda)+\tau(\Lambda)}}\non
&&\hspace{-6mm}{\dis\times\exp\left\{
\frac{1}{2\pi}\stint\biggl(x-2\lambda
t+i\psi'(\lambda)\biggr)\sign(\Lambda-\lambda)\right.}\non
&&\hspace{-6mm}{\dis\left.\vphantom{\stint}\times
\ln\Bigl\{1-\vartheta(\lambda)
\biggl(1+e^{\phi(\lambda)\sign(\lambda-\Lambda)}\biggr)\Bigr\}
\,d\lambda\right\}\Bigl(1+{\cal O}({\ln^2t}/{t})\Bigr).}
\ea
In distinction to the result \eq{ACFresult}, this asymptotic
equation takes into
account all the corrections which make a contribution into the leading
term after the averaging. Corrections to the expression \eq{MAresult0}
remain small after calculation of their vacuum expectation value.

Comparing the results \eq{ACFresult} and \eq{MAresult0} one can
observe that in the last case the constant factor
$C(\phi_{D},\phi_{A},\Lambda)$ depends on the shifted saddle point
$\Lambda$, which in turn implicitly depends on $\psi$. As before this
constant factor is not explicitly defined since it contains the
operator $\rc$.  In the framework of the old method $\rc$ depended
only on the dual fields $\phi_A$ and $\phi_D$, hence this operator did
not influence  the leading term of the asymptotics. However
now $\rc$ implicitly depends on the field $\psi$, and it might change
the asymptotics. Since we do not know explicit expression for the
operator $\rc$, we might worry about calculation of its vacuum
expectation value. However, it is easy to see that the contribution
of the constant factor $C(\phi_{D},\phi_{A},\Lambda)$ remains
constant after averaging. Indeed, remembering the decomposition
\eq{MAperser2} we have
\be{MAperser3}
C(\phi,\Lambda)=C(\phi,\lambda_0)+
\sum_{n=1}^{\infty}\frac1{n!}
\left(\frac{i}{2t}\right)^n\frac{d^{n-1}}
{d\lambda_0^{n-1}}\left[(\psi'(\lambda_0))^n
C'(\phi,\lambda_0)\right].
\ee
All the terms of this series behave as $(\psi/t)^n$. Their vacuum
expectation values together with the exponential factor are some
constants:
\be{MAtaksebe}
(0|\left(\frac{\psi}{t}\right)^ne^{t{\cal F}_e(\phi_A,\phi_D)}|0)
\sim e^{t{\cal F}_e(0,0)}({\cal O}(1)+\dots ), \qquad t\to\infty.
\ee
Thus the replacement $\lambda_0$ by $\Lambda$ in the
functional $C$ does not change the exponential and power laws of the
asymptotics of the vacuum mean value, but changes only the
common constant factor. Since in this paper we do not analyse this
factor, we can simply replace the new saddle point by the old
one in $C$. Similar arguments allow us to replace
$t_s$ by $t$ in the power law factor. However
we should keep the shifted saddle point $\Lambda$
in the exponential factor (i.e. functions $\sign(\Lambda-\lambda)$
and $\tau(\Lambda)+ \psi(\Lambda)$), and  the power $\nu(\Lambda)$
since all these objects enter the result together with $t$ and $\ln t$
respectively.

Our final answer in the case  $h<0$ can be represented as follows,
\ba{MAresult1}
&&\hspace{-11mm}
{\cal B}={\dis
C(\phi_{D},\phi_{A},\lambda_0)(2t)^{-\frac{(\nu(\Lambda)+1)^2}{2}}
e^{\psi(\Lambda)+\tau(\Lambda)}}\non
&&\hspace{-6mm}{\dis\times\exp\left\{
\frac{1}{2\pi}\stint\biggl(x-2\lambda
t+i\psi'(\lambda)\biggr)\sign(\Lambda-\lambda)\right.}\non
&&\hspace{-6mm}{\dis\left.\vphantom{\stint}\times
\ln\Bigl\{1-\vartheta(\lambda)
\biggl(1+e^{\phi(\lambda)\sign(\lambda-\Lambda)}\biggr)\Bigr\}
\,d\lambda\right\}\Bigl(1+{\cal O}(\ln^{2}t / t)\Bigr).}
\ea

In the case of positive chemical potential the situation is quite
similar. The asymptotics of ${\cal B}$ is given by the equation
\ba{MAresult2}
&&\hspace{-11mm}
{\cal B}={\dis
\tilde C(\phi_{D},\phi_{A},\lambda_0)
(2t)^{-\frac{\nu^2(\Lambda)}{2}}
e^{\psi(\Lambda_1)+\tau(\Lambda_1)}}\non
&&\hspace{-6mm}{\dis\times\exp\left\{
\frac{1}{2\pi}\int_\Gamma\biggl(x-2\lambda
t+i\psi'(\lambda)\biggr)\sign(\Lambda-\lambda)\right.}\non
&&\hspace{-6mm}{\dis\left.\vphantom{\stint}\times
\ln\Bigl\{1-\vartheta(\lambda)
\biggl(1+e^{\phi(\lambda)\sign(\lambda-\Lambda)}\biggr)\Bigr\}
\,d\lambda\right\}\Bigl(1+{\cal O}({1}/{\sqrt t})\Bigr).}
\ea
The general structure of this result is the same as for negative
chemical potential; namely, we have the exponential law, the power law, and
the constant factors. However, all these factors slightly differ from
the ones in \eq{MAresult1}. First, the integral in the exponent is taken
along the contour $\Gamma$ (see Fig.\ref{deform}). 
Secondly, we have the sum
$\tau(\Lambda_1)+\psi(\Lambda_1)$ instead of $\tau(\Lambda)+
\psi(\Lambda)$. Thirdly, the pre-exponent powers of $t$ also are 
different. The reasons of these difference were explained in the 
sections \ref{F} and \ref{PCP}.

In conclusion we would like to touch briefly the free fermionic
limit. In this limit the coupling constant $c$ goes to infinity, and
one can put all dual fields equal to zero. In this case Eqs.
\eq{MAresult1}, \eq{MAresult2} exactly reproduce the results of
\cite{IIKV}.

\section*{Summary}

We study the operator-valued RHP \eq{Fovrhp}, \eq{Fjumpmatreg}. 
This RHP allows to find
the large time and long distance asymptotics of the Fredholm
determinant, which in turn describes the temperature correlation
function of the local fields of the QNLS model out off free-fermionic point.
Although we use the
advanced technique based on the nonlinear steepest descent
method \cite{DZ}, the main idea of the scheme
applied in the present paper mostly coincides with the approach
considered in \cite {IIKV} for the free fermionic limit of the
model. However, as we saw, the operator-valued RHP is
essentially more complicated, and not only from the technical
point of view. In particular, the solution of the RHP was given up to
the function $\h\varrho(\lambda,u,v)$, which is defined as a
solution of a certain integral equation..
At the same time, in spite of this incompleteness, we were able to
obtain explicit expressions for the exponential and power laws of
the operator ${\cal B}$.

A special problem arises because of the dual fields. Due to the presence
of these quantum operators one needs to take care not only of the
asymptotics of the operator $\cal B$ itself, but also of the
asymptotics of its vacuum expectation value. This leads us to the
non-standard asymptotic analysis, which includes the shift of the
saddle point.

The remaining stage in the calculation of the correlation function is
the averaging of the results \eq{MAresult1}, \eq{MAresult2} with
respect to the auxiliary vacuum. The method for this averaging was
developed in \cite{KS2}, although in that paper the contribution of
some corrections were not considered. Nevertheless, one can
definitely state that the calculation of the vacuum expectation value
leads to nonlinear integral equations, closely related to the
equations of the Thermodynamic Bethe Anzats. The final result for the
asymptotics of the correlation function can be formulated in terms of
the solutions of these equations. We shall consider these questions
in the forthcoming publications.

\section*{Aknowledgments}

We would like to thank V.~E.~Korepin for useful discussions. This
work was supported in parts by NSF Grant N0. DMS-9801608,
RFBR Grant No. 96-01-00344 and INTAS-01-166-ext.

\appendix
\section{The Fredholm determinant}

The complete determinant representation for the correlation function
\eq{Itempcorrel} has the form
\be{A1detrep}
\langle\Psi(0,0)\Psi^\dagger(x,t)\rangle_T=
-\frac{e^{-iht}}{2 \pi}\brad
\frac{\det\left(\tilde{I}+\widetilde{V}\right)}
{\det\left(\tilde{I}-\frac{1}{2\pi}{\widetilde K}_T\right)}
\displaystyle \int_{-\infty}^{\infty}
\hb_{12}(u,v)du dv \ketd.
\ee
The integral operator $\tilde I+\tV$ acts on the real axis as
\be{A1kernel}
\left(\tilde{I}+\widetilde{V}\right)\circ f(\mu)
=f(\lambda)+\displaystyle \int_{-\infty}^{\infty}\tV(\lambda,\mu)
f(\mu)d\mu,
\ee
where $f(\lambda)$ is some trial function. The kernel
$\tV(\lambda,\mu)$ is equal to
\be{A1form}
\tV(\lambda,\mu)=\frac 1{\lambda-\mu}
\stint\,du(E_+(\lambda|u)E_-(\mu|u)-E_-(\lambda|u)E_+(\mu|u)).
\ee
Here
\ba{eplus}
&&{\dis
E_+(\lambda|u)=\frac{1}{2\pi}\frac{Z(u,\lambda)}{Z(u,u)}
\left(\frac{e^{-\phi_{A}(u)}}{u-\lambda+i0}
+\frac{e^{-\phi_{D}(u)}}{u-\lambda-i0}\right)
\sqrt{\vartheta(\lambda)}}\nona{19}
&&\hskip3cm{\dis \times
e^{\psi(u)+\tau(u)+\frac12(
\phi_{D}(\lambda)+\phi_{A}(\lambda)-\psi(\lambda)-\tau(\lambda))},}
\ea
\hskip1mm
\be{eminus}
E_-(\lambda|u)=\frac{1}{2\pi}Z(u,\lambda)
e^{\frac12(\phi_{D}(\lambda)+\phi_{A}(\lambda)
-\psi(\lambda)-\tau(\lambda))}\sqrt{\vartheta(\lambda)},
\ee

The representation \eq{A1detrep} contains also one more
Fredholm determinant, however the last one does not depend on
time $t$ and distance $x$, as well as on dual fields:
\be{A1kernel1}
\widetilde K_T(\lambda,\mu)=\frac{2c}
{(\lambda-\mu)^2+c^2}\sqrt{\vartheta(\lambda)\vartheta(\mu)}.
\ee

Starting from the kernel $\tV(\lambda,\mu)$ one can construct the
jump matrix of the operator-valued RHP, using the standard procedure
\cite{KS1}, \cite{KBI}
\be{A1jump1}
\h G^{in}(\lambda)=\h I+2\pi i
\left(\begin{array}{cc}
-E_+(\lambda|u)E_-(\lambda|v)&
E_+(\lambda|u)E_+(\lambda|v)\non
-E_-(\lambda|u)E_-(\lambda|v)&
E_-(\lambda|u)E_+(\lambda|v)
\end{array}\right).
\ee
In order to obtain the jump matrix \eq{Fjumpmat}, one need to make
the following transformation
\be{A1G}
\h G(\lambda)=\h\chi^{0}_+(\lambda)\h G^{in}(\lambda)
\left(\h\chi^{0}_-(\lambda)\right)^{-1},
\ee
where
\be{A1chi}
\h\chi^{0}(\lambda|u,v)=\h I+
\frac{e^{\psi(u)+\tau(u)}}{\lambda-u}
\left(\begin{array}{cc}
0&1\\0&0
\end{array}\right).
\ee
After this transformation we arrive at the RHP $(a)$, considered in
the section \ref{F}.

\section{The regularized integral operator}

The regularization of the RHP considered in the section \ref{F}
corresponds to new Fredholm determinant $\det(\tilde I+
\widetilde V_\epsilon)$. The kernel $\widetilde V_\epsilon(
\lambda,\mu)$ of new integral operator has just the same form as the
original one
\be{A2kernel}
\widetilde V_\epsilon(\lambda,\mu)=\frac{1}{\lambda-\mu}
\stint\,du(E^\epsilon_+(\lambda|u)E^\epsilon_-(\mu|v)-
E^\epsilon_-(\lambda|u)E^\epsilon_+(\mu|v)).
\ee
Here
\be{A2Emin}
E^\epsilon_-(\lambda|u)=\frac{1}{2\pi}Z(u,\lambda)
Z(\lambda,\lambda)\sqrt{\frac{\vartheta(\lambda)}{\Ne(\lambda)}}
e^{\frac{1}{2}(\phi_A(\lambda)+
\phi_D(\lambda)-\psi(\lambda)-\tau(\lambda))},
\ee
\vskip3mm
\be{A2Eplus}
E^\epsilon_+(\lambda|u)=E^\epsilon(\lambda|u)
E^\epsilon_-(\lambda|u),
\ee
and
\ba{A2E}
&&{\dis\hspace{-5mm}
E^\epsilon(\lambda|u)=\frac{1}{Z(u,\lambda)}
\stint\,d\xi dw\frac{e^{\psi(\xi)+\tau(\xi)}}
{\Ne(\xi)Z(\xi,\xi)}\delta_\epsilon(u-\xi)
\delta_\epsilon(w-\xi) }\non
&&{\dis\hspace{10mm}\times
Z(u,\xi)Z(w,\xi)Z(w,\lambda)
\left[\frac{e^{-\phi_D(\xi)}}{\xi-\lambda-i0}+
\frac{e^{-\phi_A(\xi)}}{\xi-\lambda+i0}\right].}
\ea
It is easy to see that the limit $\epsilon\to 0$ is well defined in
all formul\ae. In particular $\widetilde V_\epsilon\to\widetilde V$.

The original jump matrix $\h G^{in}$  has just the same structure as
the matrix \eq{A1jump1}
\be{A2jump1}
\h G^{in}(\lambda)=\h I+2\pi i
\left(\begin{array}{cc}
-E^\epsilon_+(\lambda|u)E^\epsilon_-(\lambda|v)&
E^\epsilon_+(\lambda|u)E^\epsilon_+(\lambda|v)\non
-E^\epsilon_-(\lambda|u)E^\epsilon_-(\lambda|v)&
E^\epsilon_-(\lambda|u)E^\epsilon_+(\lambda|v)
\end{array}\right).
\ee
In order to obtain the jump matrix \eq{Fjumpmatreg} one need to
make transformation similar to \eq{A1G} with matrix
\be{A2chi}
\h\chi^{0\epsilon}(\lambda|u,v)=
\h I+\stint\frac{d\xi}{\lambda-\xi}
\delta_\epsilon(u-\xi)\delta_\epsilon(v-\xi)Z(u,\xi)Z(v,\xi)
e^{\psi(\xi)+\tau(\xi)}
\left(\begin{array}{cc}
0&1\\0&0
\end{array}\right).
\ee

The expressions of the logarithmic derivatives of the regularized
Fredholm determinant in terms of the coefficients $\h b$ and $\h c$
\eq{Blogdir1}, \eq{Blogdir2} can be obtained in the same manner, as
it had been done in \cite{KS1}. The differential equations \eq{Bopeq}
also are obvious corollaries of the regularized RHP.



\end{document}